\documentclass[12pt]{article}
\usepackage{jheppub_modified_green}
\usepackage{epsfig}
\usepackage{amsfonts}
\usepackage{amssymb}
\usepackage{enumitem}
\usepackage{amsthm}
\usepackage{subfig}
\usepackage{bm}
\usepackage{hyperref}
\usepackage{mathrsfs}
\usepackage{bbm}
\usepackage{cancel}
\usepackage{xcolor}
\usepackage{accents}
\usepackage{relsize}
\usepackage{float, slashed, graphicx, amssymb, amsmath}
\usepackage{shuffle}
\usepackage{tikz}

\usepackage{tikz-feynman}
\tikzfeynmanset{compat=1.1.0}
\tikzfeynmanset{
graviton/.style={circle, draw=green!60, fill=green!5, very thick, minimum size=7mm}
}
\tikzfeynmanset{myblob/.style=
{/tikz/shape=rectangle,
/tikz/fill=red,
 /tikz/minimum size=0.3cm,
 } }
 \tikzfeynmanset{HV/.style=
{/tikz/shape=circle,
/tikz/fill={rgb:black,1;white,2},
 /tikz/minimum size=0.1cm,
 } }
  \tikzfeynmanset{GR/.style=
{/tikz/shape=ellipse,
/tikz/fill={rgb:black,1;white,2},
 /tikz/minimum size=0.3cm,
 } }
\tikzset{box/.pic={\filldraw[fill=black]  (0,0) circle (2.5pt);
				   \filldraw [fill=black] (0.5,0) circle (2.5pt);
			       \draw [line width=5pt] (0,0) -- (0.5,0);}}

\usepackage[T1]{fontenc} 

\makeatletter
\newcommand \UPlus {\mathop {\operator@font \uplus }\limits }
\makeatother
\makeatletter
\newcommand \Bigcup {\mathop {\operator@font \bigcup }\limits }
\makeatother
  \def\LabelNote#1{}
 \def\Label#1{\label{#1}%
  \smash{\hbox to0pt{\raise1ex\hbox{\tiny[#1]}\hss}}}
  
  \def\Cdot{{\cdot}}
  
\def\nn{\nonumber}

\newcommand{\black}{\color{black}}
\newcommand{\blue}{\color{blue}}

\newcommand{\cI}{\mathcal{I}}
\newcommand{\cM}{\mathcal{M}}

\newcommand{\cO}{\mathcal{O}}

\newcommand{\eps}{\epsilon}
\newcommand{\arccosh}{{\rm arccosh}}

\def\spa#1.#2{\left\langle#1\,#2\right\rangle}
\def\spb#1.#2{\left[#1\,#2\right]}

\def\be{\begin{equation}}
\def\ee{\end{equation}}
\def\bea{\begin{eqnarray}}
\def\eea{\end{eqnarray}}  
\allowdisplaybreaks

\newcommand{\neco}{\mathbb{L}}          
\newcommand{\npre}{\mathcal{N}}

\newcommand{\lamp}{\mathcal{M}}

\title{Classical gravitational scattering  from  a gauge-invariant double copy}
\author{Andreas Brandhuber,}
\author{Gang Chen,}
\author{Gabriele Travaglini}
\author{and Congkao Wen}
\affiliation{Centre for Theoretical Physics, Department of Physics and Astronomy, \\
Queen Mary University of London, Mile End Road, London E1 4NS, United Kingdom}
\emailAdd{a.brandhuber@qmul.ac.uk}
\emailAdd{g.chen@qmul.ac.uk}
\emailAdd{g.travaglini@qmul.ac.uk}
\emailAdd{c.wen@qmul.ac.uk}

\begin{document} 
\begin{flushright}
	QMUL-PH-21-18\\
	SAGEX-21-07\\
\end{flushright}


\abstract{
We propose a method to compute the  scattering angle  for classical black hole scattering 
directly from two massive particle irreducible diagrams in a heavy-mass effective field theory approach to general relativity, without the need of subtracting  iteration terms. 
The amplitudes in this effective  theory are constructed using a recently proposed novel
colour-kinematic/double copy for tree-level two-scalar, multi-graviton amplitudes, where  the BCJ numerators are  gauge invariant and local with respect to the massless gravitons. These tree amplitudes, together with graviton tree amplitudes, enter the construction of the required $D$-dimensional loop integrands and allow for a direct extraction of contributions relevant for classical physics. In particular the  soft/heavy-mass  expansions of full integrands is circumvented, and all iterating   contributions can be dropped from the get go.  We use this method to  compute the scattering  angle up to third post-Minkowskian order in four dimensions, including radiation reaction contributions, also providing the expression of the corresponding  integrand in $D$ dimensions.  
}

\maketitle
\flushbottom
\section{Introduction}

 The goal of this paper is to propose an efficient method to compute the scattering angle in a  collision of two black holes. To achieve this, we bring together a number of crucial ingredients which we now describe. 
 
 We begin by observing that in a scattering process, the heavy black holes can be represented as pointlike particles, exchanging momenta which are much smaller than their masses. In practice, in order to extract classical physics  such as a deflection angle, one rescales the momentum of the exchanged gravitons $q$ as \cite{Kosower:2018adc} $q= \hbar k$, with the wavevector $k$ being kept fixed while taking the classical limit $\hbar \to 0$.
 This motivates the use of effective field theory applied to gravity \cite{Donoghue:1994dn}, and in particular of 
 a heavy-mass  effective field theory (HEFT) \cite{Georgi:1990um, Luke:1992cs, Neubert:1993mb, Manohar:2000dt,Damgaard:2019lfh}, which is  the appropriate  tool if one wishes to describe interactions of particles where the exchanged momenta are much smaller than their masses.  Working from the outset with a simplified theory is  our first crucial ingredient  -- the great advantage is that this avoids the  soft (or heavy-mass) expansions of the integrands that would be obtained starting from Einstein gravity coupled to matter. 
 The  scattering angle is then  extracted from the computation of the elastic scattering amplitude of the two black holes, an approach that was initiated fifty years  ago in \cite{Iwasaki:1971vb} (specifically for the computation of the Newton potential), and then  pursued at 2PM order in \cite{Neill:2013wsa,Bjerrum-Bohr:2013bxa, Bjerrum-Bohr:2014zsa, Bjerrum-Bohr:2016hpa} in conjunction with the unitarity method \cite{Bern:1994zx, Bern:1994cg}. This has been applied 
successfully in several works in general relativity at 3PM \cite{Bern:2019nnu,Bern:2019crd,Parra-Martinez:2020dzs,Bjerrum-Bohr:2021din,DiVecchia:2021bdo}, including radiation \cite{Herrmann:2021lqe,Herrmann:2021tct, Jakobsen:2021smu, Jakobsen:2021lvp},   4PM  \cite{Bern:2021dqo},
and also in theories with higher-derivative interactions \cite{Brandhuber:2019qpg, Emond:2019crr, AccettulliHuber:2020oou,AccettulliHuber:2020dal,Carrillo-Gonzalez:2021mqj} to compute both corrections to Newton's potential and deflection angles. 
An  important  simplification stems from the fact that only  
terms in the four-point scattering amplitude with a  discontinuity in the $q^2$-channel need to be computed, as these contribute to long-distance effects \cite{Donoghue:1994dn, Holstein:2004dn}, which makes the unitarity-based method particularly well suited for this task. These amplitude techniques have emerged as potential alternatives to a host of other methods such as  the effective one-body formulation \cite{Buonanno:1998gg,Damour:2016gwp,Damour:2017zjx,Damour:2019lcq} and worldline theories \cite{Melville:2013qca,Kalin:2020mvi,Kalin:2020fhe,Mogull:2020sak,Dlapa:2021npj,Loebbert:2020aos}, 
which have led to a wealth of important results 
\cite{Damour:1985mt,BjerrumBohr:2002ks,BjerrumBohr:2002kt, Gilmore:2008gq,Damour:2001bu, Blanchet:2003gy, Itoh:2003fy,Foffa:2011ub,Jaranowski:2012eb, Damour:2014jta,Galley:2015kus,Damour:2015isa, Damour:2016abl, Bernard:2015njp, Bernard:2016wrg, Foffa:2012rn, Foffa:2016rgu,Porto:2017dgs,Porto:2017shd,Foffa:2019yfl,Blumlein:2020pog,Foffa:2019hrb,Blumlein:2019zku,Cheung:2020gyp,Bini:2020wpo,Blumlein:2020pyo,Blumlein:2020znm,Bini:2020uiq,Blumlein:2021txj}
also including spin effects
\cite{Porto:2005ac,Steinhoff:2010zz,Levi:2014gsa,Levi:2015msa,Vines:2017hyw,Maia:2017yok,Levi:2018nxp,Chung:2018kqs,Bern:2020buy,Bautista:2021wfy,Chung:2020rrz,Levi:2020uwu,Levi:2020lfn,Kosmopoulos:2021zoq,Chiodaroli:2021eug}.

In order to be able to compute efficiently, one needs compact  expressions for the tree amplitudes entering the unitarity cuts, and here comes the second key ingredient of our procedure. In \cite{Brandhuber:2021kpo} we have developed a systematic approach to derive compact, $D$-dimensional expressions of HEFT amplitudes  with two massive scalars and an arbitrary number of massless gravitons at leading order in an inverse mass expansion. This is 
based  on a new form of the colour-kinematics/double-copy duality  \cite{Bern:2008qj,Bern:2010ue} for heavy-mass effective field theories,  initially proposed in Yang-Mills and gravity in  \cite{Chen:2019ywi}, and further developed in \cite{Chen:2021chy}. The key advantage of this method, compared e.g.~to the earlier work of \cite{Haddad:2020tvs}, is that it provides BCJ numerators that automatically satisfy  the Jacobi relations and crossing symmetry. Furthermore, only a subset of all possible cubic graphs contribute to these numerators, which in addition are  manifestly gauge invariant (since they can be written in terms of field strengths) and also local with respect to the massless gluons or, in the double-copied theory,  gravitons. This is in contradistinction with the double copy for Yang-Mills amplitudes, where the BCJ numerators are in general not gauge invariant. Thus, our novel double copy provides us with particularly compact tree amplitudes which can be fed into the cuts to construct equally compact integrands. 

The third key simplification in our approach is of a diagrammatic nature. Specifically, we propose that the scattering angle can be computed   from a subset of all possible diagrams in the HEFT, namely  only two-massive particle irreducible (2MPI) diagrams, discarding all the others. From this 2MPI amplitude $\cM_{\rm HEFT}^{\rm 2MPI}$, one can then construct a corresponding HEFT phase $\delta_{\rm HEFT}$ by simply transforming it to impact parameter space, and finally extract from it the scattering angle $\chi$ as 
\begin{align}
\chi=-\frac{\partial}{\partial J}\text{Re}\big[\delta_{\rm HEFT}\big]\, , 
\end{align}
where $J$ is the total angular momentum of the system.

This  procedure should be contrasted with the usual eikonal method 
\cite{Levy:1969cr, Amati:1990xe, Kabat:1992tb} -- there the $S$-matrix is believed to exponentiate in impact parameter space, but starting from two loops, the computation of the eikonal phase  becomes a delicate task. Conventional unitarity tackles the computation of $S$-matrix elements, and to extract  the eikonal phase at a certain perturbative order one has to remove all terms which reconstruct the exponentiation of the phase at lower loop orders in perturbation theory. Clearly it is  highly desirable to be able to tackle the computation of the phase  directly without having to evaluate a plethora of iterating terms which pollute the phase, and this is what our proposal provides.  Important work in this direction was done in \cite{Bern:2021dqo}, where remarkably  the conservative part of the potential at 4PM  was computed from the radial action. Related ideas were presented in the recent  interesting paper \cite{Damgaard:2021ipf}, which 
proposed a different  exponential representation of the $S$-matrix inspired by the WKB formalism applied to quantum field theory, differing from the eikonal one. In particular, the authors of  \cite{Damgaard:2021ipf} were able to write down expressions of the iterating terms which have to be removed from the complete $S$-matrix elements (whose calculation is still needed) in order to compute 
matrix elements of $N$, where $S:= e^{\frac{i}{\hbar} N}$. 
This still requires knowledge of  complete amplitudes, but it cleanly  explains their structure.  
Instead, we propose an approach where by computing only a subset of all possible diagrams one can efficiently extract  the scattering angle avoiding entirely   the computation of iterating diagrams. 
 
 We test our proposal by computing the scattering angle at two loops, or 3PM, also including radiation reaction terms. Our seed tree amplitudes are $D$-dimensional, hence we can produce integrands valid in an arbitrary number of dimensions. We use integration by parts (IBP)  relations \cite{gehrmann2000differential,Smirnov:2008iw,larsen2016integration,lee2015reducing} to expand our result in a basis of seven master integrals (in the four-dimensional case), which we then integrate using the method of differential equations \cite{Henn:2013pwa} in conjunction with a study of the boundary conditions based on 
 the asymptotic expansion of Feynman integrals \cite{Beneke:1997zp,Pak:2010pt}. 
Our result for the deflection angle  agrees with  \cite{Bern:2019nnu, Antonelli:2019ytb,DiVecchia:2021bdo,Bjerrum-Bohr:2021din}, and specifically the   radiation reaction part agrees with    \cite{Damour:2020tta,DiVecchia:2020ymx,Herrmann:2021tct,Bjerrum-Bohr:2021din}.

The rest of the paper is organised as follows. Sections~\ref{sec:2} and \ref{sec:HEFTrev} contain a brief review of the kinematics of the scattering process and of the tree-level HEFT amplitudes derived in \cite{Brandhuber:2021kpo}, which will then be needed to perform our two-loop unitarity cuts, respectively.  In Section~\ref{sec:4} we discuss in detail the HEFT expansion and our diagrammatic method, in particular introducing the 2MPI diagrams needed for the computation of the scattering angle at one and two loops. Sections~\ref{sec:1loopheft}  and \ref{sec:2loopheft} describe the calculation of the HEFT amplitudes at one and two loops. In particular, in Section~\ref{sec:2loopheft} we provide the integrand of the two-loop HEFT amplitude in $D$ dimensions in terms of a basis of eight master integrals, with only  seven contributing around four dimensions. 
We evaluate the master integrals  in Section~\ref{sec:canbasis}, and  present our final result for the (integrated)  2MPI HEFT amplitude, the HEFT phase $\delta_{\rm HEFT}$ and the scattering angle in Section~\ref{sec:8}. In  Section~\ref{sec:all-loop} we present a conjecture valid in $D$ dimensions for the single diagram contributing to the probe limit for an arbitrary number of loops. Finally, we present our conclusions and give a brief outlook in Section~\ref{sec:10}.
Two appendices complete the paper: in Appendix~\ref{sec:TGAsy} we present a study of the asymptotic behaviour of our master integrals and the associated boundary conditions, with explicit examples for our one- and two-loop master integrals; and in Appendix~\ref{sec:Integrand2} we show explicitly the integrand of the four 2MPI cut diagrams contributing to the two-loop 2MPI HEFT amplitude.

\section{Kinematics of the scattering process}
\label{sec:2}

We begin by  discussing the kinematics of the $2 \to 2$ scattering process.
We   choose the particle momenta so that $p_1^2 = p_4^2 = m_1^2$, $p_2^2 = p_3^2 = m_2^2$, where  particles 1 and 2 are incoming, while  particles 4 and 3 are the outgoing deflected particles (with mass $m_1$ and $m_2$, respectively).
We   parameterise the external momenta in the centre of mass frame, hence  the momentum exchange $q$ is spacelike, $q^\mu = (0,\vec{q})$. Specifically, the momenta will be parameterised as follows: 
\begin{equation}
\label{kinematics}
\begin{array}{lr}
\begin{tikzpicture}[scale=12,baseline={([yshift=-1mm]centro.base)}]
\def\x{0}
\def\y{0}
\node at (0+\x,0+\y) (centro) {};
\node at (-3pt+\x,-3pt+\y) (uno) {$p_1$};
\node at (-3pt+\x,3pt+\y) (due) {$p_2$};
\node at (3pt+\x,3pt+\y) (tre) {$p_3$};
\node at (3pt+\x,-3pt+\y) (quat) {$p_4$};
\draw [double] (uno) -- (centro);
\draw [double] (due) -- (centro);
\draw [double] (tre) -- (centro);
\draw [double] (quat) -- (centro);
\draw [->] (-2.8pt+\x,-2pt+\y) -- (-1.8pt+\x,-1pt+\y); 
\draw [->] (1.8pt+\x,-1pt+\y) -- (2.8pt+\x,-2pt+\y) ; 
\draw [->] (-2.8pt+\x,2pt+\y) -- (-1.8pt+\x,1pt+\y)  ; 
\draw [->] (1.8pt+\x,1pt+\y) -- (2.8pt+\x,2pt+\y); 
\node at (0+\x,0+\y) [draw, fill=gray!90!black, circle, inner sep=10pt] {};
\shade [shading=radial] (centro) circle (1.5pt);
\end{tikzpicture}
&\hspace{2cm}
\begin{aligned}
p_1^\mu & := \bar{p}^\mu_1 + \tfrac{q^\mu}{2}  = (E_1,\vec{p}+\vec{q}/2)\, ,  \\
 p_2^\mu & := \bar{p}^\mu_2 - \tfrac{q^\mu}{2} = (E_2,-\vec{p}-\vec{q}/2) \, , \\
p_3^\mu & :=  \bar{p}^\mu_2 + \tfrac{q^\mu}{2} = (E_3,-\vec{p}+\vec{q}/2) \, ,  \\
p_4^\mu & := \bar{p}^\mu_1 - \tfrac{q^\mu}{2}  =  (E_4,\vec{p}-\vec{q}/2)\ .
\end{aligned}
\end{array}
\end{equation}
Since we are considering elastic scattering, we have 
\begin{align}
\label{ener}
\begin{split}
E_1 & = E_4 = \!\sqrt{m_1^2 +\vec{p}^{\, \,2}+\vec{q}^{\, \,2}/4}\ , 
\\
E_2 & = E_3 = \!\sqrt{m_2^2 +\vec{p}^{\, \, 2}+\vec{q}^{\, \,2}/4}\ , 
\end{split}
\end{align}
 where  $\vec{p} \, \Cdot \, \vec{q}=0$ due to momentum conservation.  It is also easy to see  that 
 \begin{align}
 \bar{p}_1\Cdot q = \bar{p}_2\Cdot q = 0 
 \ ,
 \end{align}
 which means that only the $\bar{p}_i$ momenta are strictly orthogonal to the momentum transfer $q$, while e.g.~$p_1 \Cdot q = q^2/2$. This will be important in the soft/heavy-mass  expansion discussed later.
  Furthermore, our Mandelstam variables are defined as
\begin{align}
\label{mandel}
s:=(p_1+p_2)^2 = (E_1+E_2)^2 := E^2  , \qquad q^2:=(p_1+p_4)^2 =  -\vec{q}^{\, \, 2} \   . 
\end{align}
We also define 
\begin{align}
 \bar{p}_1^2  = \bar{p}_4^2 := \bar{m}_1^2 = m_1^2-\frac{q^2}{4} \ , \qquad 
\bar{p}_2^2  = \bar{p}_3^2 := \bar{m}_2^2 = m_2^2-\frac{q^2}{4} \ .
\end{align}
Furthermore, it is useful to introduce the relativistic velocities of the particles via $p_i := m_i v_i$, with $i=1,2$ and $v_i^2=1$, and their
scalar product 
\begin{align}
\label{y}
y := v_1 \Cdot v_2 = \frac{p_1 \Cdot p_2}{m_1 m_2}\, , 
\end{align}
which is related to the centre of mass energy as
\begin{align}
s = m_1^2+m_2^2 + 2 m_1 m_2\,  y \ .
\end{align}
Similarly, one can define $\bar{p}_i := \bar{m}_i \bar{v}_i$ and $\bar y := \bar v_1 \Cdot \bar v_2$, with $\bar{v}_i^2=1$.
 
Finally, the deflection angle $\chi$ is related to the external kinematics
as 
\begin{equation}\label{2.6}
\sin \left( \frac{\chi}{2} \right) = \frac{ |\vec{q}\, |}{ 2 P } \ ,
\end{equation}
where $P = \sqrt{\vec{p}^{\, \, 2}+\vec{q}^{\, \, 2 }/4}$ is the common modulus of the incoming and outgoing three-momenta. This quantity
can also be expressed as
\begin{align}
\label{Pdef}
P = \frac{m_1 m_2 \sqrt{y^2-1}}{E} \ ,
\end{align}
and using this relation  we can rewrite \eqref{2.6} as
\begin{align}
\sin \left( {\chi \over 2} \right) = { |\vec{q}\, | \, E \over 2 m_1 m_2 \sqrt{y^2-1} } \ .
\end{align}
It is also  useful to introduce the conserved total angular momentum 
\begin{align}
\label{Jdef}
    J = P b\, , 
    \end{align}
    where $b$ is the asymptotic impact parameter of the scattered objects. Its inverse $J^{-1}$ is commonly used as an expansion parameter of physical observables such as the deflection angle $\chi$ (see for instance our final result \eqref{finalangle} for this quantity up to 3PM).

\section{Relevant HEFT amplitudes from the new double copy}
\label{sec:HEFTrev}
In this section we summarise the  tree-level  graviton-matter  amplitudes to leading order in an inverse mass expansion, which     enter the computation of gravitational scattering of two heavy scalars through unitarity cuts. We call these quantities   {\it ``HEFT amplitudes''}. 
These have been derived and described in detail in the companion paper \cite{Brandhuber:2021kpo} and correspond to the leading terms in the heavy-mass expansion, which are the only ones relevant for classical physics. In particular, in \cite{Brandhuber:2021kpo} it was shown that they exhibit a novel double-copy structure with manifestly gauge-invariant numerators. This feature, combined with their very compact forms, makes them ideally suited for the purpose of this paper.

The three-point YM-matter amplitude are directly obtained from the Feynman rule and the GR-matter amplitude is the square of the YM-matter amplitude,
\begin{align}\label{eq:threeAmp}
	A^{\rm YM-M}_3(2,v)
 = m \, \varepsilon_2\Cdot v  \ , && A^{\rm GR-M}_3(2,v)=m^2 ( \varepsilon_2\Cdot v)^2.
\end{align}
At 2PM order we need the two-scalar two-graviton amplitude, which  can easily be obtained  using our double copy from the corresponding YM amplitude: 
\begin{align}\label{4ptYMGR}
	A^{\rm YM-M}_4(23,v)={m\over s_{23}}\Big(\frac{v\Cdot F_2\Cdot F_3\Cdot v}{v\Cdot p_3}\Big), && A^{\rm GR-M}_4(23,v)={m^2\over s_{23}}\Big(\frac{v\Cdot F_2\Cdot F_3\Cdot v}{v\Cdot p_3}\Big)^2, 
\end{align}
where $F$  denotes the field strength $F_i^{\mu\nu}=p_i^{\mu}\varepsilon_i^{\nu}-p_i^{\nu} \varepsilon_i^{\mu}$. Up to four points, there is  only one cubic graph for each of these amplitudes.%
\footnote{\label{foot}  In the following we omit a factor of $(- i \kappa)^n$ from all graviton-matter and graviton amplitudes. These factors will be reinstated from  Section~\ref{sec:1loopheft} onward. Note that in our conventions $\kappa^2 := 32 \pi G_N$.}

At 3PM we need amplitudes up to five points. In this case there  are two cubic graphs for the colour-ordered YM-matter amplitude and three cubic graphs for the GR-matter amplitude. Then the colour-ordered YM-matter amplitude  and the GR-matter amplitude in  the novel colour-kinematics duality form are 
\begin{align}
	A^{\rm YM-M}_5(234,v)={\npre_5([[2,3],4],v)\over s_{234} s_{23}} +{\npre_5([2,[3,4]],v)\over s_{234} s_{34}}\, , 
\end{align}
and, by the double copy,  
\begin{align}
\begin{split}
\label{A5new}
A^{\rm GR-M}_5(234,v)&=
{\big[\npre_5([[2,3],4],v)\big]^2 \over s_{234} s_{23}} + {\big[\npre_5([[2,4],3],v)\big]^2 \over s_{234} s_{24}}+{\big[\npre_5([[3,4],2],v)\big]^2 \over s_{234} s_{34}} \ .
\end{split}
\end{align}
An advantage  of this double copy is that the BCJ numerator can be  written in a manifestly gauge-invariant form. For the five-point case
\begin{align}\label{eq:FiveBCJNum}
	\npre_5([[2,3],4],v)
	&= \mathbb{L}(2,3,4)\circ \Big[m {v\Cdot F_2\Cdot F_3\Cdot V_3\Cdot F_4\Cdot v\over v\Cdot p_3 v\Cdot p_{4}}\Big]\, , 
\end{align}
where $V_i^{\mu\nu}=v^{\mu}p_i^\nu$. The operator  $\mathbb{L}(2,3,4,\ldots, n-1)$ was  introduced  in \cite{Chen:2019ywi,Chen:2021chy} as a group algebra  element 
\cite{linckelmann2018block,algebra}
and  in \cite{reutenauer2003free} as a free Lie algebra element, 
\begin{align}
\neco(i_1, i_2,\ldots, i_r):= \Big[\mathbb{I}-\mathbb{P}_{(i_2i_1)}\Big]\Big[\mathbb{I}-\mathbb{P}_{(i_3i_2i_1)}\Big]\cdots \Big[ \mathbb{I}-\mathbb{P}_{(i_r\ldots i_2i_1)}\Big]\, ,
\end{align}
where  $\mathbb{P}_{(j_1 j_2 j_3 \ldots j_m)}$  denotes the cyclic permutation   $j_1\to j_2 \to j_3\to \cdots \to j_m\to j_1$. For instance 
\begin{align}
\begin{split}
	\mathbb I \circ\npre_6(2345, v)&= \npre_6(2345,v )\, , \\
	\black \mathbb P_{(432)} \circ \npre_6(2345,v)&=\npre_6(4235,v)\, .
\end{split}
\end{align}
In \cite{Brandhuber:2021kpo} we also presented the six-point HEFT amplitudes, which are  key ingredients for
computations at 4PM order which were completed only very recently for the conservative dynamics  \cite{Bern:2021dqo,Dlapa:2021npj}. Using the method of \cite{Brandhuber:2021kpo}  it is straightforward to
produce the HEFT amplitudes needed at higher PM orders. For instance, the $n$-point HEFT amplitude enters the probe-limit 
contribution at $(n-3)$-loop order (or $(n-2)$PM), as explained in Section~\ref{sec:all-loop}. 
This contribution may also   be computed from the geodesic motion of a test particle, whereas the contributions beyond the probe limit only require HEFT amplitudes up to $n-1$ points.  Hence,  all ingredients needed at 5PM are already available with the currently known HEFT amplitudes.

The HEFT amplitudes  scale universally with the mass as $m^2$, where  terms proportional to delta functions (arising from the  $i \varepsilon$ prescription) have been dropped in the expansion of the full amplitudes\footnote{This is sufficient for generic kinematics at tree level, but when using these amplitudes in cuts these terms need to be included as they contribute in loop integrations.}
-- importantly, these {\it contact terms} are captured by products of lower-point HEFT amplitudes and delta functions, and are essential to construct loop integrands. This is at the heart of our HEFT expansion and will be discussed in detail in Section~\ref{sec:4.1}.
In the following we denote  the HEFT amplitudes as
$A_n(2\ldots(n{-}1), v)$, where from now on we omit the superscript  in $A$ as in this paper we only focus on Einstein gravity. The mass dependence of the HEFT amplitude can be read off from the   label $v$ or $\bar{v}$ (e.g.~$A_n(2\ldots(n{-}1), v)\propto m^2$, and  $A_n(2\ldots(n{-}1), \bar v)\propto \bar m^2$).

We conclude this section with a few comments. Our HEFT amplitudes are valid in a generic number of dimensions $D$ and are manifestly gauge invariant, which brings a number of advantages. First, the unitarity cuts we perform are $D$-dimensional, avoiding the risk of missing terms by working in four dimensions, and lead to integrands that are valid in $D$ dimensions. Second, manifest gauge invariance allows us to obtain very compact expressions and perform intermediate $D$-dimensional state sums without introducing spurious poles at any stage.
Finally we will present all integrands in $D$ dimensions, while performing the integrations in $D=4-2\epsilon$.
 

\section{Systematics of the expansion for HEFT integrand}
\label{sec:4}

\subsection{Computational strategy}

We now  propose an approach  that addresses directly the computation of a HEFT  scattering phase $\delta_{\rm HEFT}$  from which we can derive  the bending angle as 
\begin{align}
\label{chiHEFT}
\chi = - \frac{\partial }{\partial J}\ \text{Re} \ \delta_{\rm HEFT}
\ , 
\end{align}
 where $J$ is the total angular momentum of the system introduced in \eqref{Jdef}.  This phase is defined as the sum of all  
{\it two-massive particle irreducible 
(2MPI) diagrams} contributing to the scattering amplitude%
\footnote{In the closely related context of the post-Newtonian expansion, the relevance of 2MPI diagrams had already been  noticed  in \cite{Neill:2013wsa}.}, 
    \begin{align}
\begin{split}
\delta_{\rm HEFT}&= \frac{1}{\mathcal{J}}
\int\!\frac{d^{D-2}q}{(2\pi)^{D-2}}\, e^{- i \vec{q}\Cdot \vec{b}}  \ \lamp^{\rm 2MPI}_{\rm HEFT}\, , 
\end{split}
\end{align}
where $\mathcal{J} = 4 \sqrt{(p_1\Cdot p_2)^2 - p_1^2 p_2^2} = 4m_1m_2\sqrt{y^2-1}$.
Conversely, two massive particle reducible diagrams automatically factorise  in impact parameter space, effectively trivialising the exponentiation of the classical HEFT phase. 
We note that our HEFT expansion makes this manifest at the diagrammatic level, and all iterating, two-massive-particle reducible diagrams can simply be dropped.
 
Our proposal  is very reminiscent of the non-abelian exponentiation theorem of \cite{Gatheral:1983cz,Frenkel:1984pz},  which allows the computation of the exponent $w$ of the expectation value of the Wilson loop $\langle W \rangle = e^w$ in gauge theory from a subset of all diagrams, namely those with a ``maximally non-abelian'' colour factor.
This is in complete analogy with
the Wilson loop diagrams contributing to the exponent $w$, which are also two Wilson line irreducible diagrams, 
in the sense that cutting two Wilson lines does not break the diagram into two disconnected components, see e.g.~\cite{Anastasiou:2009kna} for recent applications and \cite{Gardi:2013ita} for an extension to multiple Wilson lines.

It is useful to compare more closely our approach to that of the recent  interesting paper \cite{Damgaard:2021ipf}. As mentioned in the Introduction, in that paper it was suggested to write the $S$-matrix in an exponentiated form as $S=e^{\frac{i}{\hbar} N}$, where $N$ is a hermitian operator. 
Anticipating our story, we will find that up to two loops, the two-body matrix element of the  operator $N$ introduced there is identical to the real part of our 2MPI amplitude --  compare our \eqref{M22} to (3.27) of that paper. Our $\cM_{\rm HEFT}^{\rm 2MPI}$  has in addition an imaginary part which at 3PM arises entirely from the radiation reaction terms, but the real part is identical to the elastic matrix element of the operator $N$. 
The advantage of our approach is that it provides  a practical way  to compute $\delta_{\rm HEFT}$ directly, from which one can then extract immediately the scattering  angle using \eqref{chiHEFT}. 
We will expand on the comparison of our results to those of \cite{Damgaard:2021ipf} at the end of Section~\ref{sec:loops}.

 \subsection{Diagrammatics of the HEFT expansion for classical dynamics}
\label{sec:4.1}

We now discuss how the HEFT provides us with a systematic $\hbar$ expansion of $L$-loop amplitudes at the diagrammatic level. 
This will allow  us to cleanly isolate contributions to classical physics without the need of expanding complete amplitudes/integrands, which in general can be rather cumbersome. 
To do this expansion one rescales all the soft graviton momenta $\ell_i\to \hbar \ell_i$ and the momentum transfer $q\to \hbar q$ \cite{Kosower:2018adc}, which in turns is equivalent to a (refined) expansion in terms of  the inverse  masses $1/m_1$ and $1/m_2$ of the binary system. 

Concretely, in order to construct the loop integrands we will employ the unitarity method, where only massless graviton legs are cut. 
We only need to compute terms in the four-point amplitude which have discontinuities in the $q^2$-channel, which contribute to long-distance effects \cite{Donoghue:1994dn, Holstein:2004dn}. As we will explain, we  only need the leading-term in the heavy-mass expansion -- that is the tree-level HEFT amplitudes reviewed in 
Section~\ref{sec:HEFTrev},  as well as pure graviton amplitudes. 
For concreteness we work with massive scalar fields, and the result at leading order in the mass expansion is independent of the spin of the massive particle. 

We now describe  the expansion of the gravity-matter amplitudes in HEFT. 
The first observation is related to joining two  different HEFT amplitudes into a larger tree.  In gravity we have to sum over all possible orderings of gravitons, as implemented in~\eqref{join2heft}:
\begin{align}
\label{join2heft}
& \begin{tikzpicture}[baseline={([yshift=-0.8ex]current bounding box.center)}]\tikzstyle{every node}=[font=\small]	
\begin{feynman}
    	 \vertex (a) {\(p_2\)};
    	 \vertex [right=1.5cm of a] (f2)[HV]{H};
    	 \vertex [right=2.0cm of f2] (f3)[HV]{H};
    	 \vertex [right=1.5cm of f3] (c){$p_3$};
    	 \vertex [below=1.1cm of f2] (gma){$\boldsymbol\cdots$};
    	 \vertex [below=1.5cm of f2] (gm){$$};
    	  \vertex [below=1.1cm of f3] (gm2a){$\boldsymbol\cdots$};
    	 \vertex [below=1.5cm of f3] (gm2){$$};
    	 \vertex [left=0.5cm of gm] (g2){$\ell_{1_a}$};
    	  \vertex [right=0.5cm of gm] (g20){$\ell_{i_a}$};
    	  \vertex [left=0.5cm of gm2] (g3){$\ell_{1_b}$};
    	 \vertex [right=0.5cm of gm2] (g30){$\ell_{j_b}$};
    	  \diagram* {
(a) -- [fermion,thick] (f2)-- [fermion,thick] (f3) --  [fermion,thick] (c),
    	  (g2)--[photon,ultra thick,momentum](f2),(g20)--[photon,ultra thick,momentum'](f2), (g3)--[photon,ultra thick,momentum](f3),(g30)--[photon,ultra thick,momentum'](f3),
    	  };
    \end{feynman}  
    \end{tikzpicture}+\begin{tikzpicture}[baseline={([yshift=-0.8ex]current bounding box.center)}]\tikzstyle{every node}=[font=\small]	
\begin{feynman}
    	 \vertex (a) {\(p_2\)};
    	 \vertex [right=1.5cm of a] (f2)[HV]{H};
    	 \vertex [right=2.0cm of f2] (f3)[HV]{H};
    	 \vertex [right=1.5cm of f3] (c){$p_3$};
    	 \vertex [below=1.1cm of f2] (gma){$\boldsymbol\cdots$};
    	 \vertex [below=1.5cm of f2] (gm){$$};
    	  \vertex [below=1.1cm of f3] (gm2a){$\boldsymbol\cdots$};
    	 \vertex [below=1.5cm of f3] (gm2){$$};
    	 \vertex [left=0.5cm of gm] (g2){$\ell_{1_b}$};
    	  \vertex [right=0.5cm of gm] (g20){$\ell_{j_b}$};
    	  \vertex [left=0.5cm of gm2] (g3){$\ell_{1_a}$};
    	 \vertex [right=0.5cm of gm2] (g30){$\ell_{i_a}$};
    	  \diagram* {
(a) -- [fermion,thick] (f2)-- [fermion,thick] (f3) --  [fermion,thick] (c),
    	  (g2)--[photon,ultra thick,momentum](f2),(g20)--[photon,ultra thick,momentum'](f2), (g3)--[photon,ultra thick,momentum](f3),(g30)--[photon,ultra thick,momentum'](f3),
    	  };
    \end{feynman}  
    \end{tikzpicture} 
\end{align}
The scalar propagator  in \eqref{join2heft} is proportional to  $\dfrac{1}{(p+ q)^2 - m^2 + i \varepsilon}$, where $p=p_2$.  We can rewrite this in two ways. The first one is: 
\begin{align}
\dfrac{1}{(p+ Q)^2 - m^2 + i \varepsilon} =\frac{1}{ \, 2 p\Cdot Q+Q^2  + i \varepsilon}  \simeq \frac{1}{\, 2 p\Cdot Q  + i \varepsilon}\Big( 1 - \frac{Q^2}{2 p\Cdot Q}\Big) + \cdots \, , 
\end{align}
where $Q$ is the sum of the soft graviton momenta,  and we have performed a Taylor expansion, with  the dots indicating terms of $\cO(Q)$. 
As noted in  \cite{Parra-Martinez:2020dzs}, such a Taylor expansion does not lead to a systematic $\hbar$ expansion. 
Indeed,  since $(p+q)^2 = p^2$, it follows that $p\Cdot q = - q^2 / 2$. This implies that $p$, when dotted with other momenta, does not have a homogeneous degree in $\hbar$.

A better choice is to use  the $\bar{p} = {\bar m} {\bar v}$ variables introduced in \eqref{kinematics}, in which case the propagator  takes the  form 
 \begin{align}
 \label{eeq}
\frac{1}{(\bar{p}\pm \frac{q}{2} +Q )^2 - m^2 + i \varepsilon} =\frac{1}{ 2 \bar{p}\Cdot Q  \pm q\Cdot Q+ Q^2+ i \varepsilon} \simeq 
\frac{1}{ 2 \bar{p}\Cdot Q +  i \varepsilon} \Big(1  - \frac{\pm q\Cdot Q+ Q^2}{2 \bar{p}\Cdot Q }\Big) 
+ \cdots
\end{align}
This alternative expansion is useful because  $\bar{p}\Cdot q=0$, which follows from  the equality 
  $(\bar{p} +  \frac{q}{2})^2 =(\bar{p} -  \frac{q}{2})^2$.  
This is important when considering the combination of terms in \eqref{join2heft}, because it avoids producing terms that carry higher powers of $\hbar$. Crucially the sum of the two diagrams  in \eqref{join2heft} contains the factor  
\begin{align}
\frac{ 1}{2 \bar{p} \Cdot Q_a \, +\, i \,   \varepsilon } + \frac{ 1 }{2 \bar{p} \Cdot  Q_b  \, +\, i \, \varepsilon}\, , 
\end{align}
where $Q_a =\ell_{a_1}+ \cdots+ \ell_{a_i}:=  \ell_{a_1\cdots a_i}$ and $Q_b =  \ell_{b_1} + \cdots+ \ell_{b_i} := \ell_{b_1\cdots b_i }$. Using now $Q_b = - Q_a + q$ and $\bar{p}\Cdot q=0$, we can rewrite this as%
\footnote{This type of argument has been used in several other works, see for instance \cite{Kabat:1992tb,Saotome:2012vy, Akhoury:2013yua,Bern:2020uwk,Bjerrum-Bohr:2021din}.}
\begin{align}
\begin{split}
\label{KOargument}
&\Big(\frac{ 1}{2 \bar{p} \Cdot Q_a \, +\, i \,   \varepsilon } + \frac{ 1 }{-2 \bar{p} \Cdot Q_a \, +\, i \, \varepsilon}\Big)\, A_{i+2}(1_{a}\ldots i_{a}, {\bar v_2})\, A_{j+2}(1_{b}\ldots j_{b}, {\bar v_2}) \\
&=  - (2 \pi i) \,  \delta\big({2 \bar{p} \Cdot Q_a} \big)\, A_{i+2}(1_{a}\ldots i_{a}, {\bar v_2})\, A_{j+2}(1_{b}\ldots j_{b}, { \bar v_2})\, .
\end{split}
 \end{align}
This term is of $\cO({\bar m}^3)$: two factors of ${\bar m}^2$ from the HEFT amplitudes, and one factor of ${\bar m}^{-1}$ from the delta function. 
A Taylor expansion  of the full tree-level amplitude, without taking into account  the $i \varepsilon$ prescriptions, would  miss such terms and produce only the HEFT amplitude with $i+j$ gravitons.

This procedure works recursively for our  matter-graviton amplitudes. A generic amplitude is then decomposed as 
\begin{align}
\begin{split}
\label{eq:HEFTExpand}
	&\sum_{h=1}^{n-2} \sum_{\mathbf{P} \in\mathcal P(n-2,h) }  \Big(\prod_{j=1}^{h-1} (-2\pi i) \, \delta({2 \bar m_2 \bar v_2 \Cdot \ell_{\mathbf{P}_j}} )\Big) A_{i_{1}+2}(\mathbf{P}_1,{\bar v_2}) \cdots A_{i_{h}+2} (\mathbf{P}_h,{\bar v_2}) \\
	&+\cdots\, , 
\end{split}
\end{align}
where $\mathcal P(n-2,h)$ denotes the partitions of the $n-2$ gravitons into $h$ non-empty subsets, and the summation is taken over all the partitions with $h=1, \ldots, n-2$. $\mathbf{P}_j$ denotes the $j^{\rm th}$
subset of graviton indices of a given partition $\mathbf{P}$ with length $i_j$ and total momentum $\ell_{\mathbf{P}_j}$.  
The term without any delta function ($h=1$) is the HEFT amplitude, and the dots stand for terms which are subleading in the heavy-mass expansion, which we  ignore as they correspond to quantum corrections. 
Each term in the expansion is of order $\mathcal{O}(\bar m^{h+1}_2)$ and carries a uniform power of $\hbar$, facilitating
a systematic separation of classical and hyper-classical terms in our computations.  

The expansion  \eqref{eq:HEFTExpand} produces integrands involving inverse powers of $\bar{p}_i \Cdot \sum l_i$  and $\delta({\bar{p}_i \Cdot \sum l_i})$ which is crucial to produce a  clean $\hbar$ expansion, and in particular leads to loop integrals that have a trivial dependence on $q$ through a universal power that is determined by power counting
and otherwise are non-trivial functions of the dimensionless quantity $\bar{y}:=\bar{v}_1\Cdot\bar{v}_2$ only. 
This choice is also natural from the point of view of the Fourier transform to impact parameter space, if  the ($D-2$)-dimensional subspace of $q$ is  defined through $\bar{p}_i \Cdot q = 0$ and not via $p_i \Cdot q = \cO(q^2)$ as often done.

\subsection{Tree-level examples} 
We now give a few concrete examples of the HEFT expansion. 
 The first case we consider is the 
 four-point tree-level amplitude, expanded  in the heavy-mass limit. The  exact four-point amplitude with two massive scalars is 
\begin{align}
	A_4^{(\text{GR})}=-{(p_2\Cdot F_1\Cdot F_2\Cdot p_2)^2\over \ell_{12}^2\,  p_2\Cdot \ell_1\,  p_2\Cdot \ell_2}\, .
\end{align}
Its  expansion  in terms of ${1/\bar m_2}$ is
\begin{align}
\label{4ptHEFTex}
	A_4^{(\text{GR})}\rightarrow  -(i \pi ) \, \bar{m}_2^3 \, \delta({\bar{v}_2\Cdot\ell_{1}} )(\bar{v}_2\Cdot \varepsilon_1)^2 (\bar{v}_2\Cdot \varepsilon_2)^2+
	{\bar{m}_2^2\over \ell_{12}^2}\Big(\frac{\bar{v}_2\Cdot F_1\Cdot F_2\Cdot \bar{v}_2}{\bar{v}_2\Cdot \ell_2}\Big)^2+\cdots
\end{align}
where we have defined $\bar{p}_i = \bar{m}_i \bar{v}_i$ with $\bar{v}_i^2 =1$ and $i=1,2$. The delta function term is of $\cO(\bar m_2^3)$, while the HEFT amplitude is of $\cO(\bar m_2^2)$, and  the dots represent terms of $\cO( 1)$ in the mass.

Similarly, the expansion of the  five-point amplitude in the heavy-mass limit is
\begin{align}
\begin{split}
\label{5ptHEFTex}
	A_5^{(\text{GR})}&\rightarrow  (-i\pi)^2 \,   \bar{m}^4_2 \delta({ \bar{v}_2\Cdot\ell_{1}})\delta({ \bar{v}_2\Cdot\ell_{2}})(\bar{v}_2\Cdot \varepsilon_1)^2 (\bar{v}_2\Cdot \varepsilon_2)^2(\bar{v}_2\Cdot \varepsilon_3)^2\\
	&-
	{ i\pi\,  \bar{m}^3_2\over \ell_{12}^2}\delta({ \bar{v}_2\Cdot\ell_{12}})\Big(\frac{\bar{v}_2\Cdot F_1\Cdot F_2\Cdot \bar{v}_2}{\bar{v}_2\Cdot \ell_2}\Big)^2 (\bar{v}_2\Cdot \varepsilon_3)^2-
	{i\pi \, \bar{m}^3_2\over\ell_{23}^2}\delta({ \bar{v}_2\Cdot\ell_{23}})\Big(\frac{\bar{v}_2\Cdot F_2\Cdot F_3\Cdot \bar{v}_2}{\bar{v}_2\Cdot \ell_3}\Big)^2 (\bar{v}_2\Cdot \varepsilon_1)^2\\
	&-{i\pi \, \bar{m}^3_2\over\ell_{13}^2}\delta({\bar{v}_2\Cdot\ell_{13}})\Big(\frac{\bar{v}_2\Cdot F_1\Cdot F_3\Cdot \bar{v}_2}{\bar{v}_2\Cdot \ell_3}\Big)^2 (\bar{v}_2\Cdot \varepsilon_2)^2+A_5(234,\bar v_2)+\cdots\, ,
	\end{split}
\end{align}
where $A_5(234,\bar v_2)$ is given in \eqref{A5new}. It is easy  to see that   gauge invariance, locality and crossing symmetry of the gravitons are manifest.

\subsection{Loops}
\label{sec:loops}
At one loop, the amplitude integrand can be expanded as 
\begin{flalign}
\label{oneloopdiagrams}
 \bar m_1^3  \bar m_2^3 ~~~~~~~&& \bar m_1^2 \bar m_2^3~~~~~~~&& \bar m_1^3 \bar m_2^2 ~~~~~~~&& \bar m_1^2  \bar m_2^2~~~~~~~~ \nn \\
   \begin{tikzpicture}[baseline={([yshift=0.0ex]current bounding box.center)}]\tikzstyle{every node}=[font=\small]	
\begin{feynman}
    	 \vertex (a) {\(p_1\)};
    	 \vertex [right=1.0cm of a] (f2) [dot]{};
    	  \vertex [right=1.0cm of f2] (f3) [dot]{};
    	 \vertex [right=1.0cm of f3] (c){$p_4$};
    	 \vertex [above=2.0cm of a](ac){$p_2$};
    	 \vertex [right=1.0cm of ac] (ad) [dot]{};
    	 \vertex [right=1.0cm of ad] (f2c) [dot]{};
    	  \vertex [right=1.0cm of f2c](cc){$p_3$};
    	  \vertex [above=1.cm of a] (cutL);
    	  \vertex [right=3.0cm of cutL] (cutR);
    	  \vertex [right=0.5cm of ad] (att);
    	  \vertex [above=0.3cm of att] (cut20){};
    	  \vertex [below=0.3cm of att] (cut21){};
    	  \vertex [below=1.4cm of att] (cuta0){};
    	  \vertex [below=2.3cm of att] (cuta1){};
    	  \diagram* {
(a) -- [thick] (f2) -- [thick] (f3)-- [thick] (c),
    	  (ad)--[photon,ultra thick](f2), (f2c)-- [photon,ultra thick] (f3),(ac) -- [thick] (ad)-- [thick] (f2c)-- [thick] (cc), (cutL)--[dashed, red,thick] (cutR), (cut20)--[ red,thick] (cut21),(cuta0)--[ red,thick] (cuta1)
    	  };
    \end{feynman}  
    \end{tikzpicture} &&
	\begin{tikzpicture}[baseline={([yshift=0.0ex]current bounding box.center)}]\tikzstyle{every node}=[font=\small]	
\begin{feynman}
    	 \vertex (a) {\(p_1\)};
    	 \vertex [right=1.5cm of a] (f2) [HV]{H};
    	 \vertex [right=1.5cm of f2] (c){$p_4$};
    	 \vertex [above=2.0cm of a](ac){$p_2$};
    	 \vertex [right=1.0cm of ac] (ad) [dot]{};
    	 \vertex [right=1.0cm of ad] (f2c) [dot]{};
    	  \vertex [right=1.0cm of f2c](cc){$p_3$};
    	  \vertex [above=1.cm of a] (cutL);
    	  \vertex [right=3.0cm of cutL] (cutR);
    	  \vertex [right=0.5cm of ad] (att);
    	  \vertex [above=0.3cm of att] (cut20){$ $};
    	  \vertex [below=0.3cm of att] (cut21);
    	  \diagram* {
(a) -- [thick] (f2)-- [thick] (c),
    	  (ad)--[photon,ultra thick](f2), (f2c)-- [photon,ultra thick] (f2),(ac) -- [thick] (ad)-- [thick] (f2c)-- [thick] (cc), (cutL)--[dashed, red,thick] (cutR), (cut20)--[ red,thick] (cut21)
    	  };
    \end{feynman}  
    \end{tikzpicture} &&
	\begin{tikzpicture}[baseline={([yshift=0.0ex]current bounding box.center)}]\tikzstyle{every node}=[font=\small]	
\begin{feynman}
    	 \vertex (p1) {\(p_1\)};
    	 \vertex [above=2.0cm of p1](p2){$p_2$};
    	 \vertex [right=1.5cm of p2] (u1) [HV]{H};
    	 \vertex [right=1.5cm of u1] (p3){$p_3$};    	 
    	 \vertex [right=1.0cm of p1] (b1) [dot]{};
    	 \vertex [right=1.0cm of b1] (b2) [dot]{};
    	  \vertex [right=1.0cm of b2](p4){$p_4$};
    	  \vertex [above=1.cm of p1] (cutL);
    	  \vertex [right=3.0cm of cutL] (cutR);
    	  \vertex [right=0.5cm of b1] (att);
    	  \vertex [above=0.3cm of att] (cut20);
    	  \vertex [below=0.3cm of att] (cut21);
    	  \diagram* {
(p2) -- [thick] (u1)-- [thick] (p3),
    	  (b1)--[photon,ultra thick](u1), (b2)-- [photon,ultra thick] (u1),(p1) -- [thick] (b1)-- [thick] (b2)-- [thick] (p4), (cutL)--[dashed, red,thick] (cutR), (cut20)--[ red, thick] (cut21)
    	  };
    \end{feynman}  
    \end{tikzpicture} &&
    \begin{tikzpicture}[baseline={([yshift=0.0ex]current bounding box.center)}]\tikzstyle{every node}=[font=\small]	
\begin{feynman}
    	 \vertex (a) {$p_1$};
    	 \vertex [right=1.5cm of a] (f2) [HV]{H};
    	 \vertex [right=1.5cm of f2] (c){$p_4$};
    	 \vertex [above=2.0cm of a](ac){$p_2$};
    	 \vertex [right=1.5cm of ac] (ad) [HV]{H};
    	  \vertex [right=1.5cm of ad](cc){$p_3$};
    	  \vertex [above=1.cm of a] (cutL);
    	  \vertex [right=3.0cm of cutL] (cutR);
    	  \diagram* {
(a) -- [thick] (f2)-- [thick] (c),
    	   (ad)--[photon,out=240, in=-240, looseness=1.5,ultra thick](f2), (ad)-- [photon,out=-60, in=60, looseness=1.5,ultra thick] (f2),(ac) -- [thick] (ad)-- [thick] (cc), (cutL)--[dashed, red,thick] (cutR)
    	  };
    \end{feynman}  
    \end{tikzpicture}
\end{flalign}
Here the black dots represent three-point HEFT amplitudes while higher-point ones are depicted using grey blobs. The wavy lines correspond to the graviton lines, the dashed red lines denote unitarity cuts $\delta(l_i^2)$, while the continuous vertical red lines represent the delta functions $\delta(2 \bar{p}_i \Cdot \sum l_j)$ of the massive lines -- these are not unitarity cuts but arise from linear propagators as in \eqref{KOargument}. More concretely the four cut diagrams are obtained from the product of two copies of the expanded four-point amplitude \eqref{4ptHEFTex} -- one for particle 1 (lower line), the other for particle 2 (upper line).

 The first diagram  is two massive particle reducible 
and gives a  hyper-classical contribution ($\cO(\hbar^{-1})$ compared to the classical terms).
The last diagram gives rise to quantum corrections, which we will not compute in this paper.  Only the remaining two diagrams    are  relevant for classical scattering, and will be computed in Section \ref{sec:1loopheft}.

At two loops,  the integrand is expanded in terms of a  hyper-classical, classical and quantum part. The hyper-classical graphs are 
  \begin{align}
  \label{reduciblediagrams}
 \bar m_1^4 \bar m_2^4~~~~~~~~~~&&  \bar m_1^3 \bar m_2^4~~~~~~~~~~  && \bar m_1^4 \bar m_2^3 ~~~~~~~~~~  \nn\\
  \begin{tikzpicture}[baseline={([yshift=-0.4ex]current bounding box.center)}]\tikzstyle{every node}=[font=\small]	
\begin{feynman}
    	 \vertex (p1) {\(p_1\)};
    	 \vertex [right=1cm of p1] (b1) [dot]{};
    	  \vertex [right=1cm of b1] (b2) [dot]{};
    	   \vertex [right=1cm of b2] (b3) [dot]{};
    	 \vertex [right=1cm of b3] (p4){$p_4$};
    	 \vertex [above=2.0cm of p1](p2){$p_2$};
    	 \vertex [right=1.0cm of p2] (u1) [dot]{};
    	 \vertex [right=1.0cm of u1] (u2) [dot]{};
    	  \vertex [right=1.0cm of u2] (u3) [dot]{};
    	  \vertex [right=1.0cm of u3](p3){$p_3$};
    	  \vertex [above=1.cm of p1] (cutL);
    	  \vertex [right=4.0cm of cutL] (cutR);
    	  \vertex [right=0.5cm of u1] (cut1);
    	  \vertex [above=0.3cm of cut1] (cut1u);
    	  \vertex [below=0.3cm of cut1] (cut1b);
    	  \vertex [right=0.5cm of u2] (cut2);
    	  \vertex [above=0.3cm of cut2] (cut2u);
    	  \vertex [below=0.3cm of cut2] (cut2b);
    	   \vertex [right=0.5cm of b1] (cutb1);
    	  \vertex [above=0.3cm of cutb1] (cutb1u);
    	  \vertex [below=0.3cm of cutb1] (cutb1b);
    	  \vertex [right=0.5cm of b2] (cutb2);
    	  \vertex [above=0.3cm of cutb2] (cutb2u);
    	  \vertex [below=0.3cm of cutb2] (cutb2b);
    	  \diagram* {
(p1) -- [thick] (b1)-- [thick] (b2)-- [thick] (b3) -- [thick] (p4),
    	  (u1)--[photon,ultra thick](b1), (u2)-- [photon,ultra thick] (b2),(u3)-- [photon,ultra thick] (b3), (p2) -- [thick] (u1)-- [thick] (u2)-- [thick] (u3)-- [thick] (p3), (cutL)--[dashed, red,thick] (cutR), (cut1u)--[ red,thick] (cut1b),(cut2u)--[red,thick] (cut2b),(cutb1u)--[red,thick] (cutb1b),(cutb2u)--[red,thick] (cutb2b)
    	  };
    \end{feynman}  
    \end{tikzpicture}&&
   \begin{tikzpicture}[baseline={([yshift=-0.4ex]current bounding box.center)}]\tikzstyle{every node}=[font=\small]	
\begin{feynman}
    	 \vertex (p1) {\(p_1\)};
    	 \vertex [right=1cm of p1] (b1) [dot]{};
    	  \vertex [right=1.5cm of b1] (b2) [HV]{H};
    	 \vertex [right=1.5cm of b2] (p4){$p_4$};
    	 \vertex [above=2.0cm of p1](p2){$p_2$};
    	 \vertex [right=1.0cm of p2] (u1) [dot]{};
    	 \vertex [right=1.0cm of u1] (u2) [dot]{};
    	  \vertex [right=1.0cm of u2] (u3) [dot]{};
    	  \vertex [right=1.0cm of u3](p3){$p_3$};
    	  \vertex [above=1.cm of p1] (cutL);
    	  \vertex [right=4.0cm of cutL] (cutR);
    	  \vertex [right=0.5cm of u1] (cut1);
    	  \vertex [above=0.3cm of cut1] (cut1u);
    	  \vertex [below=0.3cm of cut1] (cut1b);
    	  \vertex [right=0.5cm of u2] (cut2);
    	  \vertex [above=0.3cm of cut2] (cut2u);
    	  \vertex [below=0.3cm of cut2] (cut2b);
    	   \vertex [right=0.5cm of b1] (cutb1);
    	  \vertex [above=0.3cm of cutb1] (cutb1u);
    	  \vertex [below=0.3cm of cutb1] (cutb1b);
    	  \diagram* {
(p1) -- [thick] (b1)-- [thick] (b2)-- [thick] (p4),
    	  (u1)--[photon,ultra thick](b1), (u2)-- [photon,ultra thick] (b2),(u3)-- [photon,ultra thick] (b2), (p2) -- [thick] (u1)-- [thick] (u2)-- [thick] (u3)-- [thick] (p3), (cutL)--[dashed, red,thick] (cutR), (cut1u)--[red,thick] (cut1b),(cut2u)--[red,thick] (cut2b),(cutb1u)--[red,thick] (cutb1b)
    	  };
    \end{feynman}  
    \end{tikzpicture} && \begin{tikzpicture}[baseline={([yshift=-0.4ex]current bounding box.center)}]\tikzstyle{every node}=[font=\small]	
\begin{feynman}
    	 \vertex (p1) {\(p_1\)};
    	 \vertex [above=2.0cm of p1](p2){$p_2$};
    	 \vertex [right=1cm of p2] (u1) [dot]{};
    	  \vertex [right=1.5cm of u1] (u2) [HV]{H};
    	 \vertex [right=1.5cm of u2] (p3){$p_3$};
    	 \vertex [right=1.0cm of p1] (b1) [dot]{};
    	 \vertex [right=1.0cm of b1] (b2) [dot]{};
    	  \vertex [right=1.0cm of b2] (b3) [dot]{};
    	  \vertex [right=1.0cm of b3](p4){$p_4$};
    	  \vertex [above=1.cm of p1] (cutL);
    	  \vertex [right=4.0cm of cutL] (cutR);
    	  \vertex [right=0.5cm of b1] (cut1);
    	  \vertex [above=0.3cm of cut1] (cut1u);
    	  \vertex [below=0.3cm of cut1] (cut1b);
    	  \vertex [right=0.5cm of b2] (cut2);
    	  \vertex [above=0.3cm of cut2] (cut2u);
    	  \vertex [below=0.3cm of cut2] (cut2b);
    	   \vertex [right=0.5cm of u1] (cutb1);
    	  \vertex [above=0.3cm of cutb1] (cutb1u);
    	  \vertex [below=0.3cm of cutb1] (cutb1b);
    	  \diagram* {
(p1) -- [thick] (b1)-- [thick] (b2)-- [thick] (b3)-- [thick] (p4),
    	  (b1)--[photon,ultra thick](u1), (b2)-- [photon,ultra thick] (u2),(b3)-- [photon,ultra thick] (u2), (p2) -- [thick] (u1)-- [thick] (u2)-- [thick] (p3), (cutL)--[dashed, red,thick] (cutR), (cut1u)--[red,thick] (cut1b),(cut2u)--[red,thick] (cut2b),(cutb1u)--[red,thick] (cutb1b)
    	  };
    \end{feynman}  
    \end{tikzpicture}
 \end{align}
As we will discuss in Section~\ref{tbnb}, 
these diagrams factorise in impact parameter space, and hence exponentiate. Therefore  they  do not need to be computed and we can drop them 
directly  at the diagrammatic level.

The classical pieces are obtained from the 2MPI diagrams:  
\begin{align}
\label{fans}
\bar m_1^2 \bar m_2^4~~~~~~~~~~~&& \bar m_1^4 \bar m_2^2~~~~~~~~~~~\nn\\
	\begin{tikzpicture}[baseline={([yshift=-0.4ex]current bounding box.center)}]\tikzstyle{every node}=[font=\small]	
\begin{feynman}
    	 \vertex (p1) {\(p_1\)};
    	 \vertex [right=2cm of p1] (b1) [HV]{H};
    	 \vertex [right=2cm of b1] (p4){$p_4$};
    	 \vertex [above=2.0cm of p1](p2){$p_2$};
    	 \vertex [right=1.0cm of p2] (u1) [dot]{};
    	 \vertex [right=1.0cm of u1] (u2) [dot]{};
    	  \vertex [right=1.0cm of u2] (u3) [dot]{};
    	  \vertex [right=1.0cm of u3](p3){$p_3$};
    	  \vertex [above=1.cm of p1] (cutL);
    	  \vertex [right=4.0cm of cutL] (cutR);
    	  \vertex [right=0.5cm of u1] (cut1);
    	  \vertex [above=0.3cm of cut1] (cut1u);
    	  \vertex [below=0.3cm of cut1] (cut1b);
    	  \vertex [right=0.5cm of u2] (cut2);
    	  \vertex [above=0.3cm of cut2] (cut2u);
    	  \vertex [below=0.3cm of cut2] (cut2b);
    	  \diagram* {
(p1) -- [thick] (b1) -- [thick] (p4),
    	  (u1)--[photon,ultra thick](b1), (u2)-- [photon,ultra thick] (b1),(u3)-- [photon,ultra thick] (b1), (p2) -- [thick] (u1)-- [thick] (u2)-- [thick] (u3)-- [thick] (p3), (cutL)--[dashed, red,thick] (cutR), (cut1u)--[ red,thick] (cut1b),(cut2u)--[ red,thick] (cut2b)
    	  };
    \end{feynman}  
    \end{tikzpicture}&& 	\begin{tikzpicture}[baseline={([yshift=-0.4ex]current bounding box.center)}]\tikzstyle{every node}=[font=\small]	
\begin{feynman}
    	 \vertex (p1) {\(p_1\)};
    	 \vertex [above=2.0cm of p1](p2){$p_2$};
    	 \vertex [right=2cm of p2] (u1) [HV]{H};
    	 \vertex [right=2cm of u1] (p3){$p_3$};
    	 \vertex [right=1.0cm of p1] (b1) [dot]{};
    	 \vertex [right=1.0cm of b1] (b2) [dot]{};
    	  \vertex [right=1.0cm of b2] (b3) [dot]{};
    	  \vertex [right=1.0cm of b3](p4){$p_4$};
    	  \vertex [above=1.cm of p1] (cutL);
    	  \vertex [right=4.0cm of cutL] (cutR);
    	  \vertex [right=0.5cm of b1] (cut1);
    	  \vertex [above=0.3cm of cut1] (cut1u);
    	  \vertex [below=0.3cm of cut1] (cut1b);
    	  \vertex [right=0.5cm of b2] (cut2);
    	  \vertex [above=0.3cm of cut2] (cut2u);
    	  \vertex [below=0.3cm of cut2] (cut2b);
    	  \diagram* {
(p2) -- [thick] (u1) -- [thick] (p3),
    	  (b1)--[photon,ultra thick](u1), (b2)-- [photon,ultra thick] (u1),(b3)-- [photon,ultra thick] (u1), (p1) -- [thick] (b1)-- [thick] (b2)-- [thick] (b3)-- [thick] (p4), (cutL)--[dashed, red,thick] (cutR), (cut1u)--[ red,thick] (cut1b),(cut2u)--[ red,thick] (cut2b)
    	  };
    \end{feynman}  
    \end{tikzpicture}
    \end{align}
    and
    \begin{align}\label{eq:m13m23}
 \bar m_1^3 \bar m_2^3
 \begin{tikzpicture}[baseline={([yshift=-0.2ex]current bounding box.center)}]\tikzstyle{every node}=[font=\small]	
\begin{feynman}
    	 \vertex (p1) {\(p_1\)};
    	 \vertex [right=1.2cm of p1] (b1) [dot]{};
    	  \vertex [right=1.5cm of b1] (b2) [HV]{H};
    	 \vertex [right=1.2cm of b2] (p4){$p_4$};
    	 \vertex [above=2.0cm of p1](p2){$p_2$};
    	 \vertex [right=1.2cm of p2] (u1) [HV]{H};
    	 \vertex [right=1.5cm of u1] (u2) [dot]{};
    	  \vertex [right=1.2cm of u2](p3){$p_3$};
    	  \vertex [above=1.cm of p1] (cutL);
    	  \vertex [right=3.9cm of cutL] (cutR);
    	  \vertex [right=0.75cm of u1] (cut1);
    	  \vertex [above=0.3cm of cut1] (cut1u);
    	  \vertex [below=0.3cm of cut1] (cut1b);
    	   \vertex [right=0.75cm of b1] (cutb1);
    	  \vertex [above=0.3cm of cutb1] (cutb1u);
    	  \vertex [below=0.3cm of cutb1] (cutb1b);
    	  \diagram* {
(p1) -- [thick] (b1)-- [thick] (b2) -- [thick] (p4),
    	  (u1)--[photon,ultra thick](b1), (u2)-- [photon,ultra thick] (b2),(u1)-- [photon,ultra thick] (b2), (p2) -- [thick] (u1)-- [thick] (u2)-- [thick] (p3), (cutL)--[dashed, red,thick] (cutR), (cut1u)--[ red,thick] (cut1b),(cutb1u)--[red,thick] (cutb1b)
    	  };
    \end{feynman}  
    \end{tikzpicture} &&
     \begin{tikzpicture}[baseline={([yshift=-0.2ex]current bounding box.center)}]\tikzstyle{every node}=[font=\small]	
\begin{feynman}
    	 \vertex (p1) {\(p_1\)};
    	 \vertex [right=1.2cm of p1] (b1) [HV]{H};
    	  \vertex [right=1.5cm of b1] (b2) [dot]{};
    	 \vertex [right=1.2cm of b2] (p4){$p_4$};
    	 \vertex [above=2.0cm of p1](p2){$p_2$};
    	 \vertex [right=1.2cm of p2] (u1) [dot]{};
    	 \vertex [right=1.5cm of u1] (u2) [HV]{H};
    	  \vertex [right=1.2cm of u2](p3){$p_3$};
    	  \vertex [above=1.cm of p1] (cutL);
    	  \vertex [right=3.9cm of cutL] (cutR);
    	  \vertex [right=0.75cm of u1] (cut1);
    	  \vertex [above=0.3cm of cut1] (cut1u);
    	  \vertex [below=0.3cm of cut1] (cut1b);
    	   \vertex [right=0.75cm of b1] (cutb1);
    	  \vertex [above=0.3cm of cutb1] (cutb1u);
    	  \vertex [below=0.3cm of cutb1] (cutb1b);
    	  \diagram* {
(p1) -- [thick] (b1)-- [thick] (b2) -- [thick] (p4),
    	  (u1)--[photon,ultra thick](b1), (u2)-- [photon,ultra thick] (b2),(u2)-- [photon,ultra thick] (b1), (p2) -- [thick] (u1)-- [thick] (u2)-- [thick] (p3), (cutL)--[dashed, red,thick] (cutR), (cut1u)--[ red,thick] (cut1b),(cutb1u)--[red,thick] (cutb1b)
    	  };
    \end{feynman}  
    \end{tikzpicture}&&
      \begin{tikzpicture}[baseline={([yshift=-0.2ex]current bounding box.center)}]\tikzstyle{every node}=[font=\small]	
\begin{feynman}
    	 \vertex (p1) {\(p_1\)};
    	 \vertex [right=1.cm of p1] (b1) [dot]{};
    	  \vertex [right=1.5cm of b1] (b2) [dot]{};
    	 \vertex [right=1.cm of b2] (p4){$p_4$};
    	 \vertex [above=2.0cm of p1](p2){$p_2$};
    	 \vertex [right=1.cm of p2] (u1) [dot]{};
    	 \vertex [right=1.5cm of u1] (u2) [dot]{};
    	  \vertex [right=1.cm of u2](p3){$p_3$};
    	  \vertex [above=1.6cm of p1] (cutL);
    	  \vertex [right=3.5cm of cutL] (cutR);
    	  \vertex [above=0.5cm of p1] (cut2L);
    	  \vertex [right=3.5cm of cut2L] (cut2R);
    	  \vertex [right=0.75cm of u1] (cut1);
    	  \vertex [above=0.3cm of cut1] (cut1u);
    	  \vertex [below=0.3cm of cut1] (cut1b);
    	   \vertex [right=0.75cm of b1] (cutb1);
    	  \vertex [above=0.3cm of cutb1] (cutb1u);
    	  \vertex [below=0.3cm of cutb1] (cutb1b);
    	  \vertex [above=0.65cm of cutb1] (m1)[GR]{GR};
    	  \diagram* {
(p1) -- [thick] (b1)-- [thick] (b2) -- [thick] (p4),
    	  (u1)--[photon,ultra thick](m1), (u2)-- [photon,ultra thick] (m1),(b1)--[photon,ultra thick](m1), (b2)-- [photon,ultra thick] (m1), (p2) -- [thick] (u1)-- [thick] (u2)-- [thick] (p3), (cutL)--[dashed, red,thick] (cutR),(cut2L)--[dashed, red,thick] (cut2R), (cut1u)--[red,thick] (cut1b),(cutb1u)--[red,thick] (cutb1b)
    	  };
    \end{feynman}  
    \end{tikzpicture}
    \end{align}
   There are additional graphs to consider, usually called ``radiation-reaction contributions''. These are 
  \begin{align}
  \label{rrdiagrams}
  	 \bar m_1^3 \bar m_2^3 &&
 \begin{tikzpicture}[baseline={([yshift=-0.2ex]current bounding box.center)}]\tikzstyle{every node}=[font=\small]	
\begin{feynman}
    	 \vertex (p1) {\(p_1\)};
    	 \vertex [right=1.2cm of p1] (b1) [dot]{};
    	  \vertex [right=1.5cm of b1] (b2) [dot]{};
    	 \vertex [right=1.2cm of b2] (p4){$p_4$};
    	 \vertex [above=2.0cm of p1](p2){$p_2$};
    	 \vertex [right=1.2cm of p2] (u1) [HV]{H};
    	 \vertex [right=1.5cm of u1] (u2) [HV]{H};
    	  \vertex [right=1.2cm of u2](p3){$p_3$};
    	  \vertex [above=1.cm of p1] (cutL);
    	  \vertex [right=3.9cm of cutL] (cutR);
    	  \vertex [right=0.75cm of u1] (cut1);
    	  \vertex [above=0.3cm of cut1] (cut1u);
    	  \vertex [below=0.3cm of cut1] (cut1b);
    	   \vertex [right=0.75cm of b1] (cutb1);
    	  \vertex [above=0.3cm of cutb1] (cutb1u);
    	  \vertex [below=0.3cm of cutb1] (cutb1b);
    	  \diagram* {(u1)-- [photon,ultra thick,out=-45,in=-135,looseness=0.5,min distance=0.7cm] (u2),(p1) -- [thick] (b1)-- [thick] (b2) -- [thick] (p4),
    	  (u1)--[photon,ultra thick](b1), (u2)-- [photon,ultra thick] (b2), (p2) -- [thick] (u1)-- [thick] (u2)-- [thick] (p3), (cutL)--[dashed, red,thick] (cutR), (cut1u)--[ red,thick] (cut1b),(cutb1u)--[red,thick] (cutb1b),(cutb1u)--[dashed,red,thick] (cut1b)
    	  };
    \end{feynman}  
    \end{tikzpicture} &&  \begin{tikzpicture}[baseline={([yshift=-0.2ex]current bounding box.center)}]\tikzstyle{every node}=[font=\small]	
\begin{feynman}
    	 \vertex (p1) {\(p_1\)};
    	 \vertex [right=1.2cm of p1] (b1) [HV]{H};
    	  \vertex [right=1.5cm of b1] (b2) [HV]{H};
    	 \vertex [right=1.2cm of b2] (p4){$p_4$};
    	 \vertex [above=2.0cm of p1](p2){$p_2$};
    	 \vertex [right=1.2cm of p2] (u1) [dot]{};
    	 \vertex [right=1.5cm of u1] (u2) [dot]{};
    	  \vertex [right=1.2cm of u2](p3){$p_3$};
    	  \vertex [above=1.cm of p1] (cutL);
    	  \vertex [right=3.9cm of cutL] (cutR);
    	  \vertex [right=0.75cm of u1] (cut1);
    	  \vertex [above=0.3cm of cut1] (cut1u);
    	  \vertex [below=0.3cm of cut1] (cut1b);
    	   \vertex [right=0.75cm of b1] (cutb1);
    	  \vertex [above=0.3cm of cutb1] (cutb1u);
    	  \vertex [below=0.3cm of cutb1] (cutb1b);
    	  \diagram* {(b1)-- [photon,ultra thick,out=45,in=135,looseness=0.5,min distance=0.7cm] (b2),(p1) -- [thick] (b1)-- [thick] (b2) -- [thick] (p4),
    	  (u1)--[photon,ultra thick](b1), (u2)-- [photon,ultra thick] (b2), (p2) -- [thick] (u1)-- [thick] (u2)-- [thick] (p3), (cutL)--[dashed, red,thick] (cutR), (cut1u)--[ red,thick] (cut1b),(cutb1u)--[red,thick] (cutb1b),(cutb1u)--[dashed,red,thick] (cut1b)
    	  };
    \end{feynman}  
    \end{tikzpicture}
  \end{align}
   In Section \ref{sec:1loopheft} we will compute explicitly all these 2MPI diagrams (including radiation reaction). 
 We will also show that the last cut diagram in \eqref{eq:m13m23},  which does not have a three-graviton cut, does not add any new contribution to the four-dimensional 2MPI amplitude, and hence to classical physics. 
  
 We also note that the following graph has the correct mass scaling to potentially give classical contributions
  \begin{align}
  \label{fakeclassical}
  \bar m_1^3 \bar m_2^3 &
      \begin{tikzpicture}[baseline={([yshift=-0.4ex]current bounding box.center)}]\tikzstyle{every node}=[font=\small]	
\begin{feynman}
    	 \vertex (p1) {\(p_1\)};
    	 \vertex [right=1.5cm of p1] (b1) [HV]{H};
    	  \vertex [right=1.5cm of b1] (b2) [dot]{};
    	 \vertex [right=1.5cm of b2] (p4){$p_4$};
    	 \vertex [above=2.0cm of p1](p2){$p_2$};
    	 \vertex [right=1.5cm of p2] (u1) [HV]{H};
    	 \vertex [right=1.5cm of u1] (u2) [dot]{};
    	  \vertex [right=1.5cm of u2](p3){$p_3$};
    	  \vertex [above=1.cm of p1] (cutL);
    	  \vertex [right=4.5cm of cutL] (cutR);
    	  \vertex [right=0.75cm of u1] (cut1);
    	  \vertex [above=0.3cm of cut1] (cut1u);
    	  \vertex [below=0.3cm of cut1] (cut1b);
    	   \vertex [right=0.75cm of b1] (cutb1);
    	  \vertex [above=0.3cm of cutb1] (cutb1u);
    	  \vertex [below=0.3cm of cutb1] (cutb1b);
    	  \diagram* {
(p1) -- [thick] (b1)-- [thick] (b2) -- [thick] (p4),
    	  (u2)--[photon,ultra thick](b2), (u1)-- [photon,out=240, in=-240, looseness=1.5,ultra thick] (b1),(u1)-- [photon,out=-60, in=60, looseness=1.5,ultra thick] (b1), (p2) -- [thick] (u1)-- [thick] (u2)-- [thick] (p3), (cutL)--[dashed, red, thick] (cutR), (cut1u)--[red, thick] (cut1b),(cutb1u)--[red, thick] (cutb1b)
    	  };
    \end{feynman}  
    \end{tikzpicture}
  \end{align}
but as we  argue in the next subsection it can be dropped since it is not 2MPI, similarly to the diagrams in \eqref{reduciblediagrams}.

The heavy-mass expansion can in principle be extended to compute quantum corrections, leading  to the following diagrams at two loops:
    \begin{align}
   \bar m_1^2 \bar m_2^3~~~~~~~~~~~~~~~ && \bar m_1^2 \bar m_2^3 ~~~~~~~~~~~~~~~&& m_1^2\bar m_2^2 ~~~~~~~~~~\nn\\ 	   \begin{tikzpicture}[baseline={([yshift=-0.4ex]current bounding box.center)}]\tikzstyle{every node}=[font=\small]	
\begin{feynman}
    	 \vertex (p1) {\(p_1\)};
    	  \vertex [right=2.25cm of p1] (b2) [HV]{H};
    	 \vertex [right=2.25cm of b2] (p4){$p_4$};
    	 \vertex [above=2.0cm of p1](p2){$p_2$};
    	 \vertex [right=1.5cm of p2] (u1) [dot]{};
    	 \vertex [right=1.5cm of u1] (u2) [HV]{H};
    	  \vertex [right=1.5cm of u2](p3){$p_3$};
    	  \vertex [above=1.cm of p1] (cutL);
    	  \vertex [right=4.5cm of cutL] (cutR);
    	  \vertex [right=0.75cm of u1] (cut1);
    	  \vertex [above=0.3cm of cut1] (cut1u);
    	  \vertex [below=0.3cm of cut1] (cut1b);
    	  \diagram* {
(p1) -- [thick] (b2) -- [thick] (p4),
    	  (u1)--[photon,ultra thick](b2), (u2)-- [photon,out=-120, in=80, looseness=1.5,ultra thick] (b2),(u2)-- [photon,out=-95, in=55, looseness=1.5,ultra thick] (b2), (p2) -- [thick] (u1)-- [thick] (u2)-- [thick] (p3), (cutL)--[dashed, red,thick] (cutR), (cut1u)--[red,thick] (cut1b)
    	  };
    \end{feynman}  
    \end{tikzpicture}&&    	   \begin{tikzpicture}[baseline={([yshift=-0.4ex]current bounding box.center)}]\tikzstyle{every node}=[font=\small]	
\begin{feynman}
    	 \vertex (p1) {\(p_1\)};
    	  \vertex [right=2.25cm of p1] (b2) [HV]{H};
    	 \vertex [right=2.25cm of b2] (p4){$p_4$};
    	 \vertex [above=2.0cm of p1](p2){$p_2$};
    	 \vertex [right=1.5cm of p2] (u1) [HV]{H};
    	 \vertex [right=1.5cm of u1] (u2) [HV]{H};
    	  \vertex [right=1.5cm of u2](p3){$p_3$};
    	  \vertex [above=1.cm of p1] (cutL);
    	  \vertex [right=4.5cm of cutL] (cutR);
    	  \vertex [right=0.75cm of u1] (cut1);
    	  \vertex [above=0.3cm of cut1] (cut1u);
    	  \vertex [below=0.3cm of cut1] (cut1b);
    	  \diagram* {
(p1) -- [thick] (b2) -- [thick] (p4),
    	  (u1)--[photon,ultra thick](b2), (u2)-- [photon,ultra thick] (b2),(u1)-- [photon,out=-40, in=-140, looseness=1.5,ultra thick] (u2), (p2) -- [thick] (u1)-- [thick] (u2)-- [thick] (p3), (cutL)--[dashed, red,thick] (cutR), (cut1u)--[red,thick] (cut1b),(cut1b)--[dashed,red,thick] (b2)
    	  };
    \end{feynman}  
    \end{tikzpicture}&&
     \begin{tikzpicture}[baseline={([yshift=-0.4ex]current bounding box.center)}]\tikzstyle{every node}=[font=\small]	
\begin{feynman}
    	 \vertex (p1) {\(p_1\)};
    	 \vertex [right=1.5cm of p1] (b1) [HV]{H};
    	 \vertex [right=1.5cm of b1] (p4){$p_4$};
    	 \vertex [above=2.0cm of p1](p2){$p_2$};
    	 \vertex [right=1.5cm of p2] (u1) [HV]{H};
    	  \vertex [right=1.5cm of u1](p3){$p_3$};
    	  \vertex [above=1.0cm of p1] (cutL);
    	  \vertex [right=3cm of cutL] (cutR);
       	  \diagram* {
(p1) -- [thick] (b1) -- [thick] (p4),
    	  (u1)--[photon,ultra thick, out=-60, in=60, looseness=1.5](b1), (u1)--[photon,ultra thick, out=-90, in=90, looseness=1.5](b1),(u1)-- [photon,ultra thick,out=-120, in=120, looseness=1.5] (b1), (p2) -- [thick] (u1)-- [thick] (u1)-- [thick] (p3), (cutL)--[dashed, red,thick] (cutR)
    	      	  };
    \end{feynman}  
    \end{tikzpicture} \cdots
    \end{align}
where we do not draw  diagrams which are related by crossing of  the external legs. In this paper we will not be concerned with such quantum contributions. 
    
At higher loops, more graph topologies need to be calculated, but we stress that only
 2MPI diagrams need to be evaluated. 
As an important example of higher-loop ($L>2$) diagram,  we will consider  the probe limit $ m_2\gg  m_1$. The relevant graph is 
    
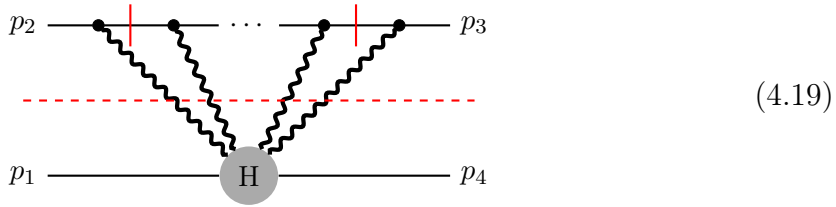
\begin{align}\label{fig:MGProbe}
	\begin{tikzpicture}[baseline={([yshift=-0.8ex]current bounding box.center)}]\tikzstyle{every node}=[font=\small]	
\begin{feynman}
    	 \vertex (a) {\(p_1\)}; 
         \vertex [right=3cm of a] (v1) [HV]{H};	 
    	 \vertex [right=3cm of v1] (b){$p_4$};
    	 \vertex [above=2.0cm of a](c){$p_2$};   	 
         \vertex [right=1.cm of c] (u1) [dot]{};
         \vertex [right=1.cm of u1] (u2) [dot]{};
         \vertex [right=1.cm of u2] (u3) []{$\cdots$};
         \vertex [right=1.cm of u3] (u4) [dot]{};
            \vertex [right=1.cm of u4] (u5) [dot]{};
    	  \vertex [right=1.cm of u5](d){$p_3$};
    	  \vertex [above=1.0cm of a] (cutL);
    	  \vertex [right=6.0cm of cutL] (cutR);
    	   \vertex [above right=0.6cm of u1] (cut1a) {}; 
    	   \vertex [below right=0.6cm of u1] (cut1b) {}; 
    	    \vertex [right =1.5cm of cut1a] (cut2a) {}; 
    	   \vertex [right =1.5cm of cut1b] (cut2b) {}; 
    	    \vertex [right =1.5cm of cut2a] (cut3a) {}; 
    	   \vertex [right =1.5cm of cut2b] (cut3b) {}; 
    	  \diagram* {
(a) -- [thick] (v1)-- [thick] (b),
    	  (u1)--[photon, ultra thick](v1), (u2)-- [photon, ultra thick] (v1),(u4)-- [photon, ultra thick] (v1),(u5)-- [photon, ultra thick] (v1), 
    	  (c) -- [thick] (u1)-- [thick](u2)-- [thick](u3)-- [thick](u4)-- [thick](u5)-- [thick]   (d), (cutL)--[dashed, red,thick] (cutR), (cut1a)--[ red,thick] (cut1b),(cut3a)--[ red,thick] (cut3b)
    	  };
    \end{feynman}  
    \end{tikzpicture}
\end{align}
which is of  $\cO(\bar m_1^2 \bar m_2^{L+2})$, for which we will present an all-loop conjecture valid in $D$ dimensions in Section \ref{sec:all-loop}.

Finally we can compare our approach to that of \cite{Damgaard:2021ipf}. It is instructive to focus on Eq.~(2.17) of that paper. In our procedure, the iterated diagrams in the second line of that equation are simply not included (hence there is no need to subtract them). Instead, the cut diagram in the first line involving two cut matter lines and one cut graviton is included in our radiation reaction diagram, and is precisely responsible for the imaginary part in the two-loop phase $\delta_{\rm  HEFT}$ computed in Section~\ref{sec:8}. Had we removed it, our procedure would have led to an identical result to the two-loop matrix element of the operator $N$ of \cite{Damgaard:2021ipf}.

\subsection{To bar or not to bar: factorisation and exponentiation}
\label{tbnb}

In the previous sections we have seen that in our HEFT expansion it is natural to use the variables $\bar{m}_i$ and $\bar{p}_i$ instead of $m_i$ and $p_i$,
and that each diagram is weighted by a unique factor $\bar{m}^i \bar{m}^j$
which allows for an immediate classification into classical, hyper-classical and quantum contributions at any loop order. This might appear unnatural
since the parameter $\bar{m}_i^2 = m_i^2 -q^2/4$ mixes classical and quantum terms. For the classical contributions this is not an issue since the expansion would only lead to subleading quantum corrections which we can drop, but expanding $\bar{m}_i$ within  hyper-classical contributions in general produces classical feed-down terms.

However, we extract the deflection angle from $\delta_{\text{HEFT}}$
that appears in the exponentiated form of the $2 \to 2$ scattering amplitude
in impact parameter space (IPS)
\begin{align}
    \tilde{S} = 1 + i \widetilde{\mathcal{M}} & = e^{i \delta_{\text{HEFT}}} \, , 
    \end{align}
    with 
    \begin{align}
    \delta_{\text{HEFT}} &= \bar{\delta}^{(0)} + \bar{\delta}^{(1)} +
    \bar{\delta}^{(2)} + \cdots
\, ,
\end{align}
where the bars indicate that the HEFT scattering phases at the respective loop orders are expressed in terms of barred variables.
Now at the level of phases only classical and possibly quantum corrections appear, and all hyper-classical contributions have been repackaged in the exponentiated form of $\widetilde{\mathcal{M}}$.
 Since we are only interested in classical physics, we can drop all quantum corrections, and at this stage  replace all quantities with bars by unbarred quantities:
\begin{align}
    \bar{m}_i, \bar{p}_i, \bar{y}= \bar{v}_1 \Cdot \bar{v}_2  \to m_i, p_i, y\, .
\end{align}
Hence in the next sections we will evaluate the appropriate sum of 2MPI diagrams that gives $\delta_{\text{HEFT}}$  using unbarred variables.

In the remainder of this section we want to show that in the HEFT expansion the exponentiation is manifest at the diagrammatic level, which is extremely useful since this allows us to ``subtract'' hyper-classical or other unwanted contributions simply by not computing them.
This is in contrast to well-known eikonal exponentiation, which usually requires the computation of complete amplitudes, even including quantum corrections \cite{DAppollonio:2015fly}. 

In order to achieve this diagrammatic rearrangement we need to align the Fourier transform to IPS with the expansion in terms of $\bar{m}_i, \bar{p}_i$. Therefore, we define the IPS form of an amplitude as
\begin{align}
\label{fourier}
\begin{split}
\widetilde{\mathcal{M}}(b) & := 
\int \frac{d^Dq}{(2 \pi)^{D-2}} \delta(2 \bar{p}_1 \Cdot q)
\delta(2 \bar{p}_2 \Cdot q) e^{i q \Cdot b} \mathcal{M}(q) \\
& = \frac{1}{\bar{\mathcal{J}}} \int \frac{d^{D-2}\vec{q}}{(2 \pi)^{D-2}} e^{-i \vec{q} \Cdot \vec{b}} \mathcal{M}(q)\, , 
\end{split}
\end{align}
where the Jacobian is $\bar{\mathcal{J}} = 4 \bar{m}_1 \bar{m}_2 \sqrt{\bar{y}^2 - 1} = 4 \sqrt{(\bar{p}_1 \Cdot \bar{p}_2)^2 - \bar{p}_1^2 \bar{p}_2^2}$.

The crucial observation is now that any two massive particle reducible diagram is a convolution integral in momentum space, which after Fourier transform to IPS becomes a simple product of sub-diagrams in IPS:
\begin{align}
\int\!\frac{d^D \ell}{(2 \pi)^D} \, (-2 \pi i)^2 \delta(\bar{p}_1 \Cdot \ell)
\delta(\bar{p}_2 \Cdot \ell) \mathcal{M}_{\text{L}}(\ell) 
\mathcal{M}_{\text{R}}(q-\ell) \, \stackrel{\text{IPS}}{\longrightarrow}\, 
- \widetilde{\mathcal{M}}_\text{L}(b) \widetilde{\mathcal{M}}_\text{R}(b)
\ ,
\end{align}
where $b = |\vec{b}\,|$. Therefore, with the choice \eqref{fourier} any massive two particle reducible diagrams factorise exactly in IPS%
\footnote{Note that in the definition of the Fourier transform usually used in the literature in the eikonal approach, the momentum transfer is taken to be orthogonal to $p_i$, which leads to differences at subleading orders in $q$ and $\hbar$ compared to our definition, but also breaks the exactness of the factorisation discussed above. Therefore, the eikonal phase and the HEFT phase are closely related but not identical, with differences starting from two loops.}.

Important examples of this are the first diagram in  \eqref{oneloopdiagrams}  and all diagrams in \eqref{reduciblediagrams}.
This makes it clear that the hyper-classical  diagram in \eqref{oneloopdiagrams} (first diagram that figure) is proportional to the square of the tree-level diagram $\widetilde{\mathcal{M}}^{(0)}$, while the first diagram in \eqref{reduciblediagrams} is a triple convolution and hence is proportional to the third power of the tree diagram. For this class of diagrams involving only trivalent vertices one can actually perform an all-loop computation. Taking into account our conventions and appropriate symmetry factors, the diagrams
can be resummed as
\begin{align}
    e^{i \widetilde{\mathcal{M}}^{(0)}} - 1 \ ,
\end{align}
which agrees of course with the well-known result for the leading eikonal
found by \cite{Kabat:1992tb}.

Focusing on the second and third diagram of \eqref{reduciblediagrams},
we see that in IPS they factorise into a tree diagram, and the sum of two irreducible one-loop diagrams (diagrams two and three in \eqref{oneloopdiagrams}) which contribute to classical physics, hence  in IPS  these reducible two-loop diagrams factorise into $\widetilde{\mathcal{M}}^{(0)} \widetilde{\mathcal{M}}^{(1)}_{\text{2MPI}}$. On the other hand the diagram
\eqref{fakeclassical} factorises in IPS into a tree diagram and the fourth diagram of \eqref{oneloopdiagrams} (which is a quantum correction), i.e.~$\widetilde{\mathcal{M}}^{(0)} \widetilde{\mathcal{M}}^{(1,\text{qu})}_{\text{2MPI}}$.

Hence, up to two-loop order we can combine the contributions from all two-loop reducible diagrams and all one-loop and tree diagrams as
\begin{align}
    e^{i \widetilde{\mathcal{M}}^{(0)} + i \widetilde{\mathcal{M}}^{(1)}_{\text{2MPI}} 
    + i \widetilde{\mathcal{M}}^{(1,\text{qu})}_{\text{2MPI}}}- 1 \ .
\end{align}
Adding now also contributions from all other two-loop 2MPI diagrams
\eqref{fans}, \eqref{eq:m13m23} and  \eqref{rrdiagrams} we
find, up to two loops, the exponentiated form of the S-matrix:
\begin{align}
    \tilde{S}=e^{i \widetilde{\mathcal{M}}^{(0)} + i \widetilde{\mathcal{M}}^{(1)}_{\text{2MPI}} + i \widetilde{\mathcal{M}}^{(1,\text{qu})}_{\text{2MPI}} + i \widetilde{\mathcal{M}}^{(2)}_{\text{2MPI}}}
\end{align}
We can now define the exponent of this equation as $\delta_{\text{HEFT}}$, 
which  is a complex quantity. We expect that in general this phase is given by the sum of 2MPI diagrams
even beyond two loops, with  the deflection angle given  as $\chi = -\partial \text{Re}(\delta_{\text{HEFT}})/\partial J$,  where  $J$ is the total angular momentum. Importantly,    $\delta_{\text{HEFT}}$ has only classical and quantum terms
but no hyper-classical terms. Since we are only interested in classical physics, we can now replace all barred quantities by unbarred ones,  and we can also drop $\widetilde{\mathcal{M}}^{(1,\text{qu})}_{\text{2MPI}}$ -- which  we actually never have to compute.

\section{The one-loop HEFT amplitude }
\label{sec:1loopheft}
In this section we  compute  the $2\to 2$  scalar  HEFT amplitude at one loop. 
In the HEFT approach, the amplitude can be expanded in powers of the masses $\bar{m}_1$ and $\bar{m}_2$.   At one loop, this expansion  is given by
 \begin{align}
 	\lamp^{(1)}_{\rm HEFT}= \lamp^{(1)}_{\bar{m}_1^3 \bar{m}_2^3}+
 	\lamp^{(1)}_{\bar{m}_1^2\bar{m}_2^3}+ \lamp^{(1)}_{\bar{m}_1^3\bar{m}_2^2} \ . 
	\end{align}
Reinserting powers of $\hbar$, one immediately sees that 
the first term corresponds to the hyper-classical part of the amplitude while the second and third to the classical part. Remaining quantum corrections are beyond the scope of this paper. 
 
 The hyper-classical part is computed from the following HEFT diagram:
 \begin{align}
 \label{eq:HCOneLoop}
 	\lamp^{(1)}_{\bar m_1^3 \bar m_2^3}&=  \begin{tikzpicture}[baseline={([yshift=0.0ex]current bounding box.center)}]\tikzstyle{every node}=[font=\small]	
\begin{feynman}
  	 \vertex (a) {\(p_1\)};
   	 \vertex [right=1.0cm of a] (f2) [dot]{};
  	  \vertex [right=1.0cm of f2] (f3) [dot]{};
  	 \vertex [right=1.0cm of f3] (c){$p_4$};
   	 \vertex [above=2.0cm of a](ac){$p_2$};
    	 \vertex [right=1.0cm of ac] (ad) [dot]{};
    	 \vertex [right=1.0cm of ad] (f2c) [dot]{};
    	  \vertex [right=1.0cm of f2c](cc){$p_3$};
    	  \vertex [above=1.cm of a] (cutL);
    	  \vertex [right=3.0cm of cutL] (cutR);
    	  \vertex [right=0.5cm of ad] (att);
    	  \vertex [above=0.3cm of att] (cut20){};
  	  \vertex [below=0.3cm of att] (cut21){};
   	  \vertex [below=1.4cm of att] (cuta0){};
   	  \vertex [below=2.3cm of att] (cuta1){};
    	  \diagram* {
(a) -- [fermion,thick] (f2) -- [fermion,thick] (f3)-- [fermion,thick] (c),
    	  (ad)--[photon,ultra thick,momentum'=\(\ell_1\)](f2), (f2c)-- [photon,ultra thick,momentum=\(\ell_2\)] (f3),(ac) -- [fermion,thick] (ad)-- [fermion,thick] (f2c)-- [fermion,thick] (cc), (cutL)--[dashed, red,thick] (cutR), (cut20)--[ red,thick] (cut21),(cuta0)--[ red,thick] (cuta1)
    	  };
    \end{feynman}  
    \end{tikzpicture}=
     i  \frac{(32 \pi  G_N)^2}{2!(4 \pi )^{D/2}}\, \int {d^{D}\ell_1\over \pi^{D/2}} \ \pi^2 \, \delta(- \bar m_2\bar{v}_2\Cdot \ell_1)\delta(\bar m_1\bar{v}_1\Cdot \ell_1)\nn\\
 	&\times \Big(\sum_{h_1,h_2} {A_3^{h_1}(\ell_1,\bar{v}_1)A_3^{h_2}(\ell_2,\bar{v}_1) 
	A_3^{-h_1}(-\ell_1,\bar{v}_2) A_3^{-h_2}(-\ell_2, \bar{v}_2)\over \ell_1^2 \ell_2^2  }\Big)\nn\\
	&=  i  \frac{(32 \pi  G_N)^2}{2(4 \pi )^{D/2}} \pi^2 {\bar m}_1^3 {\bar m}_2^3\Big({\bar y}^2-{1\over D-2}\Big)^2\int {d^{D}\ell_1\over \pi^{D/2}} \delta(\bar{v}_2\Cdot \ell_1)\delta(\bar{v}_1\Cdot \ell_1) {1 \over   \ell_1^2 \ell_2^2},
 \end{align}
 where
${\bar y} = \bar{v}_1 \Cdot \bar{v}_2$.

 As discussed earlier in \eqref{KOargument}, the top part of the figure, involving two three-point vertices, represents the sum of the two possible diagrams 
\begin{align}
\label{hyper-1loop}
& \begin{tikzpicture}[baseline={([yshift=-0.8ex]current bounding box.center)}]\tikzstyle{every node}=[font=\small]	
\begin{feynman}
    	 \vertex (a) {\(p_2\)};
    	 \vertex [right=1.0cm of a] (f2);
    	 \vertex [right=0.6cm of f2] (f3);
    	 \vertex [right=0.6cm of f3] (c){$p_3$};
    	 \vertex [below=0.8cm of f2] (g2){$\ell_1$};
    	 \vertex [below=0.8cm of f3] (g3){$\ell_2$};
    	  \diagram* {
(a) -- [fermion,thick] (f2)-- [fermion,thick] (f3) --  [fermion,thick] (c),
    	  (f2)--[photon,ultra thick](g2), (f3)--[photon,ultra thick](g3),
    	  };
    \end{feynman}  
    \end{tikzpicture}+\begin{tikzpicture}[baseline={([yshift=-0.8ex]current bounding box.center)}]\tikzstyle{every node}=[font=\small]	
\begin{feynman}
    	 \vertex (a) {\(p_2\)};
    	 \vertex [right=1.0cm of a] (f2);
    	 \vertex [right=0.8cm of f2] (f3);
    	 \vertex [right=0.6cm of f3] (c){$p_3$};
    	 \vertex [below right=1.0cm of f2] (g3){$\ell_2$};
    	 \vertex [below left=1.0cm of f3] (g2){$\ell_1$};
    	  \diagram* {
(a) -- [fermion,thick] (f2)-- [fermion,thick] (f3) --  [fermion,thick] (c),
    	  (f2)--[photon,ultra thick](g3), (f3)--[photon,ultra thick](g2)
    	  };
    \end{feynman}  
    \end{tikzpicture} 
\end{align}
Each of the three-point vertices here is $\mathcal{O}({\bar m}_2^2)$, however similarly to \cite{Kabat:1992tb} in the sum we obtain the combination 
\begin{align}
\frac{i}{2 {\bar m}_2 {\bar v}_2 \Cdot \ell_1 + i \epsilon} + \frac{i}{-2 {\bar m}_2 {\bar v}_2 \Cdot \ell_1 + i \epsilon} = 2 \pi \delta(2 {\bar m}_2 {\bar v}_2 \Cdot \ell_1)
\ . \end{align}
As a consequence, the massive propagator in \eqref{eq:HCOneLoop}  is effectively put on shell (as depicted by the continuous red line), and hence  this term  is of $\mathcal{O}({\bar m}_2^3)$.  
Note that this corresponds to the first term in \eqref{4ptHEFTex} in
the expansion of the full four-point tree amplitude. 
Furthermore, symmetrising the integrand over the two loop legs leads to the remaining delta function in \eqref{eq:HCOneLoop}.
In fact the symmetrisation argument of \cite{Kabat:1992tb} is not needed to argue for the second delta function. In order to obtain the cut integrand for the hyper-classical contribution we simply need to multiply the first term of \eqref{4ptHEFTex} with a similar factor for the lower line in the diagram \eqref{eq:HCOneLoop}, giving directly a product of two delta functions. The factor $1/2!$ in \eqref{eq:HCOneLoop} is simply the symmetry
factor required in this two-particle cut where identical particles cross the cut.

 The sum over internal graviton polarisations in \eqref{eq:HCOneLoop} was performed using%
 \footnote{For a recent discussion of completeness relations see \cite{Kosmopoulos:2020pcd}.} 
 \begin{align}
\sum_{h}\varepsilon^{*\mu_a} \varepsilon^{*\nu_a}\varepsilon^{\mu_b} \varepsilon^{\nu_b}={1\over 2}\Big[ g^{\mu_a \mu_b}g^{\nu_a \nu_b}+g^{\mu_a \nu_b}g^{\nu_a \mu_b}-{2\over (D-2)}g^{\mu_a\nu_a}g^{\mu_b\nu_b}\Big]
 \ , 
 \end{align}
  from which it follows that 
\begin{align}\label{eq:spinSum}
	\sum_{h_a}(\varepsilon^*_{-\ell_a}\Cdot v)^2 f =f|_{v}-{1\over D-2}f|_{ \eta}\, ,
\end{align}
where by  $f|_{v},f|_{ \eta}$ we denote replacing  $\varepsilon_i^{\mu}\varepsilon_j^{\nu}$ by $v^{\mu}v^{\nu}$ and  $\eta^{\mu\nu}$, respectively. 

We do not evaluate this diagram (or any other two-particle reducible diagrams), since it gives only hyper-classical contributions which we subtract simply by not evaluating them. 
From now on we will only consider diagrams that make classical contributions, therefore we can drop the distinction between barred and unbarred quantities which becomes irrelevant as the difference will
only create quantum terms, as explained in Section \ref{tbnb}.

The classical part of the amplitude is evaluated from the  2MPI diagrams as 
\begin{align}
\lamp^{\rm (1),2MPI}_{\rm HEFT} \, = \, 
\lamp^{(1)}_{m_1^2m_2^3}\, +\,  \lamp^{(1)}_{m_1^3m_2^2} 
\ , 
\end{align}
where $\lamp^{(1)}_{m_1^2 m_2^3}$  is obtained from the diagram below: 
\begin{align}\label{extrawurst1}
	\lamp^{(1)}_{m_1^2 m_2^3}&=\begin{tikzpicture}[baseline={([yshift=-0.8ex]current bounding box.center)}]\tikzstyle{every node}=[font=\small]	
\begin{feynman}
    	 \vertex (a) {\(p_1\)};
    	 \vertex [right=1.5cm of a] (f2) [HV]{H};
    	 \vertex [right=1.5cm of f2] (c){$p_4$};
    	 \vertex [above=2.0cm of a](ac){$p_2$};
    	 \vertex [right=1.0cm of ac] (ad) [dot]{};
    	 \vertex [right=1.0cm of ad] (f2c) [dot]{};
    	  \vertex [above=2.0cm of c](cc){$p_3$};
    	  \vertex [above=1.0cm of a] (cutL);
    	  \vertex [right=3.0cm of cutL] (cutR);
    	  \vertex [right=0.5cm of ad] (att);
    	  \vertex [above=0.3cm of att] (cut20){$ $};
    	  \vertex [below=0.3cm of att] (cut21);
    	  \diagram* {
(a) -- [fermion,thick] (f2)-- [fermion,thick] (c),
    	  (ad)--[photon,ultra thick,momentum'=\(\ell_1\)](f2), (f2c)-- [photon,ultra thick,momentum=\(\ell_2\)] (f2),(ac) -- [fermion,thick] (ad)-- [fermion,thick] (f2c)-- [fermion,thick] (cc), (cutL)--[dashed, red,thick] (cutR), (cut20)--[ red,thick] (cut21)
    	  };
    \end{feynman}  
    \end{tikzpicture}={1\over 2!}\frac{(32 \pi  G_N)^2}{(4 \pi )^{D/2}} (- \pi )  \times \nn\\
    &\int {d^{D}\ell_1\over \pi^{D/2}} \delta(-m_2 v_2\Cdot \ell_1)\sum_{h_1,h_2} {A_4^{h_1,h_2}(\ell_1, \ell_2,v_1) 
    A_3^{-h_1}(-\ell_1,v_2)A_3^{-h_2}(-\ell_2,v_2)\over \ell_1^2 \ell_2^2  }\, , 
\end{align}
where the dotted line represents the unitarity cut. The lower half of the graph  is the two-massive two-graviton tree amplitude in the HEFT, which is of $\mathcal{O}(m_1^2)$ and universal for all  spins of the massive particle. 
Hence the diagram at hand is  of  $\mathcal{O}(m_1^2 m_2^3)$, and  represents a classical contribution.

Using  \eqref{eq:threeAmp} and \eqref{4ptYMGR}  for the three- and four-point amplitudes, respectively, we obtain a numerator \begin{align}
\sum_{h_1,h_2}	m_2^4(\varepsilon^*_{-\ell_1}\Cdot v_2)^2 (\varepsilon^*_{-\ell_2}\Cdot v_2)^2{ \npre_4([1,2],v_1)^2\over \ell_{12}^2}\, .
\end{align}
Using  \eqref{eq:spinSum} to perform the sum over the internal polarisations, this becomes 
\begin{align}
 &	m_2^4\Big(\npre_4([1,2],v_1)^2|_{v_2,v_2}-{1\over D-2}\npre_4([1,2],v_1)^2|_{v_2, \eta}-{1\over D-2}\npre_4([1,2],v_1)^2|_{ \eta, v_2}\nn\\
	&+{1\over (D-2)^2}\npre_4([1,2],v_1)^2|_{ \eta, \eta}\Big) \ .
\end{align}
Hence the relevant integral for the classical contribution to the amplitude is
\begin{align}
\label{5.10}
	&\lamp^{(1)}_{m_1^2 m_2^3}={ m_1^2 m_2^3 G_N^2\over 2^{D-8}  \pi ^{\frac{D}{2}-2}}\Big[\big(y^2{-}\frac{1}{(D-2)^2}\big)G^{(1)}_{0}{-}\frac{q^2  \big((D-2) y^2-1\big)^2}{4 (D-2)^2}G^{(1)}_{2}{-}\frac{D-3 }{ (D-2) q^2}G^{(1)}_{-2} \Big]\ ,
\end{align}
where the  functions $G_i^{(1)}$ are defined as
\begin{align}
	G^{(1)}_i= 2\pi \int {d^D \ell_1\over \pi^{D/2}} {\delta(\mathcal D_2)\over \mathcal D_3\mathcal D_4}{1\over {\mathcal D}^i_1}\, .
\end{align}
The propagators  ${\mathcal D}_1, \ldots, {\mathcal D}_4$ are given in the following table: 
%
%
%
	\begin{align}
 \begin{array}{|c|c||c|c|}
  \hline
\mathcal D_1&\mathcal D_2&\mathcal D_3&\mathcal D_4\nn\\
\hline
	\ell _1\Cdot v_1&-\ell _1\Cdot v_2& \ell _1^2&\ell _2^2  \\
   \hline
 \end{array}	\ .
\end{align}
Using   IBP relations \cite{gehrmann2000differential,Smirnov:2008iw,larsen2016integration,lee2015reducing}  we can reduce \eqref{5.10}  to the  master integral  $G^{(1)}:= G_{0}^{(1)}$, 
getting
	\begin{align}\label{eq:MC1}
	\lamp^{(1)}_{m_1^2 m_2^3}&=2^{6-D} \pi ^{2-\frac{D}{2}}\,m_2^3m_1^2\,G_N^2 \frac{(D-3)  \Big[(2 D-3) (2 D-5)y^4-6(2 D-5) y^2+3\Big]}{(D-2)^2 \left(y^2-1\right)}\, G^{(1)} \ .	
	\end{align}
This result  is valid in $D$-dimensions, and  is in agreement  with  Eq.~(B.16) of \cite{KoemansCollado:2019ggb}, which was  obtained 
from the probe  approximation of the deflection angle of a massive particle in the background of a Schwarzschild black hole in $D$ dimensions. The master integral $G^{(1)} $ can be evaluated to 
\begin{align}
G^{(1)}=\frac{2^{5-D}\pi ^2  \left(-q^2\right)^{\frac{D-5}{2}}{\rm sec} \left(\frac{\pi  D}{2}\right)}{\Gamma \left(\frac{D}{2}-1\right)} \ ,
\end{align}
hence in four dimensions the amplitude becomes
\begin{align}
\label{1loopres}
		\lamp^{(1)}_{m_1^2 m_2^3}&=G_N^2\, m_1^2 m_2^3\frac{6 \pi ^2   \left(5 y^2-1\right)}{\sqrt{-q^2}}\, ,
\end{align}
which agrees with the results of 
\cite{Bern:2019crd, Antonelli:2019ytb, Bjerrum-Bohr:2021din}. 

 \section{The two-loop HEFT amplitude}
 \label{sec:2loopheft}
 
 In this section we give  explicit results for the 2MPI   cut diagrams contributing to the two-loop elastic HEFT amplitude, from which we then extract the deflection angle. 
 
 \subsection{First three-graviton cut: probe limit}
 \label{sec-2loop-fan}
 
The first two-loop diagram we consider is the ``fan diagram'' below. Its integrand  is: 
\begin{align}\label{fig:TwoLoop}
\lamp^{(2)}_{m_1^2 m_2^4}&=\begin{tikzpicture}[baseline={([yshift=-0.8ex]current bounding box.center)}]\tikzstyle{every node}=[font=\small]	
\begin{feynman}
    	 \vertex (a) {\(p_1\)}; 
         \vertex [right=3cm of a] (v1) [HV]{H};	 
    	 \vertex [right=3cm of v1] (b){$p_4$};
    	 \vertex [above=2.0cm of a](c){$p_2$};   	 
         \vertex [right=1.5cm of c] (u1) [dot]{};
         \vertex [right=1.5cm of u1] (u2) [dot]{};
         \vertex [right=1.5cm of u2] (u3) [dot]{};
    	  \vertex [right=1.5cm of u3](d){$p_3$};
    	  \vertex [above=1.0cm of a] (cutL);
    	  \vertex [right=6.0cm of cutL] (cutR);
    	   \vertex [right=0.75cm of u1] (cut1);
    	  \vertex [above=0.3cm of cut1] (cut1u);
    	  \vertex [below=0.3cm of cut1] (cut1b);
    	  \vertex [right=0.75cm of u2] (cut2);
    	  \vertex [above=0.3cm of cut2] (cut2u);
    	  \vertex [below=0.3cm of cut2] (cut2b);
    	  \diagram* {
(a) -- [fermion,thick] (v1)-- [fermion,thick] (b),
    	  (u1)--[photon,ultra thick,momentum'=\(\ell_1\)](v1), (u2)-- [photon,ultra thick,momentum'=\(\ell_2\)] (v1),(u3)-- [photon,ultra thick,momentum=\(\ell_3\)] (v1), (c) -- [fermion,thick] (u1)-- [fermion,thick] (u2)-- [fermion,thick] (u3)-- [fermion,thick] (d), (cutL)--[dashed, red,thick] (cutR), (cut1u)--[red,thick] (cut1b),(cut2u)--[red,thick] (cut2b),
    	  };
    \end{feynman}  
    \end{tikzpicture}\nn\\
  &={1\over 3!}\frac{(32 \pi  G_N)^3}{(4 \pi )^{D}} \pi^2 \int {d^{D}\ell_1\over \pi^{D/2}} {d^{D}\ell_3\over \pi^{D/2}} \delta(-m_2 v_2\Cdot \ell_1)\delta(m_2v_2\Cdot \ell_3 )\times\nn\\
  &\sum_{h_1,h_2,h_3} {A_5^{h_1,h_2,h_3}(\ell_1, \ell_2, \ell_3,v_1) 
  A_3^{-h_1}(-\ell_1,v_2)A_3^{-h_2}(-\ell_2,v_2)A_3^{-h_3}(-\ell_3,v_2)\over \ell_1^2 \ell_2^2 \ell_3^2 }\, ,
\end{align}
where the arguments $1,2,3$ in the numerators are a shorthand notation for $\ell_1$, $\ell_2$, $\ell_3$.

Using the novel BCJ representation \eqref{A5new} of the five-point amplitude, we can write the numerator of \eqref{fig:TwoLoop} as 
\begin{align}
	&\sum_{h_1,h_2,h_3}m_2^6(\varepsilon^*_{-\ell_1}\Cdot v_2)^2(\varepsilon^*_{-\ell_2}\Cdot v_2)^2(\varepsilon^*_{-\ell_3}\Cdot v_2)^2\nn\\ &\Big[{\npre_5([[1,2],3],v_1)^2\over \ell_{12}^2\ell_{123}^2}+{\npre_5([[1,3],2],v_1)^2\over \ell_{13}^2\ell_{123}^2}+{\npre_5([[2,3],1],v_1)^2\over \ell_{23}^2\ell_{123}^2}\Big]\, .
\end{align}
Performing the sums over internal polarisations, the numerator becomes
\begin{align}
\label{6.3}
	\sum_{X}&\Big[\npre_{5}(X,v_1)^2|_{ v_2,v_2,v_2}-{1\over D-2}\npre_{5}(X,v_1)^2|_{ v_2,v_2,\eta}-{1\over D-2}\npre_{5}(X,v_1)^2|_{ v_2,\eta,v_2}\nn\\
	&-{1\over D-2}\npre_{5}(X,v_1)^2|_{ \eta,v_2,v_2}+{1\over (D-2)^2}\npre_{5}(X,v_1)^2|_{ \eta, \eta,v_2}+{1\over (D-2)^2}\npre_{5}(X,v_1)^2|_{ \eta, \eta,v_2}\nn\\
		&+{1\over (D-2)^2}\npre_{5}(X,v_1)^2|_{ \eta, v_2,\eta}-{1\over (D-2)^3}\npre_{5}(X,v_1)^2|_{ \eta, \eta,v_2}\Big],
\end{align}
where $X$ denotes the three nested commutators $[[2,3],4]$, $[[2,4],3]$ and $[[3,4],2]$. The meaning of the three subscripts  in \eqref{6.3}
 is as in the one-loop case, with the  replacements done  with respect to $\varepsilon_{\ell_1}^{\mu}\varepsilon_{\ell_1}^{\nu},\varepsilon_{\ell_2}^{\mu}\varepsilon_{\ell_2}^{\nu},\varepsilon_{\ell_3}^{\mu}\varepsilon_{\ell_3}^{\nu}$ in the numerator.  

Using IBP reductions, we  can re-express the result in terms of a single master integral 
\begin{align}
G^{(2)}:=(2\pi)^2\int {d^{D}\ell_1\over \pi^{D/2}} {d^{D}\ell_3\over \pi^{D/2}}  {\delta(v_2 \Cdot\ell _1)\delta(v_2 \Cdot\ell _3)\over \ell _1^2 \ell _2^2 \ell _3^2}\, .
\end{align}
 All other master integrals cancel in the sum over the three nested commutators. One then arrives at the final result  for the contribution from this diagram:
\begin{align}\label{eq:MC2}
\begin{split}
	\lamp^{(2)}_{m_1^2 m_2^4}&=m_1^2m_2^{4}\frac{4^{5-D} (D-3)^2 \pi ^{3-D} G_N^3}{3 (D-2)^3 \left(y^2-1\right)^2}G^{(2)}\\
	&\Big[(3 D-4)  (D-2) (3 D-8)y^6-15 (D-2) (3 D-8) y^4+15 (3 D-8) y^2-5\Big] \, ,
\end{split}
\end{align}
where $y$ is defined in \eqref{y}.
This result is in agreement with Eq.~(B.17) of \cite{KoemansCollado:2019ggb}. The value of $G^{(2)}$ is evaluated in Appendix \ref{sec:BDV2} to
\begin{align}
	G^{(2)}\rightarrow  {4\pi ^2 \over \epsilon}\left(-{q^2\over 2}\right)^{-2 \epsilon }\, .
\end{align}
 In  four dimensions, \eqref{eq:MC2} reduces to
\begin{align}
\label{2loop-fan}
\lamp^{(2)}_{m_1^2 m_2^4}=m_1^2m_2^{4}  \frac{2\pi  G_N^3 \big(64 y^6-120 y^4+60 y^2-5\big)  \left(-q^2\right)^{-2 \epsilon }}{3 \left(y^2-1\right)^2 \epsilon }\, \, .
\end{align}

\subsection{Second three-graviton cut: beyond probe limit}
\label{zigzag-sec}

Next we consider the  two ``zig-zag''
diagrams, together with the two radiation reaction diagrams. 
These are given by 
\begin{align}
\label{z+f}
	\lamp^{(2,\rm I)}_{m_1^3 m_2^3}&= \begin{tikzpicture}[baseline={([yshift=-0.4ex]current bounding box.center)}]\tikzstyle{every node}=[font=\small]	
\begin{feynman}
    	 \vertex (p1) {\(p_1\)};
    	 \vertex [right=1.2cm of p1] (b1) [dot]{};
    	  \vertex [right=1.5cm of b1] (b2) [HV]{H};
    	 \vertex [right=1.2cm of b2] (p4){$p_4$};
    	 \vertex [above=2.0cm of p1](p2){$p_2$};
    	 \vertex [right=1.2cm of p2] (u1) [HV]{H};
    	 \vertex [right=1.5cm of u1] (u2) [dot]{};
    	  \vertex [right=1.2cm of u2](p3){$p_3$};
    	  \vertex [above=1.cm of p1] (cutL);
    	  \vertex [right=3.9cm of cutL] (cutR);
    	  \vertex [right=0.75cm of u1] (cut1);
    	  \vertex [above=0.3cm of cut1] (cut1u);
    	  \vertex [below=0.3cm of cut1] (cut1b);
    	   \vertex [right=0.75cm of b1] (cutb1);
    	  \vertex [above=0.3cm of cutb1] (cutb1u);
    	  \vertex [below=0.3cm of cutb1] (cutb1b);
    	  \diagram* {
(p1) -- [fermion,thick] (b1)-- [fermion,thick] (b2) -- [fermion,thick] (p4),
    	  (u1)--[photon,ultra thick,momentum'=\(\ell_1\)](b1), (u2)-- [photon,ultra thick,momentum=\(\ell_3\)] (b2),(u1)-- [photon,ultra thick,momentum'=\(\ell_2\)] (b2), (p2) -- [fermion,thick] (u1)-- [fermion,thick] (u2)-- [fermion,thick] (p3), (cutL)--[dashed, red,thick] (cutR), (cut1u)--[ red,thick] (cut1b),(cutb1u)--[red,thick] (cutb1b)
    	  };
    \end{feynman}  
    \end{tikzpicture}\cup  \begin{tikzpicture}[baseline={([yshift=-0.4ex]current bounding box.center)}]\tikzstyle{every node}=[font=\small]	
\begin{feynman}
    	 \vertex (p1) {\(p_1\)};
    	 \vertex [right=1.2cm of p1] (b1) [HV]{H};
    	  \vertex [right=1.5cm of b1] (b2) [dot]{};
    	 \vertex [right=1.2cm of b2] (p4){$p_4$};
    	 \vertex [above=2.0cm of p1](p2){$p_2$};
    	 \vertex [right=1.2cm of p2] (u1) [dot]{};
    	 \vertex [right=1.5cm of u1] (u2) [HV]{H};
    	  \vertex [right=1.2cm of u2](p3){$p_3$};
    	  \vertex [above=1.cm of p1] (cutL);
    	  \vertex [right=3.9cm of cutL] (cutR);
    	  \vertex [right=0.75cm of u1] (cut1);
    	  \vertex [above=0.3cm of cut1] (cut1u);
    	  \vertex [below=0.3cm of cut1] (cut1b);
    	   \vertex [right=0.75cm of b1] (cutb1);
    	  \vertex [above=0.3cm of cutb1] (cutb1u);
    	  \vertex [below=0.3cm of cutb1] (cutb1b);
    	  \diagram* {
(p1) -- [fermion,thick] (b1)-- [fermion,thick] (b2) -- [fermion,thick] (p4),
    	  (u1)--[photon,ultra thick,momentum'=\(\ell_5\)](b1), (u2)-- [photon,ultra thick,momentum=\(\ell_4\)] (b2), (u2)-- [photon,ultra thick,momentum'=\(\ell_2\)] (b1), (p2) -- [fermion,thick] (u1)-- [fermion,thick] (u2)-- [fermion,thick] (p3), (cutL)--[dashed, red,thick] (cutR), (cut1u)--[ red,thick] (cut1b),(cutb1u)--[red,thick] (cutb1b)
    	  };
    \end{feynman}  
    \end{tikzpicture}\nn\\
 &  \cup  \begin{tikzpicture}[baseline={([yshift=-0.4ex]current bounding box.center)}]\tikzstyle{every node}=[font=\small]	
\begin{feynman}
    	 \vertex (p1) {\(p_1\)};
    	 \vertex [right=1.2cm of p1] (b1) [dot]{};
    	  \vertex [right=1.5cm of b1] (b2) [dot]{};
    	 \vertex [right=1.2cm of b2] (p4){$p_4$};
    	 \vertex [above=2.0cm of p1](p2){$p_2$};
    	 \vertex [right=1.2cm of p2] (u1) [HV]{H};
    	 \vertex [right=1.5cm of u1] (u2) [HV]{H};
    	  \vertex [right=1.2cm of u2](p3){$p_3$};
    	  \vertex [above=1.cm of p1] (cutL);
    	  \vertex [right=3.9cm of cutL] (cutR);
    	  \vertex [right=0.75cm of u1] (cut1);
    	  \vertex [above=0.3cm of cut1] (cut1u);
    	  \vertex [below=0.3cm of cut1] (cut1b);
    	   \vertex [right=0.75cm of b1] (cutb1);
    	  \vertex [above=0.3cm of cutb1] (cutb1u);
    	  \vertex [below=0.3cm of cutb1] (cutb1b);
    	  \diagram* {(u1)-- [photon,ultra thick,out=-45,in=-135,looseness=0.5,min distance=0.7cm,momentum'=\(\ell_2\)] (u2),(p1) -- [fermion,thick] (b1)-- [fermion,thick] (b2) -- [fermion,thick] (p4),
    	  (u1)--[photon,ultra thick,momentum'=\(\ell_1\)](b1), (u2)-- [photon,ultra thick,momentum=\(\ell_4\)] (b2), (p2) -- [fermion,thick] (u1)-- [fermion,thick] (u2)-- [fermion,thick] (p3), (cutL)--[dashed, red,thick] (cutR), (cut1u)--[ red,thick] (cut1b),(cutb1u)--[red,thick] (cutb1b),(cutb1u)--[dashed,red,thick] (cut1b)
    	  };
    \end{feynman}  
    \end{tikzpicture} \cup  \begin{tikzpicture}[baseline={([yshift=-0.4ex]current bounding box.center)}]\tikzstyle{every node}=[font=\small]	
\begin{feynman}
    	 \vertex (p1) {\(p_1\)};
    	 \vertex [right=1.2cm of p1] (b1) [HV]{H};
    	  \vertex [right=1.5cm of b1] (b2) [HV]{H};
    	 \vertex [right=1.2cm of b2] (p4){$p_4$};
    	 \vertex [above=2.0cm of p1](p2){$p_2$};
    	 \vertex [right=1.2cm of p2] (u1) [dot]{};
    	 \vertex [right=1.5cm of u1] (u2) [dot]{};
    	  \vertex [right=1.2cm of u2](p3){$p_3$};
    	  \vertex [above=1.cm of p1] (cutL);
    	  \vertex [right=3.9cm of cutL] (cutR);
    	  \vertex [right=0.75cm of u1] (cut1);
    	  \vertex [above=0.3cm of cut1] (cut1u);
    	  \vertex [below=0.3cm of cut1] (cut1b);
    	   \vertex [right=0.75cm of b1] (cutb1);
    	  \vertex [above=0.3cm of cutb1] (cutb1u);
    	  \vertex [below=0.3cm of cutb1] (cutb1b);
    	  \diagram* {(b1)-- [photon,ultra thick,out=45,in=135,looseness=0.5,min distance=0.7cm,momentum=\(\ell_2\)] (b2),(p1) -- [fermion,thick] (b1)-- [fermion,thick] (b2) -- [fermion,thick] (p4),
    	  (u1)--[photon,ultra thick,momentum'=\(\ell_5\)](b1), (u2)-- [photon,ultra thick,momentum=\(\ell_3\)] (b2), (p2) -- [fermion,thick] (u1)-- [fermion,thick] (u2)-- [fermion,thick] (p3), (cutL)--[dashed, red,thick] (cutR), (cut1u)--[ red,thick] (cut1b),(cutb1u)--[red,thick] (cutb1b),(cutb1u)--[dashed,red,thick] (cut1b)
    	  };
    \end{feynman}  
    \end{tikzpicture}\nn\\
    &={1\over 2!}\frac{(32 \pi  G_N)^3}{(4 \pi )^{D}} \pi^2 \int {d^{D}\ell_1\over \pi^{D/2}} {d^{D}\ell_3\over \pi^{D/2}} \delta(m_1 v_1\Cdot \ell_1)\delta(m_2 v_2\Cdot \ell_3)\times\nn\\
  &\sum_{h_1,h_2,h_3} {A_3^{h_1}(\ell_1,v_1)A_4^{h_2,h_3}(\ell_2, \ell_3,v_1) 
  A_3^{-h_3}({-}\ell_3,v_2)A_4^{-h_1,-h_2}({-}\ell_1, {-}\ell_2,v_2)
  \over \ell_1^2 \ell_2^2 \ell_3^2 } \nn\\
  &\text{$\cup$  three other terms},
\end{align}
where the $\cup$ symbol denotes the operation of merging the corresponding cuts from the different cut integrands, producing an integrand which can then be evaluated with   
 IBP reductions. 
Traditionally the radiation reaction diagrams (second line of \eqref{z+f}) are treated separately from the remaining conservative term, but we find it more natural to combine this contribution together with the zig-zag diagram. 

The detailed form of the integrands of these four cut diagrams are given  separately in 
Appendix~\ref{sec:Integrand2}. 
After performing the  IBP reduction, we can express the $D$-dimensional  integrand as a function of   eight  master integrals:%
\footnote{The last master integral  $G_{1,0,1,1,1,0,0,1,1}$ has the topology of a bow-tie and can be discarded in four dimensions, see the discussion  below \eqref{6.13}. This leaves us with only seven master integrals.}
\black
\begin{align}
\label{6.9}
\begin{split}
\lamp^{(2,\rm I)}_{m_1^3 m_2^3}&={1\over 2!}\frac{(32 \pi  G_N)^3}{4(4 \pi )^{D}}\Bigg[\frac{q^2G_{0,1,0,1,1,1,1,1,1} \left(c_{12} y^7+c_{11} y^5+c_{10} y^3+c_9 y\right)}{2(y^2-1)}\\
&+{q^2\over 2}G_{1,1,1,1,1,-1,-1,1,1} \left(c_{22} y^3+c_{21} y\right)+{q^4\over 4}G_{1,1,1,1,1,0,0,1,1} \left(c_{25} y^4+c_{24} y^2+c_{23}\right)\\
&+\frac{G_{0,1,1,0,1,0,0,1,1} \left(c_{15} y^4+c_{14} y^2+c_{13}\right)}{y^2-1}+\frac{G_{0,1,0,1,1,0,0,1,1} \left(c_4 y^6+c_3 y^4+c_2 y^2+c_1\right)}{y^2-1}\,\\
&+\frac{q^2G_{0,2,0,1,1,0,0,1,1} \left(c_{20} y^8+c_{19} y^6+c_{18} y^4+c_{17} y^2+c_{16}\right)}{2\left(y^2-1\right)^2}\\
&+\frac{q^2 G_{0,1,0,1,1,0,0,2,2} \left(c_8 y^7+c_7 y^5+c_6 y^3+c_5 y\right)}{2\left(y^2-1\right)^2}\\
&-(D-4) q^2 G_{1,0,1,1,1,0,0,1,1}(c_{26}+c_{27} y^2+c_{28}y^4)\Bigg]\, , 
\end{split}
\end{align}
where the  coefficients $c_{i}$ are given by 
\begin{align}
\begin{array}{l}
 c_1 {=}  \frac{-18 D^5+277 D^4-1721 D^3+5388 D^2-8482 D+5360}{4 (D-3)^2 (D-2)^2 (D-1)} \\
 c_2 {=}  \frac{-2 D^6+16 D^5+17 D^4-555 D^3+2194 D^2-3582 D+2176}{2 (D-3)^2 (D-2)^3 (D-1)} \\
 c_3 {=}  \frac{10 D^6-152 D^5+937 D^4-2961 D^3+4934 D^2-3868 D+896}{4 (D-3)^2 (D-2)^2 (D-1)} \\
 c_4 {=}  -\frac{(D-4)^2 \left(D^3-4 D^2+4 D+1\right)}{2 (D-3) (D-2) (D-1)} \\
 c_5  {=}  \frac{12 D^4-108 D^3+383 D^2-632 D+405}{6 (D-4) (D-3) (D-2)^3 (D-1)} \\
 c_6  {=}  \frac{-12 D^4+109 D^3-378 D^2+589 D-344}{2 (D-4) (D-3) (D-2)^3 (D-1)} \\
 c_7  {=}  \frac{5 D^4-43 D^3+134 D^2-169 D+61}{2 (D-4) (D-3) (D-2)^2 (D-1)} \\
 c_8  {=}  \frac{-3 D^3+16 D^2-24 D+5}{6 (D-4) (D-2) (D-1)} \\
 c_9 {=}  \frac{19-7 D}{6 (D-2)^3} \\
c_{10} = \frac{5 D-13}{2 (D-2)^2} \\
 c_{11}  {=}  \frac{7-3 D}{2 (D-2)} \\
c_{12}=  \frac{D-1}{6} \\
 c_{13}  {=}  \frac{15 D^6-230 D^5+1454 D^4-4852 D^3+9015 D^2-8840 D+3574}{2 (D-4) (D-3) (D-2)^3 (D-1)} \\
  c_{14}  {=}  \frac{9 D^6-122 D^5+684 D^4-2052 D^3+3529 D^2-3368 D+1424}{2 (D-4) (D-3) (D-2)^3 (D-1)} \\
 \end{array}
 \begin{array}{l}
 \label{6.10}
 c_{15}  {=}  \frac{-8 D^3+56 D^2-126 D+98}{(D-4) (D-2) (D-1)} \\
 c_{16}  {=}  \frac{6 D^5-76 D^4+389 D^3-1008 D^2+1322 D-701}{(D-3)^2 (D-2)^3 (D-1)} \\
 c_{17}  {=}  \frac{4428-36 D^5+483 D^4-2557 D^3+6694 D^2-8664 D}{3 (D-3)^2 (D-2)^3 (D-1)} \\
 c_{18} {=}  \frac{5 D^5-70 D^4+357 D^3-837 D^2+889 D-324}{(D-3)^2 (D-2)^2 (D-1)} \\
 c_{19}  {=}  \frac{245-D^6+14 D^5-65 D^4+107 D^3+42 D^2-306 D}{(D-3)^2 (D-2)^2 (D-1)} \\
 c_{20}  {=}  -\frac{2 \left(D^3-10 D+15\right)}{3 \left(D^3-6 D^2+11 D-6\right)} \\
 c_{21}  {=}  \frac{4 D^4-42 D^3+165 D^2-291 D+200}{(D-3)^2 (D-2) (D-1)} \\
 c_{22}  {=}  \frac{4 D^3-30 D^2+70 D-56}{D^3-6 D^2+11 D-6} \\
 c_{23}  {=}  \frac{2 D^5-27 D^4+143 D^3-370 D^2+466 D-226}{2 (D-3)^2 (D-2)^2 (D-1)} \\
 c_{24}  {=}  \frac{(D-2) \left(2 D^2-15 D+25\right)}{2 (D-3)^2 (D-1)} \\
 c_{25} {=} \frac{-D^3+9 D^2-24 D+22}{(D-3) (D-2) (D-1)} \\
  c_{26} {=} \frac{-2 D^4+17 D^3-50 D^2+58 D-19}{2} \\
  c_{27} {=} \frac{-2 D^4+17 D^3-58 D^2+96 D-65}{2} \\
  c_{28} {=}\, {\scriptstyle 2 D^4-17 D^3+54 D^2-77 D+42} \, .\\
\end{array}
\end{align}
The integrals $G_{i_1,i_2,i_3,i_4,i_5,i_6,i_7,i_8,i_9}$  are defined  as 
 \begin{align}\label{eq:DefG}
	G_{i_1,i_2,i_3,i_4,i_5,i_6,i_7,i_8,i_9}={(2\pi)^2\over \Gamma(i_8)\Gamma(i_9)}\int {d^{D}\ell_1\over \pi^{D/2}} {d^{D}\ell_3\over \pi^{D/2}}  {\delta^{(i_8-1)}(\mathcal D_8)\delta^{(i_9-1)}(\mathcal D_9)\over {\mathcal D}^{i_1}_1{\mathcal D}^{i_2}_2 {\mathcal D}^{i_3}_3 {\mathcal D}^{i_4}_4\mathcal D^{i_5}_5\mathcal D^{i_6}_6\mathcal D^{i_7}_7} \, ,
\end{align}
with  the propagators  
\begin{align}
\label{table}
 \begin{array}{|c|c|c|c|c|c|c||c|c|}
  \hline
\mathcal D_1&\mathcal D_2&\mathcal D_3&\mathcal D_4&\mathcal D_5&\mathcal D_6&\mathcal D_7&\mathcal D_8&\mathcal D_9 \\
\hline
	\ell _1^2&\ell _2^2&\ell _3^2&\ell_4^2&\ell_5^2&v_1 \Cdot\ell _3&v_2 \Cdot\ell_1&v_1 \Cdot\ell _1&v_2 \Cdot\ell _3
  \\
   \hline
 \end{array}	\, .
\end{align}
We use the following box topologies to denote all the  propagators that can appear:
\begin{align}
\label{6.13}
\begin{tikzpicture}[baseline={([yshift=-0.8ex]current bounding box.center)}]\tikzstyle{every node}=[font=\small]	
\begin{feynman}
    	 \vertex (p1) {\(p_1\)};
    	 \vertex [right=1.8cm of p1] (b1) [dot]{};
    	 \vertex [right=1.2cm of b1] (b2)[dot]{};
    	 \vertex [right=1.8cm of b2] (p4){$p_4$};
    	  \vertex [right=0.25cm of b1] (la)[]{};
    	   \vertex [above=0.9cm of b1] (g1)[dot]{};
    	    \vertex [above=0.9cm of b2] (g2)[dot]{};
    	 \vertex [above=1.8cm of p1](p2){$p_2$};
    	 \vertex [right=1.8cm of p2] (u1) [dot]{};
    	 \vertex [right=1.2cm of u1] (u2) [dot]{};
    	  \vertex [above=1.8cm of p4](p3){$p_3$};
    	   \vertex [right=0.25cm of u1] (lb)[]{};
    	  \diagram* {
(p1) -- [ultra thick] (b1)-- [ultra thick,edge label'=\(8\)] (b2)-- [ultra thick] (p4),
    	   (b1)-- [thick,edge label=\(1\)] (g1)-- [thick,edge label=\(5\)] (u1),(b2) --  [thick,edge label'=\(4\)] (g2)--  [thick,edge label'=\(3\)] (u2), (g1)--[thick,edge label'=\(2\)](g2),(p2) -- [ultra thick] (u1)-- [ultra thick,edge label=\(9\)] (u2)-- [ultra thick] (p3),
    	  };
    \end{feynman}  
    \end{tikzpicture}&&
	\begin{tikzpicture}[baseline={([yshift=-0.8ex]current bounding box.center)}]\tikzstyle{every node}=[font=\small]	
\begin{feynman}
    	 \vertex (p1) {\(p_1\)};
    	 \vertex [right=1.0cm of p1] (b1) [dot]{};
    	 \vertex [right=3cm of b1] (b2)[dot]{};
    	 \vertex [right=1.0cm of b2] (p4){$p_4$};
    	  \vertex [right=1.5cm of b1] (la)[]{};
    	   \vertex [right=1.5cm of b1] (b3)[dot]{};
    	 \vertex [above=1.8cm of p1](p2){$p_2$};
    	 \vertex [right=1.0cm of p2] (u1) [dot]{};
    	 \vertex [right=3cm of u1] (u2) [dot]{};
    	  \vertex [above=1.8cm of p4](p3){$p_3$};
    	   \vertex [right=1.5cm of u1] (lb)[]{};
    	    \vertex [right=1.5cm of u1] (u3)[dot]{};
    	  \diagram* {
(p1) -- [ultra thick] (b1)-- [ultra thick,edge label'=\(8\)] (b3)-- [ultra thick,edge label'=\(6\)] (b2)-- [ultra thick] (p4),
    	   (b1)-- [thick,edge label=\(5\)] (u1),(b2) --  [thick,edge label'=\(4\)] (u2), (b3)--[thick,edge label'=\(2\)](u3),(p2) -- [ultra thick] (u1)-- [ultra thick,edge label=\(7\)] (u3)-- [ultra thick,edge label=\(9\)] (u2)-- [ultra thick] (p3),
    	  };
    \end{feynman}  
    \end{tikzpicture}
\end{align}
    
Two comments on  two of our  master integrals are in order here. 
\begin{itemize}
	\item The master integral $G_{1,0,1,1,1,0,0,1,1}$ has the topology of a bow-tie diagram, which is a product of two one-loop triangles, and hence is finite  in four dimensions and proportional to
	$(-q^2)^{-2 \eps}$. It appears in  \eqref{6.9} with a coefficient proportional to $\eps$ and   hence it does not contribute to  four-dimensional classical physics. If one is interested in $D > 4$, this integral would have to be included.
	\item In principle for certain integral topologies one should  be careful about the $i\varepsilon$, specifically for the master integral $G_{0,1,0,1,1,1,1,1,1}$, which was discussed in detail in  \cite{Bjerrum-Bohr:2021vuf}. The relevant master integrals arising from  the first graph in \eqref{z+f}  are 
	\begin{align}
	G_{0,1,0,1,1,1+,1+,1,1},\quad  G_{0,1,0,1,1,1-,1-,1,1},\quad G_{0,1,0,1,1,1-,1+,1,1},\quad G_{0,1,0,1,1,1+,1-,1,1}\, .
	\end{align}
	Here the $+, -$ denotes the signature of the regulators $i\varepsilon$ in the  linearised  propagators. The differential equations of these master integrals are given as
	\begin{align}
		G_{0,1,0,1,1,1\pm,1\pm,1,1}'(y)=-\frac{2G_{0,2,0,1,1,0,0,1,1}(y)}{y^2-1}-\frac{2 y \, G_{0,1,0,1,1,1\pm,1\pm,1,1}(y)}{y^2-1}\, .
	\end{align}
	According to the solutions of the differential equations and the asymptotic behaviours in the static limit, we can decompose these master integrals as 
	\begin{align}
		G_{0,1,0,1,1,1\pm,1\pm,1,1}&={c_{\pm\pm}\over y^2-1}+G_{0,1,0,1,1,1,1,1,1}\, ,
	\end{align}
	where $G_{0,1,0,1,1,1,1,1,1}\sim 
	(y-1)^{-1/2}$ as $y\to 1$. 
	The values of $c_{\pm\pm}$ can be found in \cite{Bjerrum-Bohr:2021vuf}.  We find that the regulator-dependent parts, $c_{\pm\pm}$, cancel with each other in the integrand, while the master integral $G_{0,1,0,1,1,1,1,1,1}$ is independent of the $i\varepsilon$ prescription. Therefore, in practice we can just use $G_{0,1,0,1,1,1,1,1,1}$ in the construction. This observation  greatly simplifies the calculation and the different $i \varepsilon$ prescriptions in $G_{0,1,0,1,1,1\pm,1\pm,1,1}$ do not affect the result. Furthermore, the differential equation of $G_{0,1,0,1,1,1,1,1,1}$ takes the same form as those of $G_{0,1,0,1,1,1\pm,1\pm,1,1}$.

\end{itemize}

\subsection{Four-graviton cut}

For completeness, we also compute the four-graviton cut diagram involving a four-point graviton amplitude.
However, it will turn out that the contribution of this cut is already accounted for by the first two (zig-zag) diagrams in \eqref{z+f} as we now explain. 
The relevant cut diagram is 
\begin{align}
\label{dishwasher}
	\lamp^{(2,\rm II)}_{m_1^3 m_2^3}&=  \begin{tikzpicture}[baseline={([yshift=-0.4ex]current bounding box.center)}]\tikzstyle{every node}=[font=\small]	
\begin{feynman}
    	 \vertex (p1) {\(p_1\)};
    	 \vertex [right=1.cm of p1] (b1) [dot]{};
    	  \vertex [right=1.5cm of b1] (b2) [dot]{};
    	 \vertex [right=1.cm of b2] (p4){$p_4$};
    	 \vertex [above=2.0cm of p1](p2){$p_2$};
    	 \vertex [right=1.cm of p2] (u1) [dot]{};
    	 \vertex [right=1.5cm of u1] (u2) [dot]{};
    	  \vertex [right=1.cm of u2](p3){$p_3$};
    	  \vertex [above=1.6cm of p1] (cutL);
    	  \vertex [right=3.5cm of cutL] (cutR);
    	  \vertex [above=0.5cm of p1] (cut2L);
    	  \vertex [right=3.5cm of cut2L] (cut2R);
    	  \vertex [right=0.75cm of u1] (cut1);
    	  \vertex [above=0.3cm of cut1] (cut1u);
    	  \vertex [below=0.3cm of cut1] (cut1b);
    	   \vertex [right=0.75cm of b1] (cutb1);
    	  \vertex [above=0.3cm of cutb1] (cutb1u);
    	  \vertex [below=0.3cm of cutb1] (cutb1b);
    	  \vertex [above=0.65cm of cutb1] (m1)[GR]{GR};
    	  \diagram* {
(p1) -- [fermion,thick] (b1)-- [fermion,thick] (b2) -- [fermion,thick] (p4),
    	  (u1)--[photon,ultra thick, momentum'=\(\ell_5\)](m1), (u2)-- [photon,ultra thick,momentum=\(\ell_3\)] (m1),(b1)--[photon,ultra thick,momentum=\(\ell_1\)](m1), (b2)-- [photon,ultra thick,momentum'=\(\ell_4\)] (m1), (p2) -- [fermion,thick] (u1)-- [fermion,thick] (u2)-- [fermion,thick] (p3), (cutL)--[dashed, red,thick] (cutR),(cut2L)--[dashed, red,thick] (cut2R), (cut1u)--[red,thick] (cut1b),(cutb1u)--[red,thick] (cutb1b)
    	  };
    \end{feynman}  
    \end{tikzpicture}\nn\\
    &={1\over 2!2!}\frac{(32 \pi  G_N)^3}{(4 \pi )^{D}} \pi ^2 \int {d^{D}\ell_1\over \pi^{D/2}} {d^{D}\ell_3\over \pi^{D/2}} \delta(m_1\ell_1\Cdot v_1)\delta(m_2\ell_3\Cdot v_2)\times\nn\\
  &\hspace{-1.4cm} \sum_{h_1,h_5,h_3,h_4}{A_3^{-h_1}(-\ell_1,v_1)
  A_3^{-h_5}(-\ell_5,v_2)
  A_3^{-h_3}(-\ell_3,v_2)
  A_3^{-h_4}(-\ell_4,v_1)
  A_4^{h_1,h_5,h_3,h_4}(\ell_1, \ell_5, \ell_3, \ell_4)\over \ell_1^2 \ell_5^2 \ell_3^2 \ell_4^2 }\, .
\end{align}
After performing all the intermediate, $D$-dimensional state sums we arrive at the (cut) integrand:
\begin{align}
	\lamp^{(2,\rm II)}_{m_1^3 m_2^3}&=\frac{(32 \pi  G_N)^3}{2(4 \pi )^{D}}\pi^2\Bigg[\frac{-q^2 G_{1,0,1,1,1,0,0,1,1} \left(\left(D^2-8 D+12\right) y^2+(D-2)^2 y^4+2\right)}{2 (D-2)^2}\nn\\
	&+\frac{(D-3)^2 G_{1,-2,1,1,1,0,0,1,1}}{(D-2)^2 (-q^2)}+\frac{(D-3)^2 G_{1,-1,1,1,1,0,0,1,1}}{2 (D-2)^2}\nn\\
	&+\frac{(3-D) G_{1,0,1,1,1,-2,0,1,1}}{D-2}+\frac{(3-D) G_{1,0,1,1,1,0,-2,1,1}}{D-2}\nn\\
	&+\frac{q^4 G_{1,1,1,1,1,0,0,1,1} \left[(D-2) y^2-1\right]^2}{4 (D-2)^2}+\frac{-2 q^2 y G_{1,1,1,1,1,-1,-1,1,1}}{D-2}\nn\\
	&+\frac{(D-3) G_{1,1,1,1,1,-4,0,1,1}}{D-2}+\frac{2 (D-4) G_{1,1,1,1,1,-2,-2,1,1}}{D-2}\nn\\
	&+\frac{(D-3) G_{1,1,1,1,1,0,-4,1,1}}{D-2}+q^2 y^2 G_{1,1,1,1,1,-2,0,1,1}+q^2 y^2 G_{1,1,1,1,1,0,-2,1,1}\Bigg]\, .
\end{align}
In the final step we have to identify any overlap with the  three-graviton cut considered in Section~\ref{zigzag-sec} in order to avoid
double counting, and isolate any new contribution if present.
It is easy to see that the terms in the last three lines are already contained in the integrands corresponding to the zig-zag diagrams \eqref{z+f}. The terms in the first two lines are new contributions that are not detected by the three-graviton cut and can all be reduced to the bow-tie master integral $G_{1,0,1,1,1,0,0,1,1}$:
\begin{align}
-	\frac{q^2  G_{1,0,1,1,1,0,0,1,1}}{4 (D-2)^4} &\Big[2 D^4 \left(y^4+y^2\right)+D^3 \left(-16 y^4-26 y^2+3\right)+D^2 \left(48 y^4+111 y^2-19\right)\nn\\
	&+D \left(-64 y^4-198 y^2+43\right)+32 y^4+129 y^2-35\Big]\, .
\end{align}
As discussed earlier  this master integral is  finite  in four dimensions and thus does not contribute to classical physics.  Hence  the four-graviton does not give any new contribution.

\subsection{Summary}
In conclusion, the complete $D$-dimensional  integrand is obtained by adding  \eqref{eq:MC2} (and the same term with $m_1$ and $m_2$ swapped) and \eqref{6.9}:
\begin{align}
\label{2loop-final-integrand}
\cM_{\rm HEFT}^{\rm (2), 2MPI} \ = \ \lamp^{(2)}_{m_1^2 m_2^4} +\lamp^{(2)}_{m_1^4 m_2^2} +\lamp^{(2,\rm I)}_{m_1^3 m_2^3}\ . 
\end{align}
As explained earlier,  the four-graviton cut in \eqref{dishwasher} does not give any new contribution  to the  integrand, hence the final result of our evaluation is given by 
\eqref{2loop-final-integrand}. 

In the next section we  evaluate the relevant integrals appearing in that equation, and will then state our (integrated) result at two loops in Section~\ref{sec:8}, where we  also derive the scattering angle up to 3PM.

\section{Canonical basis and evaluation of the master integrals}
\label{sec:canbasis}

\subsection{Generalities}
\label{sec-DE}
In this section we discuss the evaluation of the two-loop master integrals using the differential equation  approach of \cite{Kotikov:1990kg,gehrmann2000differential,Henn:2013pwa}. This leads to a canonical basis for the master integrals encountered in
Section \ref{zigzag-sec} which we will evaluate in the full soft region. 
This method was applied to the specific type of integrals which appear in the post-Minkowskian expansion of classical scattering,  initially for the potential region in 
 \cite{Parra-Martinez:2020dzs} and later for the soft region in \cite{DiVecchia:2021bdo,Herrmann:2021tct,Bjerrum-Bohr:2021vuf}.
The following discussion is valid in an arbitrary number of dimensions, while for the evaluation we will work
in $4-2 \epsilon$ dimensions except for the probe limit contributions, which we will evaluate in general dimensions.

We first set up the system of differential equations in terms of a basis of master integrals:
\begin{align}
	{d \mathcal I_i\over dy}\, =\, A_{ij} \,  \mathcal I_j\, .
\end{align}
To get a canonical basis, we need to perform a linear transformation on the integral basis. In general, if we transform the basis as 
$\mathcal I_j'=S\mathcal I_j$, the differential equation becomes
\begin{align}
	{d\mathcal I_i'\over dy}=\Big({dS\over dy}S^{-1}+SAS^{-1}\Big)_{ij}\mathcal I_j'
	\ .
\end{align}
The canonical basis is defined in such a way that the transformation matrix is  proportional  to the dimensional regularisation parameter $\epsilon$.
This can be realised in two steps: 
\begin{itemize}
	\item[{\bf 1.}] Diagonal transformation $S_0$: Since the matrix $A$  has  poles at $y=0, -1, +1$, we choose the following ansatz for the diagonal transformation, 
\begin{align}
	(S_0)_{ii}=b_{i} y^{m_{0,i}}(y-1)^{m_{1,i}}(y+1)^{m_{-1,i}}\, ,
\end{align}
where the parameters $b_i$ do not depend on $y$. The  matrix $A$ is then transformed as follows: 
\begin{align}
	A\rightarrow {dS_0\over dy}S_0^{-1}+S_0AS_0^{-1}\, .
\end{align}
We can choose the parameters in $S_0$ such that the   matrix $A$ is transformed into 
\begin{align}\label{eq:matrixADiag}
	A\rightarrow A_0+\epsilon A_1\, ,
\end{align} 
for general dimensions $D=d_0-2\epsilon$, 
with  $A_0$ and $A_1$ being  independent of $\epsilon$.
	\item[{\bf 2.}] Non-diagonal transformation $S_{1}$: Here $S_1$ is independent of $\epsilon$  and  satisfies the following equation:
	\begin{align}\label{eq:threeDiffEq}
		{dS_{1}\over dy}+S_{1}A_0&=0\, .
	\end{align} 
	Then it is easy to see that the new integral basis 
	\begin{align}
	\cI^\prime=S_1S_0 \cI 
	\end{align}
	is a canonical basis, which satisfies
	\begin{align}
	{d\cI^\prime\over dy} \, = \, \epsilon \, A' \, \cI^\prime, \qquad  \text{where}\qquad A'= \, S_1A_1S_1^{-1}.
	\end{align}
\end{itemize}
The formal solution of this equation is 
\begin{align}
\mathcal{I}'(y)=(\mathbb{P}e^{\epsilon \int_C A'dy}) \mathcal{I}'(y_0) \ .
\end{align}

\subsection{Canonical integral basis} 
\label{sec:canonintba}

We only need to consider the following seven master integrals:  
\begin{align}
	\mathcal{I}_1=G_{0,1,0,1,1,0,0,1,1}=\begin{tikzpicture}[baseline={([yshift=-0.8ex]current bounding box.center)}]\tikzstyle{every node}=[font=\small]	
\begin{feynman}
    	 \vertex (p1) {\(p_1\)};
    	 \vertex [right=0.8cm of p1] (b1) [dot]{};
    	 \vertex [right=0.5cm of b1] (b2)[dot]{};
    	 \vertex [right=0.8cm of b2] (p4){$p_4$};
    	 \vertex [above=0.8cm of p1](p2){$p_2$};
    	 \vertex [right=0.8cm of p2] (u1) [dot]{};
    	 \vertex [right=0.5cm of u1] (u2) [dot]{};
    	  \vertex [above=0.8cm of p4](p3){$p_3$};
    	  \diagram* {
(p1) -- [ultra thick] (b1)-- [ultra thick] (b2)-- [ultra thick] (p4),
    	   (b1)-- [thick] (u1),(b2) --  [thick] (u2), (u1)--[thick](b2),(p2) -- [ultra thick] (u1)-- [ultra thick] (u2)-- [ultra thick] (p3),
    	  };
    \end{feynman}  
\end{tikzpicture},&&\mathcal{I}_2={q^2\over - 2}G_{0,2,0,1,1,0,0,1,1}=\begin{tikzpicture}[baseline={([yshift=-0.8ex]current bounding box.center)}]\tikzstyle{every node}=[font=\small]	
\begin{feynman}
    	 \vertex (p1) {\(p_1\)};
    	 \vertex [right=0.8cm of p1] (b1) [dot]{};
    	 \vertex [right=0.5cm of b1] (b2)[dot]{};
    	 \vertex [right=0.8cm of b2] (p4){$p_4$};
    	  \vertex [right=0.25cm of b1] (la)[]{};
    	   \vertex [above=0.4cm of la] (g2)[]{$\circ$};
    	 \vertex [above=0.8cm of p1](p2){$p_2$};
    	 \vertex [right=0.8cm of p2] (u1) [dot]{};
    	 \vertex [right=0.5cm of u1] (u2) [dot]{};
    	  \vertex [above=0.8cm of p4](p3){$p_3$};
    	   \vertex [right=0.25cm of u1] (lb)[]{};
    	  \diagram* {
(p1) -- [ultra thick] (b1)-- [ultra thick] (b2)-- [ultra thick] (p4),
    	   (b1)-- [thick] (u1),(b2) --  [thick] (u2), (u1)--[thick](b2),(p2) -- [ultra thick] (u1)-- [ultra thick] (u2)-- [ultra thick] (p3),
    	  };
    \end{feynman}  
    \end{tikzpicture},\nn\\
	\mathcal{I}_3={q^2\over -2} G_{0,1,0,1,1,0,0,2,2}=\begin{tikzpicture}[baseline={([yshift=-0.8ex]current bounding box.center)}]\tikzstyle{every node}=[font=\small]	
\begin{feynman}
    	 \vertex (p1) {\(p_1\)};
    	 \vertex [right=0.8cm of p1] (b1) [dot]{};
    	 \vertex [right=0.5cm of b1] (b2)[dot]{};
    	 \vertex [right=0.8cm of b2] (p4){$p_4$};
    	  \vertex [right=0.25cm of b1] (la)[]{$\circ$};
    	 \vertex [above=0.8cm of p1](p2){$p_2$};
    	 \vertex [right=0.8cm of p2] (u1) [dot]{};
    	 \vertex [right=0.5cm of u1] (u2) [dot]{};
    	  \vertex [above=0.8cm of p4](p3){$p_3$};
    	   \vertex [right=0.25cm of u1] (lb)[]{$\circ$};
    	  \diagram* {
(p1) -- [ultra thick] (b1)-- [ultra thick] (b2)-- [ultra thick] (p4),
    	   (b1)-- [thick] (u1),(b2) --  [thick] (u2), (u1)--[thick](b2),(p2) -- [ultra thick] (u1)-- [ultra thick] (u2)-- [ultra thick] (p3),
    	  };
    \end{feynman}  
    \end{tikzpicture},&&\mathcal{I}_4=G_{0,1,1,0,1,0,0,1,1}=\begin{tikzpicture}[baseline={([yshift=-0.8ex]current bounding box.center)}]\tikzstyle{every node}=[font=\small]	
\begin{feynman}
    	 \vertex (p1) {\(p_1\)};
    	 \vertex [right=0.8cm of p1] (b1) [dot]{};
    	 \vertex [right=0.5cm of b1] (b2)[dot]{};
    	 \vertex [right=0.8cm of b2] (p4){$p_4$};
    	 \vertex [above=0.8cm of p1](p2){$p_2$};
    	 \vertex [right=0.8cm of p2] (u1) [dot]{};
    	 \vertex [right=0.5cm of u1] (u2) [dot]{};
    	  \vertex [above=0.8cm of p4](p3){$p_3$};
    	  \diagram* {
(p1) -- [ultra thick] (b1)-- [ultra thick] (b2)-- [ultra thick] (p4),
    	   (b1)-- [thick] (u1),(b1) --  [thick,out=65,in=115,looseness=0.5,min distance=0.4cm] (b2), (u2)--[thick](b2),(p2) -- [ultra thick] (u1)-- [ultra thick] (u2)-- [ultra thick] (p3),
    	  };
    \end{feynman}  
\end{tikzpicture},\nn\\
	\mathcal{I}_5={q^4\over 4} G_{1,1,1,1,1,0,0,1,1}=\begin{tikzpicture}[baseline={([yshift=-0.8ex]current bounding box.center)}]\tikzstyle{every node}=[font=\small]	
\begin{feynman}
    	 \vertex (p1) {\(p_1\)};
    	 \vertex [right=0.8cm of p1] (b1) [dot]{};
    	 \vertex [right=0.5cm of b1] (b2)[dot]{};
    	 \vertex [right=0.8cm of b2] (p4){$p_4$};
    	  \vertex [right=0.25cm of b1] (la)[]{};
    	   \vertex [above=0.4cm of b1] (g1)[dot]{};
    	    \vertex [above=0.4cm of b2] (g2)[dot]{};
    	 \vertex [above=0.8cm of p1](p2){$p_2$};
    	 \vertex [right=0.8cm of p2] (u1) [dot]{};
    	 \vertex [right=0.5cm of u1] (u2) [dot]{};
    	  \vertex [above=0.8cm of p4](p3){$p_3$};
    	   \vertex [right=0.25cm of u1] (lb)[]{};
    	  \diagram* {
(p1) -- [ultra thick] (b1)-- [ultra thick] (b2)-- [ultra thick] (p4),
    	   (b1)-- [thick] (u1),(b2) --  [thick] (u2), (g1)--[thick](g2),(p2) -- [ultra thick] (u1)-- [ultra thick] (u2)-- [ultra thick] (p3),
    	  };
    \end{feynman}  
    \end{tikzpicture},&&
	\mathcal{I}_6={q^2\over -2} G_{1,1,1,1,1,{-}1,{-}1,1,1}=\begin{tikzpicture}[baseline={([yshift=-0.8ex]current bounding box.center)}]\tikzstyle{every node}=[font=\small]	
\begin{feynman}
    	 \vertex (p1) {\(p_1\)};
    	 \vertex [right=0.8cm of p1] (b1) [dot]{};
    	 \vertex [right=0.5cm of b1] (b2)[dot]{};
    	 \vertex [right=0.8cm of b2] (p4){$p_4$};
    	  \vertex [right=0.25cm of b1] (la)[]{};
    	     \vertex [above=0.2cm of la] (n1)[]{$\mathcal{D}_6$};
    	   \vertex [above=0.4cm of b1] (g1)[dot]{};
    	    \vertex [above=0.4cm of b2] (g2)[dot]{};
    	 \vertex [above=0.8cm of p1](p2){$p_2$};
    	 \vertex [right=0.8cm of p2] (u1) [dot]{};
    	 \vertex [right=0.5cm of u1] (u2) [dot]{};
    	  \vertex [above=0.8cm of p4](p3){$p_3$};
    	   \vertex [right=0.25cm of u1] (lb)[]{};
    	    \vertex [above=0.4cm of n1] (n2)[]{$\mathcal{D}_7$};
    	  \diagram* {
(p1) -- [ultra thick] (b1)-- [ultra thick] (b2)-- [ultra thick] (p4),
    	   (b1)-- [thick] (u1),(b2) --  [thick] (u2), (g1)--[thick](g2),(p2) -- [ultra thick] (u1)-- [ultra thick] (u2)-- [ultra thick] (p3),
    	  };
    \end{feynman}  
    \end{tikzpicture},\nn\\
    \mathcal{I}_7={q^2\over -2} G_{0,1,0,1,1,1,1,1,1}=\begin{tikzpicture}[baseline={([yshift=-0.8ex]current bounding box.center)}]\tikzstyle{every node}=[font=\small]	
\begin{feynman}
    	 \vertex (p1) {\(p_1\)};
    	 \vertex [right=0.8cm of p1] (b1) [dot]{};
    	 \vertex [right=0.5cm of b1] (b2)[dot]{};
    	 \vertex [right=0.8cm of b2] (p4){$p_4$};
    	  \vertex [right=0.25cm of b1] (la)[]{};
    	   \vertex [right=0.25cm of b1] (b3)[dot]{};
    	 \vertex [above=0.8cm of p1](p2){$p_2$};
    	 \vertex [right=0.8cm of p2] (u1) [dot]{};
    	 \vertex [right=0.5cm of u1] (u2) [dot]{};
    	  \vertex [above=0.8cm of p4](p3){$p_3$};
    	   \vertex [right=0.25cm of u1] (lb)[]{};
    	    \vertex [right=0.25cm of u1] (u3)[dot]{};
    	  \diagram* {
(p1) -- [ultra thick] (b1)-- [ultra thick] (b2)-- [ultra thick] (p4),
    	   (b1)-- [thick] (u1),(b2) --  [thick] (u2), (b3)--[thick](u3),(p2) -- [ultra thick] (u1)-- [ultra thick] (u2)-- [ultra thick] (p3),
    	  };
    \end{feynman}  
    \end{tikzpicture}.&&
    \end{align}
where circles on lines denote squared propagators and for $\cI_6$ we have also indicated the presence of two numerator factors. We also set the master integral $ G_{1,0,1,1,1,0,0,1,1}$ to zero as it  does not contribute to classical four-dimensional physics.

We set up the system of differential equations as 
\begin{align}
	{d \mathcal I_i\over dy}\, =\, A_{ij} \,  \mathcal I_j\ ,
\end{align}
where
\begin{align}
	A=\left(
\begin{array}{ccccccc}
 \frac{(11-3 D) y}{y^2-1} & -\frac{2 y}{y^2-1} & -\frac{1}{(D-4) \left(y^2-1\right)} & 0 & 0 & 0 & 0 \\
 0 & \frac{y}{1-y^2} & \frac{1}{2-2 y^2} & 0 & 0 & 0 & 0 \\
 3 (D-4)^3 & \frac{2 (D-4)^2 \left( y^2+{5-D\over D-4}\right)}{y^2-1} & \frac{(D-4) y}{y^2-1} & 0 & 0 & 0 & 0 \\
 0 & 0 & 0 & \frac{(D-3) y}{y^2-1} & 0 & 0 & 0 \\
 0 & \frac{3 y}{y^2-1} & \frac{1}{2 (D-4) \left(y^2-1\right)} & \frac{2 \left(2 D^2-13 D+21\right) y}{(D-4) \left(y^2-1\right)^2} & \frac{(D-5) y}{y^2-1} & -\frac{2 (D-4)}{y^2-1} & 0 \\
 \frac{3 (D-4)}{4} & 2 & 0 & \frac{4 D^2-26 D+42}{(D-4) \left(y^2-1\right)} & \frac{D-4}{2} & -\frac{(D-4) y}{y^2-1} & 0 \\
 0 & -\frac{2}{y^2-1} & 0 & 0 & 0 & 0 & -\frac{2 y}{y^2-1} \\
\end{array}
\right)\, .
\end{align}
 Following the two steps outlined 
in Section \ref{sec-DE}, we arrive at
\begin{align}
	\mathcal{I}'_1&=16 \epsilon ^4\sqrt{y^2-1} \, \mathcal{I}_1 \ , &\mathcal{I}'_2&=-8 \epsilon ^3\sqrt{y^2-1} \, \mathcal{I}_2 \ , &\mathcal{I}'_3&=4 \epsilon ^2 \left(\mathcal{I}_3-4 y \mathcal{I}_2 \epsilon \right) \ , \nn \\
	\mathcal{I}'_4&=\frac{4 \epsilon ^2 \left(8 \epsilon ^2-6 \epsilon +1\right)}{\sqrt{y^2-1}} \, \mathcal{I}_4 \ , &\mathcal{I}'_5&=16 \epsilon ^4\sqrt{y^2-1} \, \mathcal{I}_5 \ , &\mathcal{I}'_6&=16 \epsilon ^4  \, \mathcal{I}_6 \ , \nn\\
	\mathcal{I}'_7&=16 \epsilon ^4 \left(y^2-1\right) \,  \mathcal{I}_7 \ .
\end{align}
The canonical basis satisfies the following differential equation,
\begin{align}\label{eq:diffEq}
	{d \mathcal I'_i\over dy}\, =\,  \epsilon\,  A'_{ij}  \mathcal I'_j\, ,
\end{align}
where
\begin{align}
	A' = \left(
\begin{array}{cccccccc}
 \frac{6 y}{y^2-1} & 0 & \frac{2}{\sqrt{y^2-1}} & 0 & 0  & 0 & 0 \\
 0 & -\frac{2 y}{y^2-1} & \frac{1}{\sqrt{y^2-1}} & 0 & 0  & 0 & 0 \\
 -\frac{6}{\sqrt{y^2-1}} & -\frac{4}{\sqrt{y^2-1}} & 0 & 0  & 0 & 0 & 0 \\
 0 & 0 & 0 & -\frac{2 y}{y^2-1} & 0 &  0 & 0 \\
 0 & -\frac{4 y}{y^2-1} & -\frac{1}{\sqrt{y^2-1}} & -\frac{4 y}{y^2-1}  & -\frac{2 y}{y^2-1} & \frac{4}{\sqrt{y^2-1}} & 0 \\
 -\frac{3}{2 \sqrt{y^2-1}} & -\frac{4}{\sqrt{y^2-1}} & 0 & -\frac{4}{\sqrt{y^2-1}}  & -\frac{1}{\sqrt{y^2-1}} & \frac{2 y}{y^2-1} & 0 \\
 0 & \frac{4}{\sqrt{y^2-1}} & 0 &  0 & 0 & 0 & 0 \\
\end{array}
\right).
\end{align}
At two loops, the study of the boundary conditions is well documented in the literature \cite{Parra-Martinez:2020dzs, DiVecchia:2021bdo,Bjerrum-Bohr:2021vuf}, albeit for different bases and using slightly different methods.  
As discussed in Section~\ref{zigzag-sec} the bow-tie topology would have to be included in the system of differential equations if we want to work in general dimensions. In this paper we focus on $D=4-2\eps$, and hence we can
drop the this topology.

The final step is now to provide appropriate boundary conditions for the solutions to this system of differential equations. A  systematic method is provided in Appendix~\ref{sec:TGAsy}, based on the study of  the asymptotic expansion of Feynman integrals \cite{Beneke:1997zp,Pak:2010pt} in the static limit $y\rightarrow 1$. Using the results of Appendix~\ref{sec:TGAsy}, we find  
\begin{align}
\begin{split}
\label{7.16}
	\mathcal{I}'_{1}(y)&\rightarrow \left(-{q^2\over 2}\right)^{-2\epsilon}\Big(64 \pi ^2 \sqrt{2}\sqrt{y-1} \epsilon ^3+\mathcal O(y-1,\epsilon ^4)\Big) \ , \\
	\mathcal{I}'_{2}(y)&\rightarrow \left(-{q^2\over 2}\right)^{-2\epsilon}\Big(16 i \pi  \epsilon + 16 \pi  \epsilon ^2 (-i \log [32 (y-1)]-2 i \gamma +\pi )+\mathcal O(y-1,\epsilon ^3)\Big) \ , \\
\mathcal{I}'_{3}(y)&\rightarrow \left(-{q^2\over 2}\right)^{-2\epsilon}\Big(32 \pi ^2 \epsilon ^2-64 i \sqrt{2} \pi  \sqrt{y-1} \epsilon ^2+\mathcal O(y-1,\epsilon ^3)\Big) \ , \\
\mathcal{I}'_{4}(y)&\rightarrow \left(-{q^2\over 2}\right)^{-2\epsilon}\Big(-16 i \pi  \epsilon+16 i \pi  \epsilon ^2 \left(\log [32 \left(y-1\right)]+2 \gamma +i \pi \right)+\mathcal O(y-1,\epsilon ^4)\Big) \ , \\
\mathcal{I}'_{5}(y)& \rightarrow 0+\mathcal O(\sqrt{y-1}) \ ,
\\
\mathcal{I}'_{6}(y)&\rightarrow 0+\mathcal O(y-1) \ , \\
\mathcal{I}'_{7}(y)&\rightarrow 0+\mathcal O(\sqrt{y-1}) \ .
\end{split}
\end{align}
With these boundary conditions, we finally obtain the solutions of the canonical basis from the differential equations 
\begin{align}
\begin{split}
\mathcal{I}'_{1}&=2 \epsilon ^3 \arccosh(y) \left(\mathcal{I}_{3,2}'-4 \, \arccosh(y) \mathcal{I}_{2,1}'\right)+\mathcal{O}(\epsilon ^4) \ , \\
	\mathcal{I}'_{2}&=\epsilon  \mathcal{I}_{2,1}'+\epsilon ^2 \left(\mathcal{I}_{2,2}'-\log \left(y^2-1\right) \mathcal{I}_{2,1}'\right)+\mathcal{O}(\epsilon ^3) \ , \\
	\mathcal{I}'_{3}&=\epsilon ^2 \left(\mathcal{I}_{3,2}'-4 \, \arccosh(y) \mathcal{I}_{2,1}'\right)+\mathcal{O}(\epsilon ^3) \ , \\
	\mathcal{I}'_{4}&=-\epsilon  \mathcal{I}_{2,1}'+\epsilon ^2 \left(\log \left(y^2-1\right) \mathcal{I}_{2,1}'-\mathcal{I}_{2,2}'\right)+\mathcal{O}(\epsilon ^3) \ , \\
	\mathcal{I}'_{5}&=-\epsilon ^3 \arccosh(y) \left(\mathcal{I}_{3,2}'-4 \,  \arccosh(y) \mathcal{I}_{2,1}'\right)+\mathcal{O}(\epsilon ^4) \ , \\
	\mathcal{I}'_{6}&=\mathcal{O}(\epsilon ^4) \ , \\
	\mathcal{I}'_{7}&=4 \epsilon ^2 \arccosh(y) \mathcal{I}_{2,1}'+4 \epsilon ^3 \arccosh(y) \left(\mathcal{I}_{2,2}'-\log \left(y^2-1\right) \mathcal{I}_{2,1}'\right)+\mathcal{O}(\epsilon ^4) \ ,
\end{split}
\end{align}
where
\begin{align}
\begin{split}
    \mathcal{I}'_{2,1}&= 16\pi i\,  \left(-{q^2\over 2}\right)^{-2\epsilon} \ , \\ 
	\mathcal{I}'_{2,2}&=(16\pi^2 -64\pi i \log(2)-32\pi i\, \gamma  )\left(-{q^2\over 2}\right)^{-2\epsilon} \, ,\\ 	\mathcal{I}'_{3,2}&=32\pi^2 \left(-{q^2\over 2}\right)^{-2\epsilon} \ ,
\end{split}
\end{align}
and $\gamma$ denotes the Euler-Mascheroni constant. 

The possible boundary conditions for the master integrals are related to different  physical regions (e.g.~the potential or the soft region), see for instance \cite{efremov1980factorization,Beneke:1997zp,collins1998proof} and Appendix \ref{sec:TGAsy}. 
These regions are related to different  asymptotic behaviours of the integrals in a given limit (e.g.~the static limit $y\to 1$ or the ultrarelativistic limit $y\to \infty$). In general,  a master integral  has several regions corresponding to a given limit.  For a set of master integrals  of a system of  differential equations, the asymptotic behaviour is also  constrained by the system, and therefore 
 the regions for the master integrals are usually not independent.  Here we related the independent regions for the master integrals to the physical regions.  After applying the differential equation constraints, we have two independent regions characterised by the boundary values%
 \footnote{Here LD stands for leading.}
\begin{align}
	\mathcal{I}^{(R_2)}_{2,\rm{LD}}:\quad  \mathcal{I}'_{2,1}\epsilon+\mathcal{I}'_{2,2}\epsilon^2 &&\mathcal{I}^{(R_1)}_{3,\rm{LD}}: \quad \mathcal{I}'_{3,2}\epsilon^2.
\end{align}
We  find that $\mathcal{I}^{(R_1)}_{3,\rm{LD}}$ corresponds to  the potential region and $\mathcal{I}^{(R_2)}_{2,\rm{LD}}$  to  the radiation  region. We will  also see in the next section that the real part of the two regions  generates the conservative   and radiation reaction contributions to the scattering angle, respectively.   An interesting observation is that the values of the leading terms in the two regions are all determined from three master integrals $\mathcal{I}_2^\prime$, $\mathcal{I}_3^\prime$ and $\mathcal{I}_4^\prime$, where  in the static limit the latter can be expressed in terms of the former two.

\section{Final result and scattering angle from the HEFT amplitude}
\label{sec:8}

We now combine the results of Sections \ref{sec:2loopheft} and \ref{sec:canbasis} and provide an expression for the  scattering angle up to two loops. 
 
To begin with,  we express the 2MPI contribution to the two-loop amplitude at order $\cO(m_1^3 m_2^3)$ directly in terms of the canonical basis, 
\begin{align}
\begin{split}
\lamp^{(2,\rm 2MPI)}_{m_1^3 m_2^3}&=	m_1^3 m_2^3\frac{(32 \pi  G_N)^3}{2!4(4 \pi )^{D}}\Bigg[\frac{\left(1-2 y^2\right)^2 \mathcal{I}'_5}{64 \sqrt{y^2-1}  \epsilon ^4}+\frac{y \left(2 y^2-3\right) \left(1-2 y^2\right)^2 \mathcal{I}'_7}{256 \left(y^2-1\right)^2 \epsilon ^4}+\frac{\left(2 y^2+1\right) \mathcal{I}'_1}{32 \sqrt{y^2-1} \epsilon ^4}\\
	&+\frac{\left(88 y^4-64 y^2-35\right) \mathcal{I}'_4}{192 \sqrt{y^2-1}\epsilon ^3}+\frac{\left(68 y^6-100 y^4-8 y^2+43\right) \mathcal{I}'_2}{192 (y^2-1)^{3/2} \epsilon ^3}\\
	&+\frac{y \left(36 y^6-114 y^4+132 y^2-55\right) \mathcal{I}'_3}{384 \left(y^2-1\right)^2 \epsilon ^3}+\frac{y \mathcal{I}'_6}{8 \epsilon ^4}\Bigg]+\mathcal{O}(\epsilon) \ . 
\end{split}
\end{align}
Next, using the expressions for the master integrals from \eqref{7.16}, we obtain the following result for this
contribution, which also includes the radiation-reaction part:
\begin{align}
\label{res2loops}
\begin{split}
\textrm {Re}[	\lamp^{(2,\rm 2MPI)}_{m_1^3 m_2^3}]&=	-G_N^3{m_1^3 m_2^3 }\frac{8 \pi   \left(-q^2\right)^{-2 \epsilon }}{\epsilon }\\
&\Bigg[\frac{\left(5 y^2-8\right) \left(1-2 y^2\right)^2 }{6 \left(y^2-1\right)^{3/2}}-\frac{y \left(2 y^2-3\right) \left(1-2 y^2\right)^2  \arccosh(y)}{2 \left(y^2-1\right)^2}\\
	&+\frac{y \left(55-6 y^2 \left(6 y^4-19 y^2+22\right)\right) }{6 \left(y^2-1\right)^2}+\frac{\left(4 y^4-12 y^2-3\right) \arccosh(y)}{ \sqrt{y^2-1}}\Bigg]\, .
	\end{split}
\end{align}
The   first line corresponds to  the radiation-reaction terms, while the second  to the conservative part. 
The imaginary part of the 2MPI HEFT amplitude is 
\begin{align}
\begin{split}
\label{imagpart}
\textrm {Im}[	\lamp^{(2,\rm 2MPI)}_{m_1^3 m_2^3}]	&{=} 4(4\pi) ^{4-D} G_N^3 \frac{\left(-q^2\right)^{-2 \epsilon }}{\epsilon }\Bigg[\frac{ 4 \left(4 y^2 \left(y^2-3\right)-3\right) \arccosh(y)^2}{\sqrt{y^2-1}  }\\
	&{-}\Big({1\over\epsilon}{-}\log [16e^{2\gamma} \left(y^2{-}1\right)]\Big)\left(1{-}2 y^2\right)^2 \left({5 y^2{-}8\over 3 (y^2{-}1)^{3/2}}{+}{ \left(3{-}2 y^2\right)  
	y \, \arccosh(y)\over  (y^2{-}1)^{2}}\right)\\
	&{+}\frac{140 y^6{-}220 y^4{+}127 y^2{-}56}{9(y^2{-}1)^{3/2}  }{+}\frac{(97{-}88 y^6{+}240 y^4{-}240 y^2)  y \, \arccosh(y)}{3 \left(y^2{-}1\right)^{2}  }\Bigg]\, .
	\end{split}
\end{align}
The terms in the HEFT amplitude  of order $\cO(m_1^4 m_2^2)$ and $\cO(m_1^4 m_2^2)$, 
which contribute only to the  real part of the amplitude, 
have been computed in 
Section~\ref{sec-2loop-fan}, see \eqref{2loop-fan}. Combining them with \eqref{res2loops}, we arrive at  the complete 2MPI two-loop HEFT amplitude:
\begin{align}
\label{M22} 
\begin{split}
\text{Re} [ \lamp^{(2,\rm 2MPI)} ]&=
G_N^3m_1^2 m_2^2 (m_1^2 + m_2^2)\,  \frac{\left(-q^2\right)^{-2 \epsilon }}{\epsilon} \, \frac{2\pi   \left(64 y^6-120 y^4+60 y^2-5\right)  }{3 \left(y^2-1\right)^2  }\\
&- 8 \pi \,  G_N^3{m_1^3 m_2^3 }\frac{  \left({-}q^2\right)^{-2 \epsilon }}{\epsilon }\Bigg[\left(1{-}2 y^2\right)^2 \left(\frac{\left(5 y^2{-}8\right) }{6 \left(y^2{-}1\right)^{3/2}}{-}\frac{y \left(2 y^2{-}3\right)\arccosh(y)}{2 \left(y^2{-}1\right)^2}\right)\\
	& + \frac{y \left(55-6 y^2 \left(6 y^4-19 y^2+22\right)\right) }{6 \left(y^2-1\right)^2}+\frac{\left(4 y^4-12 y^2-3\right) \arccosh(y)}{ \sqrt{y^2-1}}\Bigg]\, .
\end{split}
\end{align}
The second line is the radiation reaction part. We observe that \eqref{M22} is identical to the matrix element of the operator $N$ introduced in \cite{Damgaard:2021ipf}, see Eq.~(3.30) of that paper. 

Finally, we  compute the scattering angle. First we introduce the Fourier transform to  impact parameter space: 
\begin{align}
\begin{split}
\delta_{\rm HEFT}^{(L)}&= \frac{1}{4m_1m_2\sqrt{y^2-1}}
\int\!\frac{d^{D-2}q}{(2\pi)^{D-2}}\, e^{- i \vec{q}\Cdot \vec{b}}  \, \lamp^{(L,\rm 2MPI)}_{\rm HEFT}\, . 
\end{split}
\end{align}
At tree level we find 
\begin{align}
\begin{split}
\delta_{\rm HEFT}^{(0)}&= - \frac{1}{4m_1m_2\sqrt{y^2-1}}
\int\!\frac{d^{D-2}q}{(2\pi)^{D-2}}\, e^{- i \vec{q}\Cdot \vec{b}} (32 \pi  G_N) {m_1^2 m_2^2\over D-2}\Big[(D-2)y^2 -1\Big]{1\over q^2}\\
&=\frac{G_N m_1 m_2  \left(\pi b^2\right)^{\epsilon } \big[y^2 (2-2 \epsilon )-1\big] \Gamma (-\epsilon )}{\sqrt{y^2-1} (1- \epsilon )}\\
& = 
G_N J^{2 \eps}\frac{ ( \pi s)^\epsilon (m_1 m_2)^{1-2\eps}    \big[y^2 (2-2 \epsilon )-1\big] }{(y^2-1)^{\frac{1}{2} + \eps} }\frac{\Gamma (-\epsilon )}{1- \epsilon }\\ &
\stackrel{\eps\to 0}{\longrightarrow} 
- \frac{G_N J^{2 \eps}}{\eps} m_1 m_2 \frac{2y^2-1}{\sqrt{y^2-1} }
\, ,
\end{split}
\end{align}
where we used $J = P b$, with  $P$ being  defined in \eqref{Pdef}. 

At one loop, by performing the Fourier transform we find \begin{align}
\begin{split}
\delta_{\rm HEFT}^{(1)}&=  \frac{1}{4m_1m_2\sqrt{y^2-1}}
\int\!\frac{d^{D-2}q}{(2\pi)^{D-2}}\, e^{- i \vec{q}\Cdot \vec{b}}  (m_1^2 m_2^3+m_2^2 m_1^3)\frac{3  \pi ^2 G_N^2  \left(5 y^2-1\right)}{\sqrt{-q^2}}\ \\
&= {G_N^2\over J}{3 \pi \over 4   \sqrt{s}}m_1^2 m_2^2 (m_1 + m_2)  \left(5 y^2-1\right)\, .
\end{split}
\end{align}
Finally, at two loops our result~is 
\begin{align}
\text{Re}\big[\delta_{\rm HEFT}^{(2)}\big]&=  \frac{1}{4m_1m_2\sqrt{y^2-1}}
\int\!\frac{d^{D-2}q}{(2\pi)^{D-2}}\, e^{- i \vec{q}\Cdot \vec{b}}  \text{Re}\big[ \lamp^{(2,\rm 2MPI)} \big]  \ \nn\\
  &= \frac{1}{J^2 }\frac{m_1m_2\sqrt{y^2-1}}{2\pi  s}
 \Big(\text{Re}\big[ \lamp^{(2,\rm 2MPI)}\big]{ \epsilon\over\left(-q^2\right)^{-2 \epsilon }}  \Big)_{\eps = 0} \, , 
\end{align}
where $\text{Re}\big[ \lamp^{(2,\rm 2MPI)}\big]$ is given in \eqref{res2loops}.
The scattering  angle is  then obtained from 
\eqref{chiHEFT}, and our final result up to two loops, or 3PM,  is
\begin{align}
\label{finalangle}
\begin{split}
	\chi &= \blue \frac{G_N}{J}\ \black \frac{2 m_1 m_2 \left(2 y^2-1\right)}{\sqrt{y^2-1}}+ \blue {G_N^2\over J^2}\ \black {3 \pi\over 4   \sqrt{s}}m_1^2 m_2^2 (m_1 + m_2)   \left(5 y^2-1\right)\\
	&+ \blue \frac{G_N^3}{J^3 }\ \black \frac{m_1m_2\sqrt{y^2-1}}{\pi  s}\Bigg\{m_1^2 m_2^2 (m_1^2 + m_2^2) \frac{2\pi   \left(64 y^6-120 y^4+60 y^2-5\right) }{3 \left(y^2-1\right)^2  }\\
&+{m_1^3 m_2^3 }(-8 \pi  )\bigg[\frac{\left(5 y^2-8\right) \left(1-2 y^2\right)^2 }{6 \left(y^2-1\right)^{3/2}}-\frac{y \left(2 y^2-3\right) \left(1-2 y^2\right)^2  \arccosh(y)}{2 \left(y^2-1\right)^2}\\
	&+\frac{y \left(55-6 y^2 \left(6 y^4-19 y^2+22\right)\right) }{6 \left(y^2-1\right)^2}+\frac{\left(4 y^4-12 y^2-3\right) \arccosh(y)}{ \sqrt{y^2-1}}\bigg]\Bigg\}\, .
\end{split}
\end{align}
This is in agreement with the result of \cite{Bern:2019nnu, Antonelli:2019ytb,DiVecchia:2021bdo,Bjerrum-Bohr:2021din}. The radiation reaction part (third line of \eqref{finalangle})  was also computed in \cite{Damour:2020tta,DiVecchia:2020ymx, Bjerrum-Bohr:2021din} and our result agrees with theirs. 

We conclude this section with two comments.  
We  note that the imaginary part of our 2MPI amplitude, quoted earlier in \eqref{imagpart},  is very similar (but not identical)  to that of  \cite{DiVecchia:2021bdo,Bjerrum-Bohr:2021din}. There is no reason for these  quantities to agree completely since there are subtle differences in how iterating terms are removed in \cite{DiVecchia:2021bdo,Bjerrum-Bohr:2021din}.  However the 
relation~\cite{DiVecchia:2021bdo}
\begin{align}
\lim_{\eps \to 0} \big[ \text{Re}\,   \delta_{\rm HEFT}^{(2)}\big]_{rr} \ = \ - \lim_{\eps\to 0}\big[  \pi \eps \, \text{Im} \,  \delta_{\rm HEFT}^{(2)}\big] \, , 
\end{align}
where $rr$ stands for the radiation reaction part, does hold also in our case, with  our $\big[ \text{Re}\,   \delta_{\rm HEFT}^{(2)}\big]_{rr}$ being identical to that of \cite{DiVecchia:2021bdo,Bjerrum-Bohr:2021din}.

The appearance  of the imaginary part \eqref{imagpart} 
in the 3PM amplitude is a very  interesting fact,   related to radiation emission. 
The radial action (from which the deflection angle is extracted)  is 
believed 
to be related  to the matrix element of the hermitian operator $N$ introduced in \cite{Damgaard:2021ipf}. As discussed in Section \ref{sec:loops}, the real part of our  $\delta_{\rm HEFT}$ is  identical to this matrix element -- with the extra imaginary part only arising from the first diagram in  (2.17) of that reference (the triple cut involving two massive particles and a radiation graviton). Hence at 3PM order we  just need to drop the imaginary part of   $\delta_{\rm HEFT}$ to obtain the matrix element of $N$ and hence the radial action. 

Beyond 3PM the situation is more involved. It is reasonable to expect that the correct prescription  in our HEFT approach is again to relate $\delta_{\rm HEFT}$ to a specific matrix element of $N$, generalising (2.10) of \cite{Damgaard:2021ipf} to the next PM order. Our expectation is  that our approach again automatically implements many of the subtractions implied by the definition of $N$ but will require the subtraction of a larger set of radiation cut diagrams.

 \section{All-loop conjecture for the classical contribution  in the probe limit}
 \label{sec:all-loop}
In this section we present a conjecture for the all-loop structure of the leading terms in the 2MPI amplitude in the probe limit. 
Consider the following ``fan'' diagram at $L$ loops: 
\begin{align}\label{fig:MCr}
	\begin{tikzpicture}[baseline={([yshift=-0.8ex]current bounding box.center)}]\tikzstyle{every node}=[font=\small]	
\begin{feynman}
    	 \vertex (a) {\(p_1\)}; 
         \vertex [right=3cm of a] (v1) [HV]{H};	 
    	 \vertex [right=3cm of v1] (b){$p_4$};
    	 \vertex [above=2.0cm of a](c){$p_2$};   	 
         \vertex [right=1.cm of c] (u1) [dot]{};
         \vertex [right=1.cm of u1] (u2) [dot]{};
         \vertex [right=1.cm of u2] (u3) []{};
         \vertex [right=1.cm of u3] (u4) [dot]{};
            \vertex [right=1.cm of u4] (u5) [dot]{};
    	  \vertex [right=1.cm of u5](d){$p_3$};
    	  \vertex [above=1.0cm of a] (cutL);
    	  \vertex [right=6.0cm of cutL] (cutR);
    	   \vertex [below=0.5cm of u3] (note)[]{$\cdots$};
    	   \vertex [above right=0.75cm of u1] (cut1a) {}; 
    	   \vertex [below right=0.75cm of u1] (cut1b) {}; 
    	    \vertex [right =1.5cm of cut1a] (cut2a) {}; 
    	   \vertex [right =1.5cm of cut1b] (cut2b) {}; 
    	    \vertex [right =1.5cm of cut2a] (cut3a) {}; 
    	   \vertex [right =1.5cm of cut2b] (cut3b) {}; 
    	  \diagram* {
(a) -- [fermion,thick] (v1)-- [fermion,thick] (b),
    	  (u1)--[photon, ultra thick,momentum'=\(\ell_1\)](v1), (u2)-- [photon, ultra thick,momentum=\(~\)] (v1),(u4)-- [photon, ultra thick,momentum'=\(~\)] (v1),(u5)-- [photon, ultra thick,momentum=\(\ell_{L+1}\)] (v1), 
    	  (c) -- [fermion,thick]    (d), (cutL)--[dashed, red,thick] (cutR), (cut1a)--[ red,thick] (cut1b),(cut2a)--[red,thick] (cut2b),(cut3a)--[ red,thick] (cut3b)
    	  };
    \end{feynman}  
    \end{tikzpicture}
\end{align}
 The lower part of the diagram  is  tree-level amplitude in  HEFT, which is of    $\mathcal{O}(m_1^2)$ while the upper is of  $\mathcal{O}(m_2^{L+2})$
 at  $L$ loops.  Therefore the diagram behaves as $\mathcal{O}(m_1^2 m_2^{L+2})$, which dominates the classical gravitational scattering in the probe limit $m_2 \gg m_1$.

A similar procedure to the one followed in previous sections at one and two loops  allows us to express the $L$-loop diagram in terms of a basis of master integrals. Leaving a systematic study of general higher-loop diagrams for the future, here we conjecture that the simple fan structure found at  one and two loops gives the leading contribution in the probe limit also at higher loops. In particular, all the internal propagators arising from the lower tree-level HEFT amplitude are absent in the master integrals. That implies that there is only one master integral,  given by 
\begin{align} 
	G^{(L)}&=
	\begin{tikzpicture}[baseline={([yshift=-0.8ex]current bounding box.center)}]\tikzstyle{every node}=[font=\small]	
\begin{feynman}
    	 \vertex (a) {\(p_1\)}; 
         \vertex [right=3cm of a] (v1) [dot]{};	 
    	 \vertex [right=3cm of v1] (b){$p_4$};
    	 \vertex [above=2.0cm of a](c){$p_2$};   	 
         \vertex [right=1.cm of c] (u1) [dot]{};
         \vertex [right=1.cm of u1] (u2) [dot]{};
         \vertex [right=1.cm of u2] (u3) []{};
         \vertex [right=1.cm of u3] (u4) [dot]{};
            \vertex [right=1.cm of u4] (u5) [dot]{};
    	  \vertex [right=1.cm of u5](d){$p_3$};
    	  \vertex [above=1.0cm of a] (cutL);
    	  \vertex [right=6.0cm of cutL] (cutR);
    	   \vertex [below=0.5cm of u3] (note)[]{$\cdots$};
    	    \vertex [above right=0.75cm of u1] (cut1a) {}; 
    	   \vertex [below right=0.75cm of u1] (cut1b) {}; 
    	    \vertex [right =1.5cm of cut1a] (cut2a) {}; 
    	   \vertex [right =1.5cm of cut1b] (cut2b) {}; 
    	    \vertex [right =1.5cm of cut2a] (cut3a) {}; 
    	   \vertex [right =1.5cm of cut2b] (cut3b) {}; 
    	  \diagram* {
(a) -- [thick] (v1)-- [thick] (b),
    	  (u1)--[thick](v1), (u2)-- [thick] (v1),(u4)-- [thick] (v1),(u5)-- [thick] (v1), 
    	  (c) -- [thick]    (d),  (cut1a)--[ red,thick] (cut1b),(cut2a)--[ red,thick] (cut2b),(cut3a)--[red,thick] (cut3b)
    	  };
    \end{feynman}  
    \end{tikzpicture}\nn\\
	&=(2\pi )^L {1\over \pi^{LD/2}}\int d\ell_1\cdots d\ell_L {\delta(\ell_1\cdot v_2)\delta(\ell_2\cdot v_2)\cdots \delta(\ell_L\cdot v_2)\over \ell_1^2\ell_2^2\cdots \ell_L^2 \ell_{L+1}^2}.
\end{align}
The needed HEFT amplitude has some simple general properties, which allow us to write down a general expression for the fan diagram at  an arbitrary number of loops~$L$. The tree-level amplitude in the HEFT as obtained from the double copy contains a factor  ${1\over (p\Cdot v)^2}$ from each heavy-mass propagator, which will generate a factor $1\over y^2-1$ after the IBP reduction, resulting at $L$ loops in a factor of  $1\over (y^2-1)^L$.  The terms that do not contain any heavy propagators will have a similar  velocity dependence as the tree-level amplitude,  which is of  $ \mathcal{O}(y^2)$ in the limit $y\rightarrow \infty$.  Therefore we conclude that the $L$-loop contribution to the classical potential in the probe limit takes the  form 
\begin{align}
\lamp_{m_1^2m_2^{L+2} }^{(L)}\sim  \frac{m_1^2m_2^{L+2} G_N^{L+1} }{ \left(y^2-1\right)^L}\times \Big[\text{degree-$(2L+2)$ polynomial in $y$}\Big] 
G^{(L)}
\, .
\end{align}
We can now compare  the tree-level amplitude 
\begin{align}
	\lamp^{(0)}_{m_1^2 m_2^2}= -  \frac{32 \pi  G_N}{ q^2\black } {m_1^2 m_2^2\over D-2}\big[(D-2)y^2 -1\big]\, , 
\end{align}
together with the one-loop and  two-loop results in \eqref{eq:MC1} and  \eqref{eq:MC2}, respectively. We  then observe some interesting pattern for the $D$-dependent factors: 
\begin{itemize}
	\item The overall prefactor is of the form $\dfrac{ (D-3)^{L}}{(D-2)^{L+1}}$. 
	\item In the degree-$(2L+2)$ polynomial in $y$, the coefficient of the $y^2$ term is proportional to $D-2+L (D-3)$. 
	\item The ratio between the $y^{2j+2}$ coefficient and the  $y^{2j}$ coefficient in the polynomial is proportional to $ D-2+L (D-3)+2j$.
\end{itemize}
Combining  these observations, we expect that the amplitude contributing to the scattering angle  in the probe limit at $L$ loops takes the form 
\begin{align}
\label{eq:AllLoopProbe}
\begin{split}
	\lamp_{m_1^2m_2^{L+2}}^{(L)}&={1\over (L+1)!}\frac{(32 \pi  G_N)^{L+1}}{2^L(4 \pi )^{LD/2}} \frac{m_1^2m_2^{L+2}(D-3)^{L}  }{ (D-2)^{L+1} \left(y^2-1\right)^L}\\
	&\times \Big[\sum_{i=0}^{L+1} c_{L,i} \prod_{j=0}^{i-1}\Big(D-2+L (D-3)+2j\Big)y^{2i} \Big]G^{(L)}\, , 
	\end{split}
\end{align}
where $c_{L,i}$ are some undetermined $D$-independent constants.  
For example, at three loops, the relevant amplitude  in the probe limit is conjectured to be 
\begin{align}
\label{9.6}
\begin{split}
\lamp_{m_1^2m_2^{5}}^{(3)}&=	{1\over 4!}\frac{(32 \pi  G_N)^{4}}{8(4 \pi )^{3D/2}}  \frac{m_1^2m_2^{5}(D-3)^{3}  }{ (D-2)^{4} \left(y^2-1\right)^3} \Big[c_{3,4}(4D-5)(4D-7)(4D-9)(4D-11)y^8\\
&+c_{3,3}(4D-7)(4D-9)(4D-11)y^6+c_{3,2}(4D-9)(4D-11)y^4\\
&+c_{3,1}(4D-11)y^2+c_{3,0} \Big] G^{(3)}\, .
\end{split}
\end{align}
We have verified this pattern up to 9PM by comparing   \eqref{eq:AllLoopProbe}   to  the scattering  angle  for a massive probe particle  in the background of a Schwarzschild black hole (see for example the Appendix of \cite{KoemansCollado:2019ggb}, and we conjecture that it holds for any loop order.

\section{Outlook}
\label{sec:10}

In this paper we have studied the $2\to 2$ scattering amplitude of two heavy particles  at one and two loops using a formalism based on heavy-mass effective field theory.  We have defined a new HEFT phase as the Fourier transform to impact parameter space of the sum of all 2MPI  diagrams in this theory. We have found that  this phase is directly related to the scattering angle by taking the derivative with respect to  the angular momentum of  the binary system.

An important question is to understand more precisely the connection of our work to the results of \cite{Bern:2021dqo,Damgaard:2021ipf}, and in particular the relation of our HEFT phase to the radial action (or the real part thereof), which our phase agrees with for the conservative part. This hints at a strong relation between the ab initio  removal of iterating terms advocated in \cite{Bern:2021dqo} and   our approach, avoiding the task of explicitly evaluating terms which  exponentiate as usually done  in the eikonal approach. It would also be interesting to compare to worldline approaches, which   tackle directly the computation of an  effective action sitting  in an exponent \cite{Kalin:2020mvi}.

An obvious task  consists in  generalising our work to higher loops. The use of our novel colour-kinematic duality \cite{Brandhuber:2021kpo} will be crucial to produce compact expressions for the tree amplitudes entering the unitarity cuts. An additional  challenge beyond 3PM is that elliptic functions appear \cite{Bern:2021dqo,Dlapa:2021npj}.
The inclusion of spin effects would also be interesting to discuss within our HEFT approach, as well as the computation of other classical observables in the spirit of \cite{Kosower:2018adc,Cristofoli:2021vyo}.    

Finally, since our approach has produced integrands valid in $D$ dimensions, it would be interesting to work out the 3PM integrated amplitude beyond the probe limit in a general number of dimensions. This requires work on the relevant master integrals, setting up the relevant differential equations around integer dimension $D>4$, and computing the corresponding boundary conditions.  We leave these questions for future work.

\section*{Acknowledgements}

We would like to thank Manuel Accettulli Huber, Clifford Cheung, Stefano De Angelis, Henrik Johansson, Jung-Wook Kim, Rodolfo Russo and Mao Zeng for  discussions and comments. We also thank Ievgen Dubovyk, Janusz Gluza, Roman Lee and Tord Riemann for  sharing their programmes AMBREv3 and LiteRed2, and Gregor K\"{a}lin for discussions  and testing of the programme  LiteRed.  This work  was supported by the Science and Technology Facilities Council (STFC) Consolidated Grants ST/P000754/1 \textit{``String theory, gauge theory \& duality''} and  ST/T000686/1 \textit{``Amplitudes, strings  \& duality''}
and by the European Union's Horizon 2020 research and innovation programme under the Marie Sk\l{}odowska-Curie grant agreement No.~764850 {\it ``\href{https://sagex.org}{SAGEX}''}.
CW is supported by a Royal Society University Research Fellowship No.~UF160350.

\newpage

\appendix

 \section{Asymptotic behaviour of master integrals  and boundary conditions}\label{sec:TGAsy}
 
 \subsection{The general methodology}
In order to find the asymptotic behaviour \cite{Beneke:1997zp} of the master integrals, it is convenient to use Feynman  parameters. A general scalar integral with a single dimensionless kinematic parameter $x$ (e.g.~a ratio of masses and Mandelstam variables) takes the following  form
\begin{align}               
\mathcal I_{a_1,a_2,\ldots, a_m}(x)	=\int_{0}^{\infty}\! d^m\alpha \ \delta (1-\sum_{i=1}^m \alpha_i)\  \prod_{j=1}^{m}{ \alpha_j^{a_j-1} \over \Gamma(a_j)}{U^{a -{L+1\over 2} D}(\alpha)\over F^{a -{L\over 2} D}(\alpha,x)}\, ,
\end{align}
where $a=\sum_{i=1}^{m} a_i$, each $a_i$ is the power of a Feynman propagator, and $F$, $U$ are the Symanzik polynomials. The numerical evaluations of the integrals are usually performed using  sector decomposition \cite{Heinrich:2008si,Hidding:2020ytt} or Mellin-Barnes representations \cite{Gluza:2007rt}. Here we use the Cheng-Wu theorem \cite{Cheng:1987ga} to remove the delta function, as follows: 
\begin{align}
\mathcal I_{a_1,a_2,\ldots, a_i,\ldots, a_m}(x)	=\int_{0}^{\infty} d\alpha_1\wedge \cdots \widehat{d\alpha_i}\cdots \wedge d\alpha_m  \ \Big(\prod_{j=1}^{m}{ \alpha_j^{a_j-1} \over \Gamma(a_j)}{U^{a -{L+1\over 2} D}(\alpha)\over F^{a -{L\over 2} D}(\alpha,x)}\Big)\Big|_{a_i\rightarrow 1}\, , 
\end{align}
and the hat denotes the omission of the corresponding integration. 
In general,  the integral can be expanded as%
\footnote{See \cite{Semenova:2018cwy} for a recent review. }
 \begin{align}
 \label{rre} 
 	\mathcal {I}(x)\xrightarrow{x\rightarrow 0} \sum_{R} x^{w_{R}} \left(\mathcal{I}_{\rm LD}^{(R)}+\cdots \right)\, , 
 \end{align}
 where the summation is over different regions $R$, and \text{LD} stands for  leading term. 
 The parameters  $w_{R}$  depend on the number of dimensions and the difference of any two $w_{R}$ is never  an integer. In the literature, each non-vanishing $\mathcal{I}_{\rm LD}^{(R)}$ is called a region, while  the dots in \eqref{rre}  denote higher-order terms in each region. 
 
 In principle, all the expansion coefficients can be written as an integral over the Feynman parameters. However, the closed integral form of $\mathcal{I}_{\rm LD}^{(R)}$ for a general Feynman integral is still not known. It was conjectured that all these integral forms can be obtained by re-parameterising the Feynman parameters $\alpha_i\rightarrow \alpha'_i(\alpha_1,\cdots, \alpha_m, x)$ and then extracting the common factor $x^{w_{R}}$ as 
 \begin{align}
 	\mathcal I(x)=x^{w_{R}} \int\!d^{m-1}\alpha' \ \mathcal R(\alpha', x)\, . 
 \end{align}
 Then for the region $R$  we have 
 \begin{align}
 \mathcal{I}_{\rm LD}^{(R)}=\int\!d^{m-1}\alpha' \ \mathcal R(\alpha',0) \, .
 \end{align}
 In practice, the method introduced in \cite{Pak:2010pt,Semenova:2018cwy} is more efficient, and we now  review it.  
The key assumption in  \cite{Pak:2010pt}  is that the leading terms in the  power expansion of the integral in the parameter $x$ can be obtained by rescaling the Feynman parameters by some powers of $x$ and expanding the integrand directly.%
 \footnote{For a recent discussion on this point see \cite{Semenova:2018cwy}.}  
In particular, the monomials in the $U$ and $F$ functions have the form $x^{r_0}\prod_{i=1}^{m}\alpha_{i}^{r_i}$. We then rescale  $\alpha_i$ as $\alpha_i\, x^{v_i}$, and  denote by $U^\prime$ and $F^\prime$ the rescaled  functions, dropping all subleading terms in $x$. Furthermore, we introduce vectors containing as components the powers of  the monomials in the $U'$ and $F'$ functions 
  \begin{align}
\mathbf{r}^{(i_1)},\,\, \mathbf{r}^{(i_2)}, \,\, \cdots,\,\, \mathbf{r}^{(i_t)},\,\, \mathbf{r}^{(j_1)}, \,\, \mathbf{r}^{(j_2)},\,\, \cdots, \,\, \mathbf{r}^{(j_s)}\, , \end{align}
 where  $t$ and $s$ denote the number of monomials in $U'$ and $F'$. More concretely, each $\mathbf{r}$ is a vector whose components are the degrees of the monomials with respect to the variables 
\begin{align}
x, \,\, \alpha_1, \,\, \alpha_2, \,\, \cdots, \,\, \widehat\alpha_i, \,\, \cdots, \,\, \alpha_m\, , 
\end{align} 
where the hat indicate omission of the corresponding variable. 
 We then scan  all  possible leading  terms (in $x$)  by choosing different monomials  in the original $U$ and $F$ functions. For each choice of leading term, the new  $U'$ and $F'$  are represented as sums over the following monomials from the original $U$ and $F$ functions:
\begin{align}
	\mathsf m^{(i_1)}_U, \mathsf m^{(i_2)}_U, \cdots, \mathsf m^{(i_t)}_U,  \mathsf m^{(j_1)}_F, \mathsf m^{(j_2)}_F, \cdots, \mathsf m^{(i_s)}_F \ .
\end{align}  
As the  terms in the $U'$ or $F'$ functions are of the same degree in $x$ after the rescaling, we have the constraint equations for the rescaling degrees 
\begin{align}\label{eq:vEq}
\begin{split}
	\mathbf{v}\cdot (\mathbf{r}^{(i_h)}-\mathbf{r}^{(i_{h-1})})&=0,  \qquad \forall h\in[2,t]\, , \\
	\mathbf{v}\cdot (\mathbf{r}^{(j_h)}-\mathbf{r}^{(j_{h-1})})&=0, \qquad \forall h\in[2,s]\, ,
\end{split}
\end{align}
where $\mathbf{v}$ is the  vector  $\{1,v_1,\cdots,\hat v_i, \cdots, v_m\}$ made of the powers of the rescalings  by  $x$, i.e.~the rescaled variables are 
\begin{align}
x^1, \,\, x^{v_1}\alpha_1, \,\, x^{v_2}\alpha_2, \,\, \cdots, \,\, \widehat\alpha_i, \,\, \cdots, \,\, x^{v_m}\alpha_m\,.
\end{align}
If there is no solution to the constraint equations, then such a region does not exist. Indeed, if the constraint equations do not fix the values of $\mathbf{v}$, then there exists some rescaling freedom such that the leading integral $\mathcal I_{\rm LD}$ has the scaling 
\begin{align}
	\mathcal I_{\rm LD}\xrightarrow{\alpha_i\rightarrow x^{v_i}\alpha_i}x^{w} \,  \mathcal I_{\rm LD} \ , 
\end{align}
where $w$ is  non-zero. In this case the  leading integral always vanishes in dimensional regularisation.
If the constraint equations fix the scaling vector $\mathbf{v}$, 
 then using the solution for $\mathbf{v}$ from \eqref{eq:vEq}  it is easy to check  that  all the other monomials in the original $U$ and $F$ functions are of higher degree in $x$ after rescaling. 
 
 We scan over all the $U'$ or $F'$ functions, and solve the constraint equations \eqref{eq:vEq} for the monomials in the  $U'$ or $F'$ function. 
Finally we are able to find all the regions: 
 \begin{align}
 	\mathcal {I}(x)\xrightarrow{x\rightarrow 0} \sum_{R} x^{w_R}\,\mathcal{I}_{\rm LD}^{(R)}\, ,
 \end{align}
 where, again, the difference of any two $w^{(R)}$ is never an integer.

\subsection{One-loop example}
As an example, we consider the one-loop box integral
\begin{align} \label{eq:box}
	 \mathcal I_{C}^{(1,4)}=\int \!\frac{d^D \ell_1}{\pi^{D/ 2}} \ \frac{1}{ \ell _1^2 \left(p_{2}+p_{3}-\ell _1\right){}^2 (2p_1\cdot \ell _1) (-2v_2\cdot  \ell _1)} \, .
\end{align}
After using Feynman parametrisation, we arrive at 
\begin{align}
	 \mathcal I_{C}^{(1,4)}= m_1^{ \left(D-7\right)} \Gamma \left(\frac{8{-}D}{2}\right) \int_{0}^{\infty}\!d^{3}\alpha\left(\alpha _3+1\right)^{4-D}\left(\alpha _1^2+\alpha _2^2-2 \alpha _2 \alpha _1 y'-2 \alpha _1 x+2 \alpha _3 x\right){}^{\frac{D-8}{2}},	 \end{align}
	 where $y'={p_1\Cdot v_2 \over m_1}$ and $x=-{p_2\Cdot p_3\over m_1^2}$.   
	 The power index vectors for the monomials 
	 \begin{align}
	 	1&&\alpha _3&&\alpha _1^2&&\alpha _2^2&&\alpha _2 \alpha _1&&\alpha _1 x&&\alpha _3 x
	 \end{align}
	 are given by 
\begin{align}
	\left(
\begin{array}{cc|ccccc}
\mathbf r^{(1)}&\mathbf r^{(2)}&\mathbf r^{(3)}&\mathbf r^{(4)}&\mathbf r^{(5)}&\mathbf r^{(6)}&\mathbf r^{(7)}\\
 0 & 0 &  0 & 0 & 0 & 1 & 1  \\
 0 & 0 &  2 & 0 & 1 & 1 & 0  \\
 0 & 0 & 0 & 2 & 1 & 0 & 0  \\
 0 & 1 &  0 & 0 & 0 & 0 & 1 
\end{array}
\right).
\end{align}
Then we scan over all the possible leading terms for the $U$ or $F$ function. For example if we choose the leading terms to be 
$1, \alpha _3 $ and $\alpha _1^2, \alpha _2^2,\alpha _2 \alpha _1$, then by \eqref{eq:vEq} the constraint equations for the rescaling degrees are 
\begin{align}
	v_3=0, && v_1=v_2,&& 2v_2=v_1+v_2.
\end{align}
These equations do not determine the rescaling degrees and hence the region with these leading terms does not exist. 

If we choose the leading terms to be $1, \alpha _3 $ and $\alpha _1^2, \alpha _2^2,\alpha _2 \alpha _1,x \alpha _3$  for the $U$ and $F$ functions, respectively, then the constraint equations are 
\begin{align}\label{eq:rescalingV}
	v_3=0, && v_1=v_2,&& 2v_2=v_1+v_2,&& 1+v_3=v_1+v_2.
\end{align}
In this case the solution exists, which is given by 
\begin{align}
	\mathbf v=\{1,{1\over 2},{1\over 2},0\}\, .
\end{align}
 It is easy to check that the chosen monomials are the leading terms under this rescaling. 

After scanning over all the possible leading terms in  the  $U$ and $F$ functions, we find that the only nontrivial solution for the rescaling  degrees  is \eqref{eq:rescalingV}. Therefore we get, for  the asymptotic behaviour,
\begin{align}
	 \mathcal I_{C}^{(1,4)}&\xrightarrow{x\rightarrow 0}\Big(\int_{0}^{\infty}d^3\alpha\, m_1^{\left(D-7\right)}\,  \Gamma \left(4-\frac{D}{2}\right) \left(\alpha _3+1\right){}^{4-D}\left(-2 \alpha _2 \alpha _1 y'+\alpha _1^2+\alpha _2^2+2 \alpha _3\right){}^{\frac{D}{2}-4}\Big)\nn\\
	 \times& \, x^{\frac{D-6}{2}}\big[1+\mathcal{O}(x)\big].
\end{align}
There is only one region for the box integral $\mathcal I_{C}^{(1,4)}$ defined in \eqref{eq:box}.

\subsection{Two-loop examples}\label{sec:BDV2}

In this section we discuss the asymptotic behaviour of the two-loop master integrals. 

 In order to evaluate our two-loop HEFT amplitude, only the results quoted in this Section are required. 
In particular, our two-loop master integrals contain both linear propagators raised to arbitrary powers and delta functions. These are implemented  by the following replacements:
\begin{align}
	{i\over \mathcal{D}^n}={i^{n-1}\over \Gamma(n)}\int_{0}^{\infty}\!d\alpha  \ \alpha^{n-1} e^{i p \alpha}, && \delta^{(n-1)}(p)={i^{n-1}\over 2\pi} \int_{-\infty}^{\infty}\!d\alpha \ \alpha^{n-1} \, e^{i p \alpha},
\end{align}
so that in  Feynman  parameterisation, the parameter corresponding to a cut propagator is  integrated over the domain  $(-\infty, \infty)$. Such integrals   can usually be transformed into contour  integrals and evaluated as  a sum  of residues.

As a first example we consider  the master integral $\mathcal{I}_1=G_{0,1,0,1,1,0,0,1,1}$. 
Its Feynman-parameterized form is
 \begin{align}
	&\mathcal{I}_1=4^{5-D}\Gamma (5-D) \left(-{q^2\over 2}\right)^{D-4}\int_{0}^{\infty} d\alpha _2d\alpha _4d\alpha _5  \int_{-\infty}^{\infty} d\alpha _8d\alpha _9\ \delta \left(\alpha _2{+}\alpha _4{+}\alpha _5{+}\alpha _8{+}\alpha _9{-}1\right) \nn\\
	&\left(\alpha _2 \alpha _4+\alpha _5 \alpha _4+\alpha _2 \alpha _5\right){}^{5-\frac{3 D}{2}} \left(\alpha _2 \alpha _8^2+\alpha _5 \alpha _8^2+\alpha _2 \alpha _9^2+\alpha _4 \alpha _9^2+8 \alpha _2 \alpha _4 \alpha _5-2 \alpha _2 \alpha _9 \alpha _8 y\right){}^{D-5}.
\end{align}
The integration over $\alpha _8,\alpha _9$ is evaluated directly, leading to
\begin{align}
	\mathcal{I}_1=\int_{\Delta} d\alpha _2d\alpha _4d\alpha _5\frac{  2^{D-2}\pi\left(-{q^2\over 2}\right)^{D-4} \Gamma (5{-}D) \left(\alpha _2 \alpha _4 \alpha _5\right){}^{D-4}\left(\alpha _4 \alpha _5+\alpha _2 \left(\alpha _4+\alpha _5\right)\right){}^{5-\frac{3 D}{2}} }{(4-D) \sqrt{\left(\alpha _4+\alpha _5\right) \alpha _2+\alpha _4 \alpha _5+\alpha _2^2 \left(1-y^2\right)}},
\end{align}
where we define the integral $\int_{\Delta}$ as 
\begin{align}
	\int_{\Delta} d\alpha _{i_1}d\alpha _{i_2}\cdots d\alpha _{i_r}:=\int_{0}^{\infty}d\alpha _{i_1}d\alpha _{i_2}\cdots d\alpha _{i_r}\ \delta(1-\sum_{j=1}^{r}\alpha_{i_j}).
\end{align}
In the static limit $y\rightarrow 1$, this master integral has one region up to  $\mathcal{O}(\sqrt{y-1})$   from the rescaling 
\begin{align}
(R_1):&&	\alpha_{2}\rightarrow \alpha_{2}(y-1)^0, \qquad \alpha_{4}\rightarrow \alpha_{4} (y-1)^0,
\qquad \alpha_{5}\rightarrow \alpha_{5}(y-1)^0.
\end{align}
Then the leading order integral in this region is 
\begin{align}
	\mathcal{I}_{1,\rm{LD}}^{(R_1)}&=\int_{\Delta} d\alpha _2d\alpha _4d\alpha _5\frac{2^{D-2}\pi \left(-{q^2\over 2}\right)^{D-4} \Gamma (5{-}D) \left(\alpha _2 \alpha _4 \alpha _5\right){}^{D-4}\left(\alpha _4 \alpha _5{+}\alpha _2 \left(\alpha _4{+}\alpha _5\right)\right){}^{5-\frac{3 D}{2}} }{(4-D) \sqrt{\left(\alpha _4+\alpha _5\right) \alpha _2+\alpha _4 \alpha _5}}\nn\\
	&=-\frac{4 \pi ^{5/2} \left(-{q^2\over 2}\right)^{D-4} \csc (\pi  D) \Gamma \left(\frac{D-3}{2}\right)^2}{\Gamma \left(\frac{3 (D-3)}{2}\right) \Gamma \left(\frac{D-2}{2}\right)}.
\end{align}
Hence we have $\mathcal{I}_1|_{y\rightarrow 1}\rightarrow  \mathcal{I}_{1,\rm{LD}}^{(R_1)}.$

As a second example, we consider the master integral $\mathcal{I}_2=-{q^2\over 2}G_{0,2,0,1,1,0,0,1,1}$. After performing some  integrations,  similarly to what was done earlier,  
we obtain 
\begin{align}
	\mathcal{I}_{2}&=\int_{\Delta} d\alpha _2d\alpha _4d\alpha _5\frac{\pi  \alpha _2^{D-4} (\alpha _4 \alpha _5)^{D-5} \left(\alpha _4 \alpha _5{+}\alpha _2 \left(\alpha _4{+}\alpha _5\right)\right){}^{6-\frac{3 D}{2}} (-{q^2\over 2})^{D-4} \Gamma (5{-}D)}{2^{3-D} \sqrt{\left(\alpha _4+\alpha _5\right) \alpha _2+\alpha _4 \alpha _5+\alpha _2^2 \left(1-y^2\right)}}.
\end{align}
In the static limit $y\rightarrow 1$, this master integral has two regions up  to  $\mathcal{O}(\sqrt{y-1})$ from the rescalings 
\begin{align}
\begin{split}
(R_1):\quad &	\alpha_{2}\rightarrow \alpha_{2}(y-1)^0, \qquad \alpha_{4}\rightarrow \alpha_{4} (y-1)^0, \qquad \alpha_{5}\rightarrow \alpha_{5}(y-1)^0, \\
(R_2):\quad &	\alpha_{2}\rightarrow \alpha_{2}(y-1)^{0}, \qquad \alpha_{4}\rightarrow \alpha_{4} (y-1),
\qquad \ \, \alpha_{5}\rightarrow \alpha_{5}(y-1).
\end{split}
\end{align}
Then the leading-order integrals in these two regions are 
\begin{align}
\begin{split}
	\mathcal{I}_{2,\text{LD}}^{(R_1)}&=\int_{\Delta} d\alpha _2d\alpha _4d\alpha _5\pi  {\alpha _2^{D-4}\over 2^{3-D}} (\alpha _4 \alpha _5)^{D-5} \left(\alpha _4 \alpha _5{+}\alpha _2 \left(\alpha _4{+}\alpha _5\right)\right){}^{\frac{11-3 D}{2}} \left({-}{q^2\over 2}\right)^{D-4} \Gamma (5{-}D)\\
	&=\frac{4 \pi ^{5/2} \left(-{q^2\over 2}\right)^{D-4} \csc (\pi  D) \Gamma \left(\frac{D-5}{2}\right) \Gamma \left(\frac{D-3}{2}\right)}{\Gamma \left(\frac{D}{2}-2\right) \Gamma \left(\frac{3 D}{2}-\frac{11}{2}\right)} \\
	\mathcal{I}_{2,\text{LD}}^{(R_2)}&=\int_{\Delta} d\alpha _2d\alpha _4d\alpha _5\frac{\pi  \alpha _2^{2-\frac{D}{2}} (\alpha _4 \alpha _5)^{D-5} \left(\alpha _4{+}\alpha _5\right){}^{6-\frac{3 D}{2}} \left(-{q^2\over 2}\right)^{D-4} (y{-}1)^{\frac{D}{2}-\frac{5}{2}} \Gamma (5{-}D)}{2^{3-D} \sqrt{\alpha _2 \left(\alpha _4+\alpha _5\right)-2 \alpha _2^2} } \\
	&=-\frac{i \pi ^2 2^{\frac{7}{2}-\frac{D}{2}} e^{\frac{i \pi  D}{2}} \left(-{q^2\over 2}\right)^{D-4} (y-1)^{\frac{D-5}{2}} \csc (\pi  D) \Gamma \left(\frac{5}{2}-\frac{D}{2}\right) \Gamma \left(\frac{D}{2}-2\right)}{\Gamma \left(D-\frac{7}{2}\right)}\, .
\end{split}
\end{align}
Thus, we find for $\mathcal{I}_2$  the following asymptotic behaviour,
\begin{align}
	\mathcal{I}_2|_{y\rightarrow 1}\rightarrow  \mathcal{I}_{2,\text{LD}}^{(R_1)}+\mathcal{I}_{2,\text{LD}}^{(R_2)}+\mathcal{O}(\sqrt{y-1}).
\end{align}
The master integral $\mathcal{I}_3$ can be expressed as 
\begin{align}
	\mathcal{I}_3=\frac{2 \left(2 (D-5) \bar{\mathcal{I}}_3-(D-4) \left(y^2+1\right) \mathcal{I}_2\right)-3 (D-4)^2 \left(y^2-1\right) \mathcal{I}_1}{2 y}, 
\end{align}
where $\bar{\mathcal{I}}_3=\left(-{q^2\over 2}\right)G_{0,1,0,2,1,0,0,1,1}.$
We only need to derive  the asymptotic behaviour for~$\bar{\mathcal{I}}_3$ 
\begin{align}
   \bar{\mathcal{I}}_3= \int_{\Delta} d\alpha _2d\alpha _4d\alpha _5 \frac{4 \pi  \left(\alpha _2 \alpha _4 \alpha _5\right){}^D \left(\alpha _4 \alpha _5+\alpha _2 \left(\alpha _4+\alpha _5\right)\right){}^{6-\frac{3 D}{2}} \left(-q^2\right)^{D-4} \Gamma (5-D)}{\alpha _2^5 \alpha _4^4 \alpha _5^5 \sqrt{\left(\alpha _4+\alpha _5\right) \alpha _2+\alpha _4 \alpha _5-\alpha _2^2 \left(y^2-1\right)}}\, . 
\end{align}
There are two regions for this master integral
\begin{align}
\begin{split}
(R_1):\quad &	\alpha_{2}\rightarrow \alpha_{2}(y-1)^0, \qquad \alpha_{4}\rightarrow \alpha_{4} (y-1)^0, \qquad \alpha_{5}\rightarrow \alpha_{5}(y-1)^0, \\
(R_2):\quad &	\alpha_{2}\rightarrow \alpha_{2}(y-1)^{0}, \qquad \alpha_{4}\rightarrow \alpha_{4} (y-1),
\qquad \ \, \alpha_{5}\rightarrow \alpha_{5}(y-1).
\end{split}
\end{align}
For the first region, we have 
\begin{align}
    \bar{\mathcal{I}}^{( R_1)}_{3,\rm{LD}}&=\frac{2 \pi ^{3/2} \left(-{q^2\over 2}\right)^{D-4} \Gamma (5-D) \Gamma \left(\frac{D-5}{2}\right) \Gamma \left(\frac{D-3}{2}\right) \Gamma (D-3)}{\Gamma \left(\frac{D}{2}-1\right) \Gamma \left(\frac{3 D}{2}-\frac{11}{2}\right)}\, ,
\end{align}
while for the second 
\begin{align}
   \bar{\mathcal{I}}^{( R_2)}_{3,\rm{LD}}= \frac{\pi ^2 i^{D+1} 2^{\frac{1}{2}-\frac{3 D}{2}} \left(-q^2\right)^{D-4} (y-1)^{\frac{D-3}{2}} \csc (\pi  D) \Gamma \left(\frac{3}{2}-\frac{D}{2}\right) \Gamma \left(\frac{D}{2}-1\right)}{\Gamma \left(D-\frac{7}{2}\right)}\, .
\end{align}
We can then derive  the asymptotic behaviour for the master integral
\begin{align}
\begin{split}
	 \bar{\mathcal{I}}_3|_{y\rightarrow 1}&\rightarrow  \bar{\mathcal{I}}^{( R_1)}_{3,\rm{LD}}+ \bar{\mathcal{I}}^{( R_2)}_{3,\rm{LD}}+\mathcal{O}(y-1) \, .
\end{split}
\end{align}
The Feynman  parameterisation of the master integral is 
\begin{align}
   \mathcal{I}_4= -\int_{\Delta} d\alpha _2d\alpha _3d\alpha _5\frac{4 \pi  \left(\alpha _2 \alpha _3 \alpha _5\right){}^D \left(\alpha _2 \left(\alpha _3+\alpha _5\right)\right){}^{5-\frac{3 D}{2}} \left(-q^2\right)^{D-4} \Gamma (5-D)}{\alpha _2^4 \alpha _3^4 \alpha _5^4 (D-4) \sqrt{\alpha _2 \left(\alpha _2+\alpha _3+\alpha _5-\alpha _2 y^2\right)}}\, .
\end{align}
The master integral has only one region 
\begin{align}
\begin{split}
(R_2):\quad &	\alpha_{2}\rightarrow \alpha_{2}(y-1)^{0}, \qquad \alpha_{4}\rightarrow \alpha_{4} (y-1),
\qquad \ \, \alpha_{5}\rightarrow \alpha_{5}(y-1)\, .
\end{split}
\end{align}
In this region, we have 
\begin{align}
 \mathcal{I}^{(R_2)}_{4,\rm{LD}}=-\frac{\pi ^2 \left(-{q^2\over 2}\right)^{D-4} \left(2-2y\right)^{\frac{D-3}{2}} \csc (\pi  D) \Gamma \left(\frac{3}{2}-\frac{D}{2}\right) \Gamma \left(\frac{D}{2}-1\right)}{2^{D-5} \Gamma \left(D-\frac{5}{2}\right)}\, .
\end{align}
Then the asymptotic behaviour for $\mathcal{I}_4$ is  
\begin{align}
\begin{split}
	\mathcal{I}_4|_{y\rightarrow 1}&\rightarrow\mathcal{I}^{(R_2)}_{4,\rm{LD}}+\mathcal{O}(y-1)\, .
\end{split}
\end{align}

Summarising, for all of our master integrals we find a total of  six regions:
\begin{align}
    \mathcal{I}^{(R_1)}_{1,\rm{LD}}&&\mathcal{I}^{(R_1)}_{2,\rm{LD}}&&\mathcal{I}^{(R_2)}_{2,\rm{LD}}&&\mathcal{I}^{(R_1)}_{3,\rm{LD}}&&\mathcal{I}^{(R_2)}_{3,\rm{LD}},&&\mathcal{I}^{(R_2)}_{4,\rm{LD}}.
\end{align}
Other master integrals vanish in  the static limit. 

According to the canonical basis in \eqref{7.16} and the  differential equation in \eqref{eq:diffEq}, the regions $\mathcal{I}^{(R_2)}_{4,\rm{LD}}$ and $\mathcal{I}^{(R_2)}_{2,\rm{LD}}$ are  related.  Moreover the regions $\mathcal{I}^{(R_1)}_{1,\rm{LD}}$,  $\mathcal{I}^{(R_1)}_{2,\rm{LD}}$ and $\mathcal{I}^{(R_3)}_{2,\rm{LD}}$ are vanishing in the static limit. Hence only two regions 
remain:
\begin{align}
   \mathcal{I}^{(R_2)}_{2,\rm{LD}},&&\mathcal{I}^{(R_1)}_{3,\rm{LD}},
\end{align}
where $\mathcal{I}^{(R_1)}_{3,\rm{LD}}$ is the potential region and the other one is the radiation reaction region. 

The relevant master integral for  the probe limit  can be evaluated directly as it does not depend on $y$,  
\begin{align}
\begin{split}
	G^{(2)}&=-\frac{4 \pi ^{5/2}({-q^2\over 2})^{D-4} \csc (\pi  D) \Gamma \left(\frac{D-3}{2}\right)^2}{\Gamma \left(\frac{3 (D-3)}{2}\right) \Gamma \left(\frac{D-2}{2}\right)}\xrightarrow{\epsilon\rightarrow 0}\frac{4\pi ^2  \left(-q^2\right)^{-2 \epsilon }}{\epsilon }.
\end{split}
\end{align}

\newpage

\section{Integrand of the 2MPI two-loop HEFT amplitude}\label{sec:Integrand2}

In this appendix we list the integrands for the four 2MPI cut  graphs of the two-loop HEFT amplitude in \eqref{z+f}. The integrand basis is given
by the following 71 monomials according to the definitions \eqref{eq:DefG} and \eqref{table}:%
\footnote{Since here we are describing integrands, the integrations implied in \eqref{eq:DefG} are suppressed.} 
\begin{align*}
 G_1&=G_{-1,1,-1,1,1,2,2,1,1} & G_2&=G_{-1,1,0,1,1,0,2,1,1} & G_3&=G_{-1,1,1,-1,1,0,4,1,1} \\
 G_4&=G_{-1,1,1,0,1,-1,3,1,1} & G_5&=G_{-1,1,1,0,1,0,2,1,1} & G_6&=G_{-1,1,1,1,1,-2,2,1,1} \\
 G_7&=G_{-1,1,1,1,1,0,0,1,1} & G_8&=G_{0,1,-1,1,1,2,0,1,1} & G_9&=G_{0,1,0,1,1,-1,1,1,1} \\
 G_{10}&=G_{0,1,0,1,1,0,0,1,1} & G_{11}&=G_{0,1,0,1,1,1,-1,1,1} & G_{12}&=G_{0,1,0,1,1,1,1,1,1} \\
 G_{13}&=G_{0,1,1,-1,1,-1,3,1,1} & G_{14}&=G_{0,1,1,-1,1,0,2,1,1} & G_{15}&=G_{0,1,1,0,1,-2,2,1,1} \\
 G_{16}&=G_{0,1,1,0,1,-1,1,1,1} & G_{17}&=G_{0,1,1,0,1,0,0,1,1} & G_{18}&=G_{0,1,1,0,1,0,2,1,1} \\
 G_{19}&=G_{0,1,1,1,1,-3,1,1,1} & G_{20}&=G_{0,1,1,1,1,-2,0,1,1} & G_{21}&=G_{0,1,1,1,1,-1,-1,1,1} \\
 G_{22}&=G_{0,1,1,1,1,-1,1,1,1} & G_{23}&=G_{0,1,1,1,1,0,-2,1,1} & G_{24}&=G_{0,1,1,1,1,0,0,1,1} \\
 G_{25}&=G_{1,1,-1,1,-1,4,0,1,1} & G_{26}&=G_{1,1,-1,1,0,2,0,1,1} & G_{27}&=G_{1,1,-1,1,0,3,-1,1,1} \\
 G_{28}&=G_{1,1,-1,1,1,0,0,1,1} & G_{29}&=G_{1,1,-1,1,1,2,-2,1,1} & G_{30}&=G_{1,1,0,1,-1,2,0,1,1} \\
 G_{31}&=G_{1,1,0,1,-1,3,-1,1,1} & G_{32}&=G_{1,1,0,1,0,0,0,1,1} & G_{33}&=G_{1,1,0,1,0,1,-1,1,1} \\
 G_{34}&=G_{1,1,0,1,0,2,-2,1,1} & G_{35}&=G_{1,1,0,1,0,2,0,1,1} & G_{36}&=G_{1,1,0,1,1,-2,0,1,1} \\
 G_{37}&=G_{1,1,0,1,1,-1,-1,1,1} & G_{38}&=G_{1,1,0,1,1,0,-2,1,1} & G_{39}&=G_{1,1,0,1,1,0,0,1,1} \\
 G_{40}&=G_{1,1,0,1,1,1,-3,1,1} & G_{41}&=G_{1,1,0,1,1,1,-1,1,1} & G_{42}&=G_{1,1,1,-1,-1,2,2,1,1} \\
 G_{43}&=G_{1,1,1,-1,0,0,2,1,1} & G_{44}&=G_{1,1,1,-1,1,-2,2,1,1} & G_{45}&=G_{1,1,1,-1,1,0,0,1,1} \\
 G_{46}&=G_{1,1,1,0,-1,2,0,1,1} & G_{47}&=G_{1,1,1,0,0,-1,1,1,1} & G_{48}&=G_{1,1,1,0,0,0,0,1,1} \\
 G_{49}&=G_{1,1,1,0,0,1,-1,1,1} & G_{50}&=G_{1,1,1,0,0,1,1,1,1} & G_{51}&=G_{1,1,1,0,1,-3,1,1,1} \\
 G_{52}&=G_{1,1,1,0,1,-2,0,1,1} & G_{53}&=G_{1,1,1,0,1,-1,-1,1,1} & G_{54}&=G_{1,1,1,0,1,-1,1,1,1} \\
 G_{55}&=G_{1,1,1,0,1,0,-2,1,1} & G_{56}&=G_{1,1,1,0,1,0,0,1,1} & G_{57}&=G_{1,1,1,1,-1,0,0,1,1} \\
 G_{58}&=G_{1,1,1,1,-1,2,-2,1,1} & G_{59}&=G_{1,1,1,1,0,-2,0,1,1} & G_{60}&=G_{1,1,1,1,0,-1,-1,1,1} \\
 G_{61}&=G_{1,1,1,1,0,0,-2,1,1} & G_{62}&=G_{1,1,1,1,0,0,0,1,1} & G_{63}&=G_{1,1,1,1,0,1,-3,1,1} \\
 G_{64}&=G_{1,1,1,1,0,1,-1,1,1} & G_{65}&=G_{1,1,1,1,1,-4,0,1,1} & G_{66}&=G_{1,1,1,1,1,-2,-2,1,1} \\
 G_{67}&=G_{1,1,1,1,1,-2,0,1,1} & G_{68}&=G_{1,1,1,1,1,-1,-1,1,1} & G_{69}&=G_{1,1,1,1,1,0,-4,1,1} \\
 G_{70}&=G_{1,1,1,1,1,0,-2,1,1} & G_{71}&=G_{1,1,1,1,1,0,0,1,1} &  
\end{align*}

The integrands for the four different cut diagrams  in \eqref{z+f} are
\begin{align}
\sum_{i=1}^{71}e^{(a)}_i {G}_i,  &&\sum_{i=1}^{71}e^{(b)}_i {G}_i,&&\sum_{i=1}^{71}e^{(c)}_i {G}_i,&&\sum_{i=1}^{71}e^{(d)}_i {G}_i
\end{align}
where the coefficients  are given in $D$ dimensions by
\begin{align*}
\begin{array}{cccc}
 e^{(a)} &e^{(b)}& e^{(c)} & e^{(d)} \\
 0 & \frac{\left[(D-2) y^2-1\right]^3}{16 (D-2)^3} & 0 & 0 \\
 0 & \frac{-(D-2)^3 y^4+(D-2) (D-1) y^2-1}{4 (D-2)^3} & 0 & 0 \\
 0 & 0 & 0 & \frac{(D-3) \left[(D-2) y^2-1\right]^2}{16 (D-2)^3} \\
 0 & 0 & 0 & \frac{(D-3) y \left[(D-2) y^2-1\right]}{4 (D-2)^2} \\
 0 & 0 & 0 & -\frac{(D-3) \left[(D-2) y^2-1\right]}{4 (D-2)^3} \\
 0 & \frac{(D-3) \left[(D-2) y^2-1\right]}{4 (D-2)^2} & 0 & \frac{(D-3) \left[(D-2) y^2-1\right]}{4 (D-2)^2} \\
 0 & \frac{1-(D-2) y^2}{2 (D-2)^2} & 0 & \frac{1-(D-2) y^2}{2 (D-2)^2} \\
 0 & \frac{-(D-2)^3 y^4+(D-2) (D-1) y^2-1}{4 (D-2)^3} & 0 & 0 \\
 0 & -\frac{y^3}{2}-\frac{(D-4) y}{2 (D-2)^2} & 0 & 0 \\
 0 & \frac{1}{2} \left(y^2+\frac{D-4}{(D-2)^3}\right) & 0 & 0 \\
 0 & -\frac{y^3}{2}-\frac{(D-4) y}{2 (D-2)^2} & 0 & 0 \\
 0 & \frac{y \left[(D-2) y^2-1\right]^2 \left(-{q^2\over 2}\right)}{2 (D-2)^2} & 0 & 0 \\
 0 & 0 & 0 & \frac{(D-3) y \left[(D-2) y^2-1\right]}{4 (D-2)^2} \\
 0 & 0 & 0 & -\frac{(D-3) \left[(D-2) y^2-1\right]}{4 (D-2)^3} \\
 0 & 0 & 0 & \frac{(D-3) y^2}{D-2} \\
 0 & 0 & 0 & -\frac{2 (D-3) y}{(D-2)^2} \\
 0 & 0 & 0 & \frac{1}{2} \left(y^2+\frac{D-4}{(D-2)^3}\right) \\
 0 & 0 & 0 & -\frac{\left[(D-2) y^2-1\right]^2 \left(-{q^2\over 2}\right)}{2 (D-2)^2} \\
 0 & \frac{(D-3) y}{D-2} & 0 & \frac{(D-3) y}{D-2} \\
 0 & \frac{3-D}{(D-2)^2} & 0 & \frac{3-D}{(D-2)^2} \\
 0 & \frac{(D-4) y}{D-2} & 0 & \frac{(D-4) y}{D-2} \\
 ~~~~~~~~~~0~~~~~~~~~~ & y \left(\frac{1}{D-2}-y^2\right) \left(-{q^2\over 2}\right)& ~~~~~~~~~~0~~~~~~~~~~ & y \left(\frac{1}{D-2}-y^2\right) \left(-{q^2\over 2}\right) \\
\end{array}
 \end{align*}
 \begin{align*}
\begin{array}{cccc}
 0 & \frac{1}{(D-2)^2} & 0 & \frac{1}{(D-2)^2} \\
 0 & \frac{\left[(D-2) y^2-1\right] \left(-{q^2\over 2}\right)}{(D-2)^2} & 0 & \frac{\left[(D-2) y^2-1\right] \left(-{q^2\over 2}\right)}{(D-2)^2} \\
 0 & 0 & \frac{(D-3) \left[(D-2) y^2-1\right]^2}{16 (D-2)^3} & 0 \\
 0 & 0 & -\frac{(D-3) \left[(D-2) y^2-1\right]}{4 (D-2)^3} & 0 \\
 0 & 0 & \frac{(D-3) y \left[(D-2) y^2-1\right]}{4 (D-2)^2} & 0 \\
 0 & \frac{1-(D-2) y^2}{2 (D-2)^2} & \frac{1-(D-2) y^2}{2 (D-2)^2} & 0 \\
 0 & \frac{(D-3) \left[(D-2) y^2-1\right]}{4 (D-2)^2} & \frac{(D-3) \left[(D-2) y^2-1\right]}{4 (D-2)^2} & 0 \\
 0 & 0 & -\frac{(D-3) \left[(D-2) y^2-1\right]}{4 (D-2)^3} & 0 \\
 0 & 0 & \frac{(D-3) y \left[(D-2) y^2-1\right]}{4 (D-2)^2} & 0 \\
 0 & 0 & \frac{1}{2} \left(y^2+\frac{D-4}{(D-2)^3}\right) & 0 \\
 0 & 0 & -\frac{2 (D-3) y}{(D-2)^2} & 0 \\
 0 & 0 & \frac{(D-3) y^2}{D-2} & 0 \\
 0 & 0 & -\frac{\left[(D-2) y^2-1\right]^2 \left(-{q^2\over 2}\right)}{2 (D-2)^2} & 0 \\
 0 & \frac{1}{(D-2)^2} & \frac{1}{(D-2)^2} & 0 \\
 0 & \frac{(D-4) y}{D-2} & \frac{(D-4) y}{D-2} & 0 \\
 0 & \frac{3-D}{(D-2)^2} & \frac{3-D}{(D-2)^2} & 0 \\
 0 & \frac{\left[(D-2) y^2-1\right] \left(-{q^2\over 2}\right)}{(D-2)^2} & \frac{\left[(D-2) y^2-1\right] \left(-{q^2\over 2}\right)}{(D-2)^2} & 0 \\
 0 & \frac{(D-3) y}{D-2} & \frac{(D-3) y}{D-2} & 0 \\
 0 & y \left(\frac{1}{D-2}-y^2\right) \left(-{q^2\over 2}\right) & y \left(\frac{1}{D-2}-y^2\right) \left(-{q^2\over 2}\right) & 0 \\
 \frac{\left[(D-2) y^2-1\right]^3}{16 (D-2)^3} & 0 & 0 & 0 \\
 \frac{-(D-2)^3 y^4+(D-2) (D-1) y^2-1}{4 (D-2)^3} & 0 & 0 & 0 \\
 \frac{(D-3) \left[(D-2) y^2-1\right]}{4 (D-2)^2} & 0 & 0 & \frac{(D-3) \left[(D-2) y^2-1\right]}{4 (D-2)^2} \\
 \frac{1-(D-2) y^2}{2 (D-2)^2} & 0 & 0 & \frac{1-(D-2) y^2}{2 (D-2)^2} \\
 \frac{-(D-2)^3 y^4+(D-2) (D-1) y^2-1}{4 (D-2)^3} & 0 & 0 & 0 \\
 -\frac{y^3}{2}-\frac{(D-4) y}{2 (D-2)^2} & 0 & 0 & 0 \\
 \frac{1}{2} \left(y^2+\frac{D-4}{(D-2)^3}\right) & 0 & 0 & 0 \\
 -\frac{y^3}{2}-\frac{(D-4) y}{2 (D-2)^2} & 0 & 0 & 0 \\
 \frac{y \left[(D-2) y^2-1\right]^2 \left(-{q^2\over 2}\right)}{2 (D-2)^2} & 0 & 0 & 0 \\
\end{array}
 \end{align*}
\begin{align*}
\begin{array}{cccc}
 \frac{(D-3) y}{D-2} & 0 & 0 & \frac{(D-3) y}{D-2} \\
 \frac{3-D}{(D-2)^2} & 0 & 0 & \frac{3-D}{(D-2)^2} \\
 \frac{(D-4) y}{D-2} & 0 & 0 & \frac{(D-4) y}{D-2} \\
 y \left(\frac{1}{D-2}-y^2\right) \left(-{q^2\over 2}\right)& 0 & 0 & y \left(\frac{1}{D-2}-y^2\right) \left(-{q^2\over 2}\right) \\
 \frac{1}{(D-2)^2} & 0 & 0 & \frac{1}{(D-2)^2} \\
 \frac{\left[(D-2) y^2-1\right] \left(-{q^2\over 2}\right)}{(D-2)^2} & 0 & 0 & \frac{\left[(D-2) y^2-1\right] \left(-{q^2\over 2}\right)}{(D-2)^2} \\
 \frac{1-(D-2) y^2}{2 (D-2)^2} & 0 & \frac{1-(D-2) y^2}{2 (D-2)^2} & 0 \\
 \frac{(D-3) \left[(D-2) y^2-1\right]}{4 (D-2)^2} & 0 & \frac{(D-3) \left[(D-2) y^2-1\right]}{4 (D-2)^2} & 0 \\
 \frac{1}{(D-2)^2} & 0 & \frac{1}{(D-2)^2} & 0 \\
 \frac{(D-4) y}{D-2} & 0 & \frac{(D-4) y}{D-2} & 0 \\
 \frac{3-D}{(D-2)^2} & 0 & \frac{3-D}{(D-2)^2} & 0 \\
 \frac{\left[(D-2) y^2-1\right] \left(-{q^2\over 2}\right)}{(D-2)^2} & 0 & \frac{\left[(D-2) y^2-1\right] \left(-{q^2\over 2}\right)}{(D-2)^2} & 0 \\
 \frac{(D-3) y}{D-2} & 0 & \frac{(D-3) y}{D-2} & 0 \\
 y \left(\frac{1}{D-2}-y^2\right) \left(-{q^2\over 2}\right) & 0 & y \left(\frac{1}{D-2}-y^2\right) \left(-{q^2\over 2}\right) & 0 \\
 \frac{D-3}{D-2} & \frac{D-3}{D-2} & \frac{D-3}{D-2} & \frac{D-3}{D-2} \\
 2-\frac{4}{D-2} & 2-\frac{4}{D-2} & 2-\frac{4}{D-2} & 2-\frac{4}{D-2} \\
 -2 y^2 \left(-{q^2\over 2}\right) & -2 y^2 \left(-{q^2\over 2}\right) & -2 y^2 \left(-{q^2\over 2}\right) & -2 y^2 \left(-{q^2\over 2}\right) \\
 \frac{4 y \left(-{q^2\over 2}\right)}{D-2} & \frac{4 y \left(-{q^2\over 2}\right)}{D-2} & \frac{4 y \left(-{q^2\over 2}\right)}{D-2} & \frac{4 y \left(-{q^2\over 2}\right)}{D-2} \\
 \frac{D-3}{D-2} & \frac{D-3}{D-2} & \frac{D-3}{D-2} & \frac{D-3}{D-2} \\
 -2 y^2 \left(-{q^2\over 2}\right)& -2 y^2 \left(-{q^2\over 2}\right) & -2 y^2 \left(-{q^2\over 2}\right) & -2 y^2 \left(-{q^2\over 2}\right) \\
 \frac{\left[(D-2) y^2-1\right]^2 \left(-{q^2\over 2}\right)^2}{(D-2)^2} & \frac{\left[(D-2) y^2-1\right]^2 \left(-{q^2\over 2}\right)^2}{(D-2)^2} & \frac{\left[(D-2) y^2-1\right]^2 \left(-{q^2\over 2}\right)^2}{(D-2)^2} & \frac{\left[(D-2) y^2-1\right]^2 \left(-{q^2\over 2}\right)^2}{(D-2)^2} \\
\end{array}
\end{align*}

 \newpage

\bibliographystyle{JHEP}
\bibliography{ScatEq}

\end{document}